\documentclass[12pt,a4paper]{article} 
\pdfoutput=1
\usepackage{style}
\usepackage{rotating}
\usepackage{lscape}
\usepackage{bbold}
\usepackage{multirow}
\usepackage{mathrsfs}
\usepackage{amsmath}
\usepackage{amssymb}
\usepackage{amsthm}
\usepackage{psfrag}
\usepackage{graphicx}
\usepackage{hyperref}
\usepackage{feynmp}
\usepackage{color}
\usepackage{bm}

\newcommand{\be}{\begin{equation}}
\newcommand{\ee}{\end{equation}}
\newcommand{\beqa}{\begin{eqnarray}}
\newcommand{\eeqa}{\end{eqnarray}}

\renewcommand\O{\Omega}
\newcommand\g{\gamma}
\newcommand\D{\Delta}

\newcommand\s{\sigma}
\renewcommand\t{\theta}
\renewcommand\a{\alpha}
\renewcommand\b{\beta}
\renewcommand\l{\lambda}
\newcommand{\T}{\Theta}

\newcommand{\HH}{{\cal H}}

\newcommand\x{{\bf x}}
\renewcommand\k{{\bf k}}
\newcommand\q{{\bf q}}

\renewcommand{\P}{\mathcal{P}}

\def\e{{\rm e}}
\def\d{\partial}
\newcommand{\bseq}{\begin{subequations}}
\newcommand{\eseq}{\end{subequations}}
\newcommand{\ctg}{\mathop{\rm ctg}\nolimits}
\newcommand{\tg}{\mathop{\rm tg}\nolimits}

\newcommand{\ch}{\mathop{\rm ch}\nolimits}
\newcommand{\sh}{\mathop{\rm sh}\nolimits}
\renewcommand{\ln}{\mathop{\rm ln}\nolimits}

\newcommand{\Tr}{{\rm Tr}}

\renewcommand{\D}{\mathcal{D}}

\renewcommand{\L}{\Lambda}

\renewcommand{\k}{{\bf k}}

\newcommand\vk{\varkappa}

\newcommand{\QQ}{\mathring{Q}}

\def\Xint#1{\mathchoice
      {\XXint\displaystyle\textstyle{#1}}%
      {\XXint\textstyle\scriptstyle{#1}}%
      {\XXint\scriptstyle\scriptscriptstyle{#1}}%
      {\XXint\scriptscriptstyle\scriptscriptstyle{#1}}%
      \!\int}
   \def\XXint#1#2#3{{\setbox0=\hbox{$#1{#2#3}{\int}$}
        \vcenter{\hbox{$#2#3$}}\kern-.5\wd0}}
   
   \def\dashint{\Xint-}

\relax

\title{
Non-perturbative 
probability 
distribution function for cosmological 
counts in cells 
}

\author[a,b,c]{Mikhail M. Ivanov\footnote{\texttt{ivanov@ias.edu}}}
\author[a]{Alexander A. Kaurov\footnote{\texttt{kaurov@ias.edu}}}
\author[b,d,c]{Sergey Sibiryakov\footnote{\texttt{sergey.sibiryakov@cern.ch}}}

\affiliation[a]{School of Natural Sciences, Institute for Advanced Study, \\
1 Einstein Drive, Princeton, NJ 08540, United States}
\affiliation[b]{Institute of Physics, Laboratory of Particle Physics and Cosmology (LPPC), 
\'Ecole Polytechnique F\'ed\'erale de Lausanne (EPFL), \normalsize\it CH-1015, Lausanne, Switzerland}
\affiliation[c]{Institute for Nuclear Research of the
Russian Academy of Sciences, \\ 
\normalsize \it  60th October Anniversary Prospect, 7a, 117312
Moscow, Russia}
\affiliation[d]{Theory Department, CERN, \\
1 Esplanade des Particules, CH-1211 Gen\`eve 23, Switzerland}

\abstract{
We present a non-perturbative calculation of the 1-point probability 
distribution function (PDF) for the spherically-averaged 
matter density field. 
The PDF is represented as a path integral and 
is evaluated using the saddle-point
method. It factorizes into an exponent given by a
spherically symmetric saddle-point solution and a prefactor produced
by fluctuations. The exponent encodes the leading
sensitivity of the PDF    
to the dynamics of gravitational clustering 
and statistics of the initial conditions.
In contrast, the prefactor has only a weak dependence on
cosmology. 
It splits into a monopole contribution which is evaluated
exactly, and a factor corresponding to aspherical fluctuations. 
The latter is crucial for the consistency of the
calculation: neglecting it would make the PDF incompatible with
translational invariance. We compute the aspherical prefactor using a
combination of analytic and numerical techniques. We demonstrate 
the factorization of spurious enhanced contributions of large bulk
flows and their cancellation due to the equivalence
principle. We also identify the sensitivity 
to the short-scale physics and argue that it must be
properly renormalized. The uncertainty associated with the
renormalization procedure gives an estimate of the theoretical
error. For zero redshift, the precision varies from sub percent for
moderate density contrasts to tens of percent at the
tails of the distribution. It improves at higher redshifts. 
We compare our
results with N-body simulation data and find an excellent agreement.
}

\begin{document}

\begin{flushright}
CERN-TH-2018-242\\
INR-TH-2018-027
\end{flushright}

\maketitle

\section{Introduction}
\label{sec:intro}

Current and planned cosmological surveys are going to map the
large-scale structure (LSS)  
of the universe with
unprecedented precision at a wide range of scales and redshifts. These data will
potentially carry a wealth of information on cosmological parameters,
the initial conditions of the universe, the properties of dark matter
and dark energy. Extracting this information requires accurate
quantitative understanding of matter clustering in the non-linear regime,
both in the standard $\L$CDM cosmology, as well as its extensions.

The direct approach relies on numerical N-body simulations that have
made an impressive progress in the last decades. However, reaching the
required level of accuracy still remains computationally 
expensive~\cite{Schneider:2015yka}. Moreover, while the N-body
methods have been well adapted to the $\L$CDM cosmology, their
modification to include the effects of new physics is often extremely
demanding. This calls for development of the analytic approaches to
LSS. Being perhaps less powerful than N-body simulations in the
description of the $\L$CDM cosmology, the analytic approach provides 
more flexibility in going beyond it and a deeper insight in the
relevance of different physical processes. Hence, analytic
and N-body 
methods are complementary to each other.

The most developed analytic approach to LSS is the cosmological perturbation
theory, where the evolution equations for the density and velocity fields
are solved iteratively treating the density contrast as a small
quantity. The correlation functions of
cosmological observables are then evaluated by averaging over the initial
conditions \cite{Bernardeau:2001qr}. 
An intensive research in this direction in recent years has clarified
various physical effects.
The developments include understanding the role of the equivalence principle
in the cancellation of the 
so-called `IR-divergences'
\cite{Scoccimarro:1995if,Kehagias:2013yd,Peloso:2013zw,Blas:2013bpa,
Creminelli:2013mca,Horn:2014rta,Blas:2015qsi}, accurate treatment of
the effects of large 
bulk flows on baryon acoustic oscillations 
\cite{Senatore:2014via,Baldauf:2015xfa,Blas:2016sfa,Ivanov:2018gjr}
and systematic accounting for the contribution of non-linear density
inhomogeneities at short scales along the lines of effective
field theory (EFT) 
\cite{Baumann:2010tm,Carrasco:2012cv,Pajer:2013jj,Baldauf:2014qfa,
Abolhasani:2015mra,Baldauf:2016sjb}. As a result of this progress a
sub-percent-level precision has been achieved 
in perturbative calculation of the matter
power spectrum and bispectrum for comoving wavenumbers\footnote{Here
  $h\approx 0.7$ is defined through the value of the present-day Hubble
  parameter, 
$$H_0=h\cdot 100\frac{\rm km}{{\rm s}\cdot{\rm Mpc}}\;.$$ The
precision cited above refers to the quantities at zero
redshift, $z=0$. At higher redshifts relevant for actual surveys the
precision is further improved and the range of wavenumbers accessible
to perturbative methods increases.} $k\lesssim 0.1
h/\rm{ Mpc}$. 

In this paper we show that the analytic approach can be rigorously extended
beyond perturbation theory. The non-perturbative observable that we 
are going to consider is counts-in-cells statistics (see
e.g.~\cite{Peebles:1980}). 

The counts-in-cells method amounts to splitting the cosmic density field 
into cells in position space and taking an aggregate 
of this field 
inside each cell. 
In the case of discrete tracers one counts the number of objects 
inside each cell.
The distribution of cells over the relevant variable
reveals statistical properties of the underlying field.
In this paper we discuss the 
1-point probability distribution function (PDF) of finding a certain
average matter density
in a sphere of a given fixed radius $r_*$.
The deviation of this spherically-averaged density from the mean
density of the universe does not need to be small, and thus the desired PDF
cannot be calculated within perturbation theory.

Formally, the count-in-cells statistics include
information from all n-point functions of the density field 
in a compressed way which facilitates measurements,
but looses the information encoded in the shape dependence of the n-point
correlators. 
Therefore, it is complementary to perturbative methods in the information
content.

The counts-in-cells statistics are one of the classic observables 
in LSS. The distribution of galaxies in 2-dimensional angular cells on the sky
was first measured  by
E.~Hubble \cite{Hubble:1934me}, who noticed that it 
is close to log-normal. For the total matter
density this has been recently tested in \cite{Clerkin:2016kyr,Gruen:2017xjj}. 
The log-normal distribution was also suggested as a model for the
1-point PDF in the case of three-dimensional cells \cite{Coles:1991if}
and has been quite successful in describing both N-body simulations 
\cite{Kofman:1993mx,Kayo:2001gu} and observational data
\cite{Wild:2004me,Hurtado-Gil:2017dbm}. 
However, as pointed out in \cite{Bernardeau:1994zd,Bernardeau:1994aq},
this success appears to be accidental and is due to the specific shape
of the power spectrum at mildly non-linear scales. 
Recent
high-accuracy N-body simulations performed in \cite{Klypin:2017jjg}
revealed significant deviations of the measured PDF from the
log-normal fit.

Pioneering  
calculations of the counts-in-cells
PDF from first principles 
were performed in Refs.~\cite{Bernardeau:1992zw,Bernardeau:1994zd}
using insights from perturbation theory. This study was extended
beyond perturbation theory in
Refs.~\cite{Valageas:1997sg,Valageas:1998xr,Valageas:2001zr,Valageas:2001td}, where it
was argued  
that the most probable dynamics producing a given overdensity in a
spherical cell respects the symmetry of the problem, i.e. it is given
by a spherical collapse. 
Recently, these calculations were revisited in the context of
the Large Deviation Principle (LDP)~\cite{Bernardeau:2015khs}.
In particular, Ref.~\cite{Uhlemann:2015npz} introduced the 
logarithmic density transformation to avoid certain problems
associated with the application of LDP directly to the density PDF
\cite{Bernardeau:2013dua}. 
This formalism has been applied to joint PDF of densities
in two cells \cite{Bernardeau:1995ty,Bernardeau:2015tha,Uhlemann:2016liz}
and to biased tracers \cite{Uhlemann:2017hgi}. 
An alternative approach to the counts-in-cells statistics
developed in \cite{Sheth:1998ew,BetancortRijo:2001ge,Lam:2007qw}
 is based on the
Lagrangian-space description of LSS.
Ref.~\cite{Pajer:2017ulp} recently derived 
1-point PDF in a toy model of $(1+1)$ dimensional universe. 
Counts-in-cells
statistics were suggested as promising probes of primordial
non-Gaussianity \cite{Matarrese:2000iz,Uhlemann:2017tex} and as a
suitable tool to analyze the future 21 cm intensity mapping 
data~\cite{Leicht:2018ryx}.

In this paper we pursue the path-integral approach to counts-in-cells
pioneered in \cite{Matarrese:2000iz,Valageas:2001zr,Valageas:2001td}. In this approach
the calculation of the 1-point PDF closely resembles a calculation of instanton
effects in quantum field theory (QFT). 
Following Ref.~\cite{Blas:2015qsi} we introduce a formal parameter
characterizing the overall amplitude of the matter power spectrum and
argue that it plays a role of the coupling constant in the
theory. When the coupling is small, the path integral defining the 
1-point PDF can be
evaluated in the saddle-point (`semiclassical') approximation. Thereby
the PDF factorizes into the exponential part given by the leading
saddle-point configuration and a prefactor coming from integration
over small fluctuations around the saddle-point solution. We confirm
the assertion \cite{Valageas:2001zr,Valageas:2001td} that the saddle-point
configuration corresponds to the spherically symmetric dynamics. In
this way we recover the well-known result 
\cite{Valageas:2001zr,BetancortRijo:2001ge,
Lam:2007qw,Bernardeau:2015khs,Uhlemann:2015npz}
for the leading exponential part of the PDF. 
Our key result is computation of 
the prefactor due to aspherical
perturbations around the spherical collapse which has not been done in
the previous works. 
We demonstrate that this `aspherical prefactor' 
is crucial for the consistency of the saddle-point calculation. In
particular, it is required to ensure that the mean
value of the density contrast vanishes. 

In the QFT analogy, evaluation of the aspherical prefactor amounts to
a 1-loop computation in a non-trivial background. As such, it is
instructive in several respects. First, it shows how the vanishing of
the mean density contrast is related to the translational invariance of
the theory,  spontaneously broken by the position of the cell. Second,
the sector of dipole perturbations exhibits `IR divergences' at
intermediate steps of the calculation associated to large bulk
flows. We show that the equivalence principle ensures cancellation of
these divergences. We devise a procedure to isolate the IR-enhanced
contributions and cancel them analytically, prior to any numerical
evaluation. Finally, the contributions of high multipoles are
sensitive to short-distance dynamics and must be
renormalized. Unfortunately, it is impossible to
unambiguously fix the renormalization procedure from first principles.
We
isolate the `UV-divergent' part of the prefactor and consider two
models for its renormalization, differing by the dependence of the
corresponding counterterm on the density contrast. Both models use as
input the value of the counterterm for the 1-loop power spectrum, and
thus do not introduce any new fitting parameters. We suggest to use
the difference between the two models as an estimate of the
theoretical uncertainty introduced by renormalization. This uncertainty is less
than percent in the range of moderate cell densities, 
$\rho_{\rm cell}/\rho_{\rm univ}\in[0.5,2]$, where $\rho_{\rm univ}$
is the average density 
of the universe, and degrades to 30\% for extreme values $\rho_{\rm
  cell}/\rho_{\rm univ}=0.1$ or $\rho_{\rm cell}/\rho_{\rm univ}=10$ at $z=0$. 

To verify our approach we ran a suite of N-body
simulations\footnote{The details of the simulations are described in
  Appendix~\ref{app:Nbody}.}  
using 
the \texttt{FastPM} code
\cite{Feng:2016yqz}. The numerical studies are performed for the
following cosmology: a flat $\L$CDM with $\Omega_m = 0.26$,
$\Omega_b=0.044$, $h = 0.72$, $n_s=0.96$, Gaussian initial conditions,
$\sigma_8=0.794$. This is the same choice as in
Ref.~\cite{Uhlemann:2016liz}
which used the counts-in-cells distribution extracted from the Horizon
run 4 simulation \cite{Kim:2011ab};
it facilitates a direct comparison between our results and those 
of~\cite{Uhlemann:2016liz}. Throughout the paper the 
linear power spectrum is computed with the Boltzmann code
\texttt{CLASS}~\cite{Blas:2011rf}.  

The predictions of our method are found to be in complete agreement
with the results of N-body simulations. First, the 1-point PDF
clearly exhibits the
semiclassical scaling. The 
aspherical prefactor extracted from the N-body data shows a very weak
dependence on redshift or the radius of the cell, as predicted by
theory. Second, the data fall inside the range spanned by our
theoretical uncertainty. Remarkably, one of the counterterm models
matches the data within the accuracy of the simulations
throughout the whole range of
available densities, $\rho_{\rm cell}/\rho_{\rm univ}\in[0.1,10]$, at all
redshifts and for different cell radii.   

The paper is organized as follows. 
In Sec.~\ref{sec:spherical} we introduce the path integral
representation of the 1-point PDF,
identify its saddle point and demonstrate the factorization of the
PDF into the leading exponent and prefactor. We evaluate the
leading exponential part. 
In Sec.~\ref{sec:prefactor} we evaluate explicitly the prefactor due
to spherically symmetric perturbations and discuss the general
properties of the aspherical prefactor. We compare the theoretical
expectations with the prefactor extracted from the N-body data and provide
simple fitting formulas for it. The rest of the paper is devoted to
the calculation of the aspherical prefactor from first principles. 
In Sec.~\ref{sec:aspt} we compute the aspherical prefactor 
at small values of the density
contrast using
perturbation theory.  
In Sec.~\ref{sec:mainEq} we derive the set of equations describing the
prefactor in the non-perturbative regime of large density contrasts
and present an algorithm for its numerical evaluation. 
In Sec.~\ref{sec:dipole} we modify the algorithm for the sector of
dipole perturbations in order to explicitly factor out and cancel the
IR-enhanced contributions. 
In Sec.~\ref{sec:WKB} we compute the contributions of high multipoles
using the
Wentzel--Kramers--Brillouin (WKB) approximation.
In Sec.~\ref{sec:results} we present our numerical 
results for the aspherical prefactor, discuss the 
contribution of short-distance physics and its renormalization. 
Section~\ref{sec:comp} contains a summary of our results and
discussion.

Several appendices contain supplementary material. 
Appendix~\ref{sec:notations} summarizes our conventions. 
Appendix~\ref{app:Nbody} is devoted to the details of our N-body simulations.
In Appendix~\ref{sec:SC} we review the dynamics of spherical collapse in
Einstein--de Sitter (EdS) and $\L$CDM universes. In
Appendix~\ref{app:determinant} we derive a useful formula for the
determinant of matrices of a special form.
Appendix~\ref{sec:LCDM} contains equations for the aspherical
prefactor in $\L$CDM cosmology. Some technical aspects of the WKB
calculation of the high-multipole contributions are discussed in 
Appendix~\ref{app:WKB}.
Appendix~\ref{sec:aspy} contains details of our numerical
procedure.
In Appendix~\ref{app:ln} we comment on the log-normal
model for the counts-in-cells statistics.

\section{Path integral for counts-in-cells PDF}
\label{sec:spherical}

\subsection{Spherical collapse saddle point}
\label{sec:2.1}

Consider the density contrast
averaged over a spherical cell of radius $r_*$,
\be
\label{eq:wf}
\bar \delta_W =  \int \frac{d^3x}{r_*^3} \,\tilde{W}(r/r_*)\, 
\delta(\x)=\int_\k W(kr_*) \delta(\k)\,,
\ee
where $\delta({\bf x})\equiv \frac{\delta\rho({\bf x})}{\rho_{\rm
    univ}}$, $\tilde 
W(r/r_*)$ is a window function, $W(kr_*)$ is its
Fourier transform, and we have introduced the notation
$ \int_\k \equiv \int \frac{d^3k}{(2\pi)^3}$. We will soon specify the
window function to be top-hat in the position space, which is the
standard choice for counts-in-cells statistics. 
However, it is instructive to see how far one can proceed
without making any specific assumptions about $\tilde W$, apart from
it being spherically symmetric. The window function is normalized as
\be 
\label{wnorm}
\int \frac{d^3x}{r_*^3} \,\tilde{W}(r/r_*)=1\,.
\ee
We are interested in the 1-point PDF $\P(\delta_*)$ describing the
probability that the random variable $\bar\delta_W$ takes a given
value $\delta_*$.
Due to translational invariance, the 1-point
statistics do not depend on the position of the cell. 
Thus, without loss of generality we center the cell at 
the
origin, $\x=0$. 

We assume that the initial conditions for the density perturbations at
some large redshift $z_i$ are adiabatic and Gaussian, so that their statistical
properties are fully determined by the 2-point cumulant,
\be 
\langle\delta_{i}(\k)\delta_{i}(\k')\rangle = 
(2\pi)^3\delta^{(3)}_{\rm D}(\k+\k')\, g^2(z_i)\,P(k)\,,
\ee
where $\delta_{\rm D}^{(3)}$ is the 3-dimensional Dirac
delta-function. Here 
$P(k)$ is the \textit{linear} power spectrum at redshift zero and
$g(z)$ is the linear growth factor\footnote{The growth factor is
  commonly denoted by $D(z)$ in the LSS literature. We prefer the
  notation $g(z)$ to emphasize the analogy with a coupling constant in
QFT.}.
The latter is normalized to be 1 at $z=0$. Nevertheless, it is
convenient to keep $g^2$ explicitly in the formulas and treat it as a
small free parameter. The rationale behind this approach is to use
$g^2$ as a book-keeping parameter that characterizes the overall
amplitude of the power spectrum and thereby controls the saddle-point
evaluation of the PDF, just like a coupling constant
controls the semiclassical expansion in QFT (cf.~\cite{Blas:2015qsi}). The
true physical expansion parameter in our case is the smoothed density
variance at the scale $r_*$, as will become clear shortly.

Instead of working directly with the initial density field $\delta_i$,
it is customary to rescale it to redshift $z$ using the linear growth
factor, 
\be
\delta_L(\k,z)  = \frac{g(z)}{g(z_i)}\delta_i(\k)\,.
\ee
We will refer to $\delta_L$ as the `linear density field' in what
follows and will omit the explicit $z$-dependence to simplify notations.

The desired PDF is given by the following path integral
\cite{Valageas:2001zr,Matarrese:2000iz},
\be 
\label{eq:pdf2}
\mathcal{P}(\delta_*)={\cal N}^{-1}
\int\mathscr{ D}\delta_L 
\exp\left\{-\int_\k\frac{|\delta_L(\k)|^2}{2 g^2P(k)}\right\} 
\,\delta^{(1)}_{\rm D}\big(\delta_* - \bar \delta_W[\delta_L]\big)\,,
\ee
where different linear density perturbations are weighted with the
appropriate Gaussian weight. The Dirac delta-function ensures that
only the configurations that produce the average density contrast
$\delta_*$ are retained in the integration. Note that we have written
$\bar\delta_W$ as a functional of the linear density field,
$\bar\delta_W[\delta_L]$. In general, this functional is complicated
and its evaluation requires knowing non-linear dynamics that map
initial linear perturbations onto the final non-linear density field
$\delta(\x)$.   
The normalization factor in \eqref{eq:pdf2} is 
\be 
\mathcal{N}=
\int \mathscr{D}\delta_L 
\exp\bigg\{-\int_\k\frac{|\delta_L(\k)|^2}{2g^2P(k)}\bigg\}\,.
\ee

It is convenient to rewrite the delta-function constraint 
using the inverse Laplace transform, 
\be 
\label{eq:pdfLaplace}
\mathcal{P}(\delta_*)={\cal N}^{-1}
\int_{-i\infty}^{i\infty}\frac{d\l}{2\pi i g^2}
\int \mathscr{ D}\delta_L \exp\bigg\{-\frac{1}{g^2}\bigg[
\int_\k\frac{|\delta_L(\k)|^2}{2P(k)}-\l\big(
\delta_* - \bar \delta_W[\delta_L]\big)\bigg]\bigg\}\,.
\ee 
where we introduced the Lagrange multiplier $\lambda$.
Our goal is to compute the above integral by the steepest-decent method. 
We expect the result to take the form,
\be
\label{eq:scaling}
\mathcal{P}(\delta_*)= \exp\left\{-\frac{1}{g^2}\left(\a_0+\a_1 g^2 + \a_2 g^4+...\right)\right\} \,.
\ee
The leading term $\alpha_0$ corresponds to the exponent of the integrand in \eqref{eq:pdfLaplace} evaluated on the saddle-point configuration.
The first correction
$\a_1 g^2$ stems from the Gaussian integral around the saddle point.
It gives rise to a $g$-independent prefactor\footnote{In fact, we will see that $\alpha_1$ also has a term $\sim \ln g$ which introduces an overall factor $1/g$ in the PDF.} in the PDF.
As we discuss below, the evaluation of $\a_1$ corresponds to a one-loop 
calculation in the saddle-point background.
Higher loops give further corrections $\alpha_2 g^4$ etc., which can be rewritten as $O(g^2)$ corrections to the prefactor.
We will not consider them in this paper.


We are looking for a saddle point of the integral \eqref{eq:pdfLaplace} 
in the limit
$g^2\to 0$.   
Taking variations of the expression in the exponent w.r.t. $\delta_L$
and $\l$, we obtain the equations for the saddle-point
configuration\footnote{We write the variational derivatives w.r.t. the
linear density field as an ordinary partial derivative
$\d/\d\delta_L(\k)$ to avoid proliferation of deltas.},   
\begin{subequations}
\label{eq:sp}
\begin{align}
\label{sp1}
& \frac{\delta_L(\k)}{P(k)}+\lambda \frac{\d
  \bar\delta_W}{\d \delta_L(\k)}=0\,,\\ 
\label{sp2}
& \bar\delta_W[\delta_L]=\delta_* \,.
\end{align} 
\end{subequations}
Now comes a crucial observation: a spherically symmetric Ansatz for
$\delta_L(\k)$ goes through these equations. Let us prove this. The
check is non-trivial only for Eq.~(\ref{sp1}). Clearly, if the linear field is
spherically symmetric, the first term in (\ref{sp1}) depends only on
the absolute value $k$ of the momentum. We need to show that this is
also the case for the second term. To this end, expand the variational
derivative,
\be
\label{eq:varder}
\frac{\d\bar \delta_W}{\d\delta_L(\k)}=\int\frac{d^3x}{r_*^3}
\,\tilde W(r/r_*)\,\frac{\d\delta(\x)}{\d\delta_L(\k)}\;.
\ee
Due to rotational invariance of dynamics, the derivative
$\d\delta(\x)/\d\delta_L(\k)$, evaluated on a spherically symmetric
linear density configuration, is a rotationally invariant function of
the vectors $\x$ and $\k$. Thus, it depends only on the lengths $x$,
$k$ and the scalar product $(\k\x)$. Upon integration with a
spherically symmetric window function $\tilde W$, only the dependence
on the absolute value of the momentum $k$ survives. This completes the
proof.    

The previous observation greatly simplifies the solution of the
saddle-point equations (\ref{eq:sp}). It implies that we can search for
the saddle point among spherically symmetric configurations. For such
configurations there exists a simple mapping between the linear and
non-linear density fields prior to shell-crossing, see
Appendix~\ref{sec:SC}. This mapping relates the non-linear density
contrast averaged over a cell of radius $r$,
\be 
\label{deltabar}
\bar \delta(r) \equiv  \frac{3}{r^3}\int_0^r dr_1\,r_1^2\, \delta(r_1)\;,
\ee
with the linear averaged density
\be
\label{deltaLbar}
\bar\delta_L(R)\equiv\frac{3}{R^3}\int_0^R dR_1\,R_1^2\,\delta_L(R_1)
\ee
at the radius
\be
\label{eq:LagR}
R=r\big(1+\bar\delta(r)\big)^{1/3}\;.
\ee
In the last expression one recognizes the Lagrangian radius of the
matter shell whose Eulerian radius is $r$. The mapping then gives
$\bar\delta_L(R)$ as a function of $\bar\delta(r)$ and vice versa,
\be
\label{eq:scmap}
\bar\delta_L(R)=F\big(\bar\delta(r)\big)~~~~~\Longleftrightarrow~~~~~
\bar\delta(r)=f\big(\bar\delta_L(R)\big)\;.
\ee
Evaluation of the functions $F$ or $f$ 
requires
an inversion of an elementary analytic
function (in EdS cosmology) or solution of a first-order ordinary differential
equation (in $\L$CDM). Both operations are easily performed using
standard computer packages. 
Curiously, the mapping (\ref{eq:scmap}) is almost independent
of 
cosmology (EdS vs. $\L$CDM)\footnote{At sub-percent level, see
  Fig.~\ref{fig:F} and 
  the discussion in the next subsection.}.

The existence of the mapping (\ref{eq:scmap}) allows us to compute
the variational derivative in Eq.~(\ref{sp1}) explicitly for
spherically symmetric\footnote{To avoid confusion, let us stress that
  we do not intend to restrict the path integral (\ref{eq:pdfLaplace})
to spherical configurations. This restriction is used only to find the
saddle point.} $\delta_L(k)$. Assuming that the non-linear density
field $\delta(r)$ has not undergone shell-crossing, we transform
the expression for $\bar\delta_W$ as follows,
\be
\label{wminus1}
\begin{split}
\bar\delta_W =& \frac{4\pi}{r_*^3}\int dr\,r^2\,\tilde{W}(r/r_*)
\big(1+\delta(r)\big) - 1\\
=&\frac{4\pi}{r_*^3}\int dR\,R^2
\tilde W\Big(R\big(1+f\big(\bar\delta_L(R)\big)\big)^{-1/3}/r_*\Big)-1
\,.
\end{split}
\ee
Taking into account that 
\be
\label{eq:delLbar}
\bar\delta_L(R)=\int_\k\frac{3j_1(kR)}{kR}\,\delta_L(k)\;,
\ee
where $j_1$ is the spherical Bessel function (see
Appendix~\ref{sec:notations} for conventions), we obtain,
\be
\frac{\d\bar\delta_W}{\d\delta_L(k)}=
-\frac{4\pi}{r_*^4k}\int dR\,R^2\,
\tilde W'\big(R(1+f)^{-1/3}/r_*\big)\,
\frac{f'}{(1+f)^{4/3}}\,j_1(kR)\;,
\ee
where primes denote differentiation of the functions w.r.t. their
arguments. Here $f$ and $f'$ are functions of $\bar\delta_L(R)$ and
hence functionals of $\delta_L(k)$. Substituting this expression into
(\ref{sp1}) we obtain,
\be 
\label{dL1}
\delta_L(k)=\lambda  P(k)\frac{4\pi}{r_*^4k}
\int dR\, R^2\frac{\tilde{W}'\big(R(1+f)^{-1/3}/r_*\big)\,f'\,j_1(kR)}
{(1+f)^{4/3}}\,.
\ee
This is a non-linear integral equation for $\delta_L(k)$ which can, in
principle, be solved numerically. Together with Eq.~(\ref{sp2}) that fixes
the value of the Lagrange multiplier $\l$ through the overall
normalization of $\delta_L(k)$, they form a complete system of
equations determining the saddle-point linear density. For a generic
window function $\tilde W$ the solution of this system appears
challenging. We are now going to see that Eq.~(\ref{dL1}) gets
drastically simplified for top-hat $\tilde W$.

\subsection{Leading exponent for top-hat window function}
\label{sec:SP}

From now on we specify to the case of a top-hat window function in
position space,
\be
\label{thx}
\tilde{W}_{\rm th}(r/r_*)=\frac{3}{4\pi}
\Theta_{\rm H}\left(1-\frac{r}{r_*}\right)~~~~~ \Longleftrightarrow~~~~~
W_{\rm th}(kr_*)= \frac{3j_1(kr_*)}{kr_*}\,,
\ee
where
$\Theta_{\text{H}}$ stands for the Heaviside
theta-function. As the derivative of $\tilde W_{\rm th}$ is
proportional to the Dirac delta-function, the integral in (\ref{dL1})
localizes to $R=R_*$, where
\be
\label{Rstar}
R_*=r_*(1+\delta_*)^{1/3}\;.
\ee 
After a straightforward calculation Eq.~(\ref{dL1}) simplifies to 
\be 
\label{dL4}
\delta_L(k)=-\frac{\lambda}{C} P(k)\,W_{\rm th}(k R_*)\,
\ee
with
\be
\label{eq:G1}
C=F'(\delta_*)+\frac{\bar\delta_L(R_*)-\delta_L(R_*)}{1+\delta_*}\;.
\ee
Here $F$ is the spherical-collapse mapping function introduced in
(\ref{eq:scmap}) and in deriving (\ref{dL4}), (\ref{eq:G1}) we have
used the relation,
\[
F'(\delta_*)=\frac{1}{f'\big(\bar\delta_L(R_*)\big)}\;.
\]
One observes that (\ref{dL4}) fixes the $k$-dependence of the
saddle-point configuration. We now use Eq.~(\ref{sp2}) where we act
with the function $F$ on both sides. This yields,
\be
\bar\delta_L(R_*)=F(\delta_*)\;.
\ee
Combining it with Eqs.~(\ref{dL4}), (\ref{eq:delLbar}) gives an
equation for the Lagrange multiplier,
\be
\label{eq:lambda1}
\l=-\frac{F(\delta_*)}{\sigma^2_{R_*}}C\,,
\ee
where
\be
\label{sigmaRstar}
\sigma^2_{R_*} \equiv \int_{\k}\,P(k)\,|W_{\rm th}(kR_*)|^2\,
\ee
is
the linear density variance filtered at the scale $R_*$.
Note that it depends on $\delta_*$ through the corresponding
dependence of $R_*$, see Eq.~(\ref{Rstar}).

Substituting (\ref{eq:lambda1}) back into (\ref{dL4}) we arrive at the
final expression for the saddle-point linear density, which will be
denoted with an overhat,
\be 
\label{eq:deltahatlin}
\hat{\delta}_L(k)=\frac{F(\delta_*)}{\sigma^2_{R_*}} P(k)\,W_{\rm th}(kR_*)\,.
\ee
In Lagrangian position space the linear density reads,
\be 
\label{eq:deltahatlinprofile}
\hat{\delta}_L(R)=\frac{F(\delta_*)}{\sigma^2_{R_*}} \hat\xi(R)\,.
\ee 
where we introduced the profile function
\be 
\label{xi}
\hat\xi(R) \equiv  \frac{1}{2\pi^2} 
\int dk\,k^2\frac{\sin(kR)}{kR}W_{\rm th}(kR_*)P(k)\,.
\ee
Note that it coincides with the 2-point correlation function smeared
with the top-hat filter.
In what follows we will also need the saddle-point value of the
Lagrange multiplier. This is obtained by substituting
(\ref{eq:deltahatlin}) into (\ref{eq:G1}), (\ref{eq:lambda1}). The
result is,
\be 
\label{eq:lambda}
\hat\l=-\frac{F(\delta_*)}{\sigma^2_{R_*}}\hat C~,~~~~~~~
 \hat C(\delta_*)=F'(\delta_*)+\frac{F(\delta_*)}{1+\delta_*}
\left(1-\frac{\xi_{R_*}}{\sigma^2_{R_*}}\right)\,,
 \ee
where we have denoted $\xi_{R_*}\equiv\hat\xi(R_*)$.
Finally, substituting the saddle-point configuration into the
expression (\ref{eq:pdfLaplace}) for the PDF we obtain the leading
exponential behavior,
\be
\label{leadingexp}
\P(\delta_*)\propto \exp\left\{-\frac{F^2(\delta_*)}{2 g^2 \sigma^2_{R_*}}\right\}\,.
\ee
We observe 
that the PDF exhibits a characteristic `semiclassical'
scaling in the limit $g^2 \to 0$.

\begin{figure}[t]
\begin{center}
\includegraphics[width=0.47\textwidth]{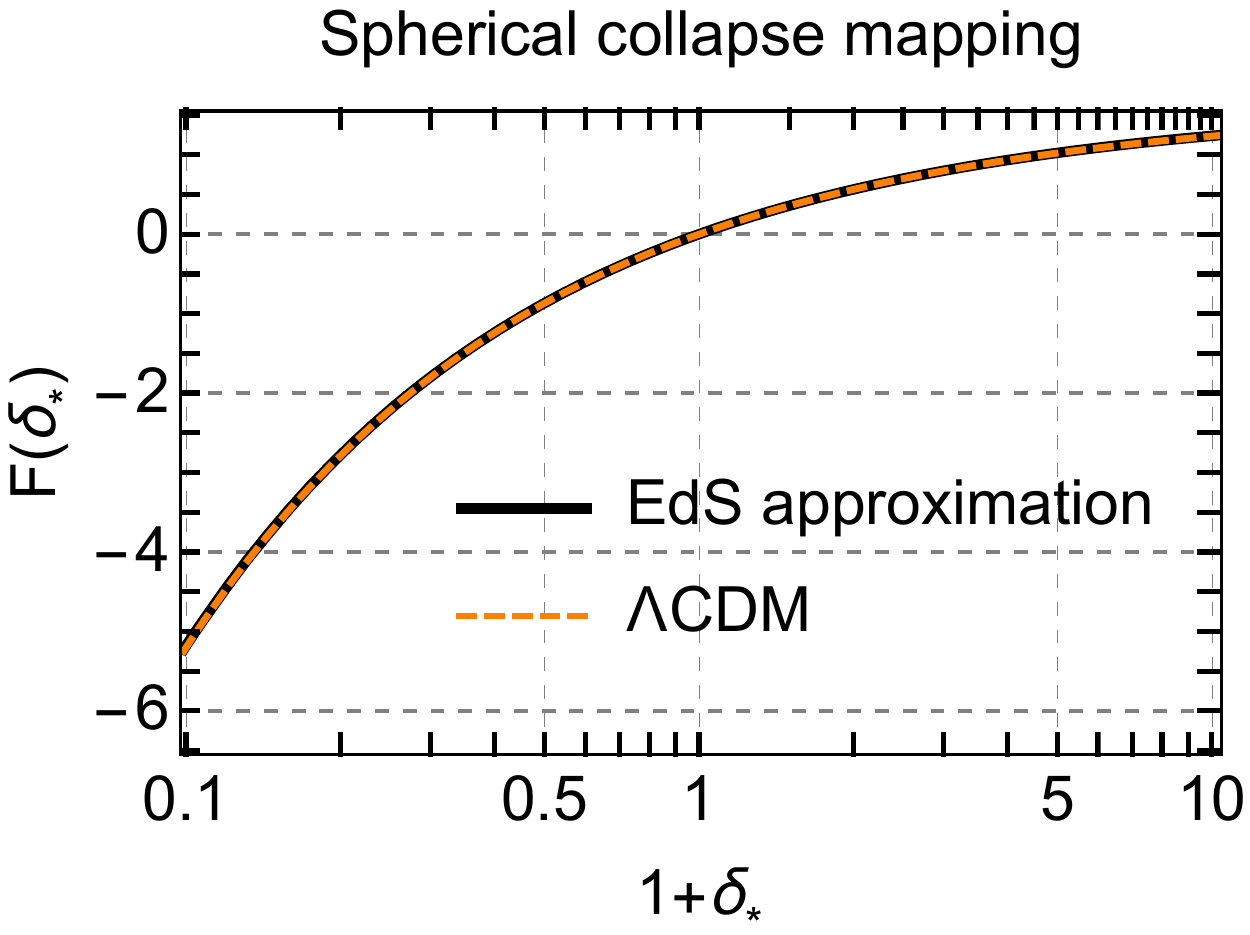}
\includegraphics[width=0.52\textwidth]{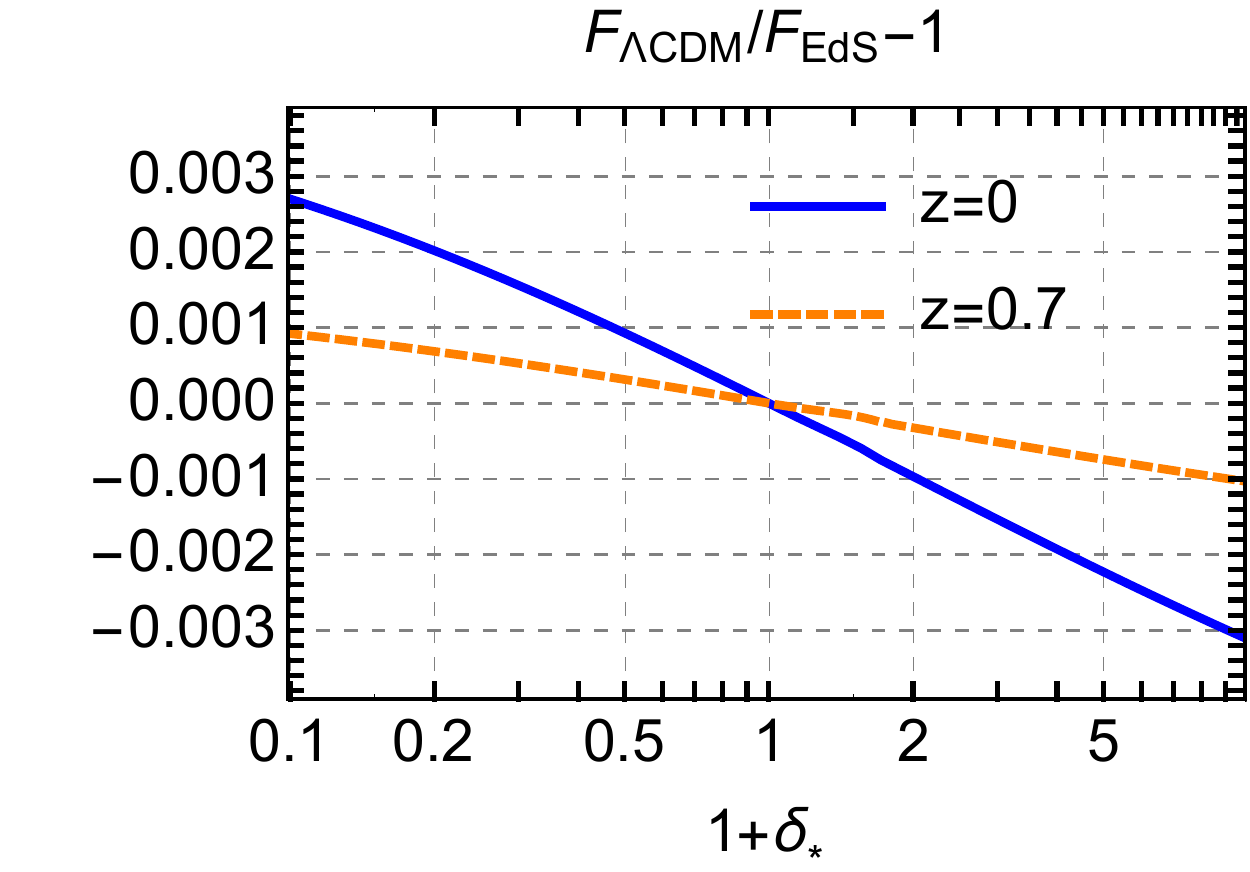}
\end{center}
\caption{\label{fig:F}
Left panel: the function $F$ mapping spherically-averaged non-linear
density contrast into its linear counterpart within the spherical
collapse dynamics. The results are shown for an EdS universe and
$\L$CDM cosmology at $z=0$. The two curves practically coincide.  
Right panel: the relative difference between $F_{\L{\rm CDM}}$ and
$F_{\rm EdS}$ at two values of the redshift.
}
\end{figure}

Let us take a closer look at the various ingredients that define the
saddle-point configuration. We start with the function
$F(\delta_*)$. It is determined exclusively by the dynamics of
spherical collapse and does not depend at all on the statistical
properties of the perturbations. We have computed it using the
procedure described in Appendix~\ref{sec:SC} for the cases of an EdS
universe 
($\Omega_m=1, \Omega_\Lambda=0$) 
and the reference $\L$CDM cosmology ($\Omega_m=0.26,
\Omega_\Lambda=0.74$). 
The results are shown in 
Fig.~\ref{fig:F}, left panel. The dependence on cosmology is very
weak, so that the curves essentially overlay. In the EdS case the
mapping is redshift-independent. Its behavior for small values of the
argument is,
\bseq
\label{scpert}
\be
\label{scpert1}
F_{\rm EdS}(\delta_*)=\delta_* -\frac{17}{21}\delta_*^2 
+ \frac{2815}{3969}\delta_*^3+{\cal O}(\delta_*^4)\;,
\ee
whereas its asymptotics at large over/underdensities are
\begin{align}
\label{scpert2}
&F_{\rm EdS} \to 1.686 &\text{at}\quad \delta_*\to \infty\,,\\
\label{scpert3}
&F_{\rm EdS}\sim -(1+\delta_*)^{-3/2} & \text{at}\quad \delta_*\to -1\,.
\end{align}
\eseq
For $\L$CDM this function has a very mild redshift
dependence illustrated in the right panel of Fig.~\ref{fig:F}, which
shows the relative difference between $F_{\L{\rm CDM}}$ and $F_{\rm
  EdS}$. This difference is maximal for $z=0$, where it reaches a few
per mil at the edges of the considered range of
$\delta_*$. However, $F$ enters in the exponent of the PDF (see
(\ref{leadingexp})) and a few per mil inaccuracy in it would generate
a few percent relative error at the tails of the PDF. 
For these reasons we will use the exact 
$\L$CDM
mapping whenever the function $F$ appears 
in the leading 
exponent. 
In all other instances the EdS approximation provides sufficient accuracy. 

\begin{figure}[t]
\begin{center}
\includegraphics[width=0.50\textwidth]{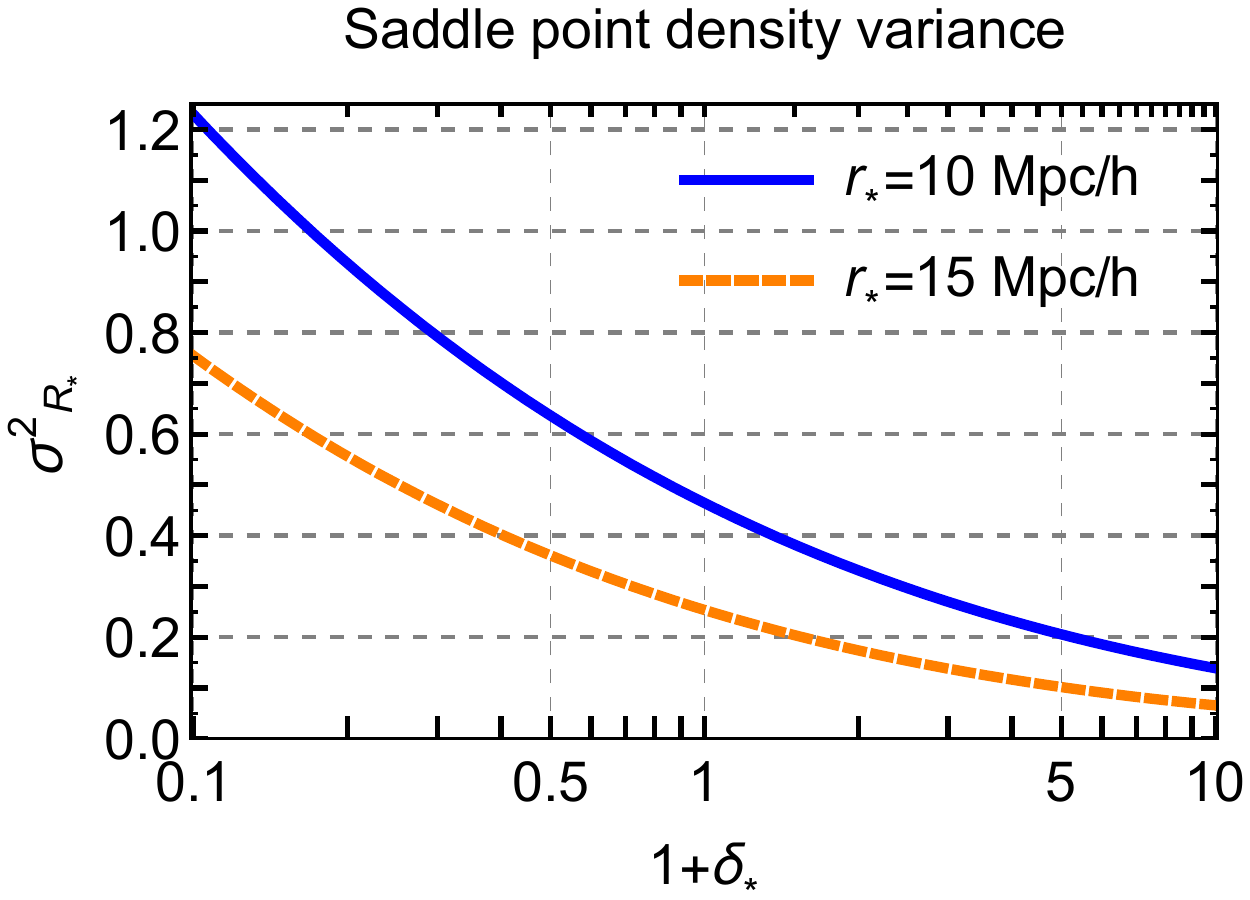}
\includegraphics[width=0.48\textwidth]{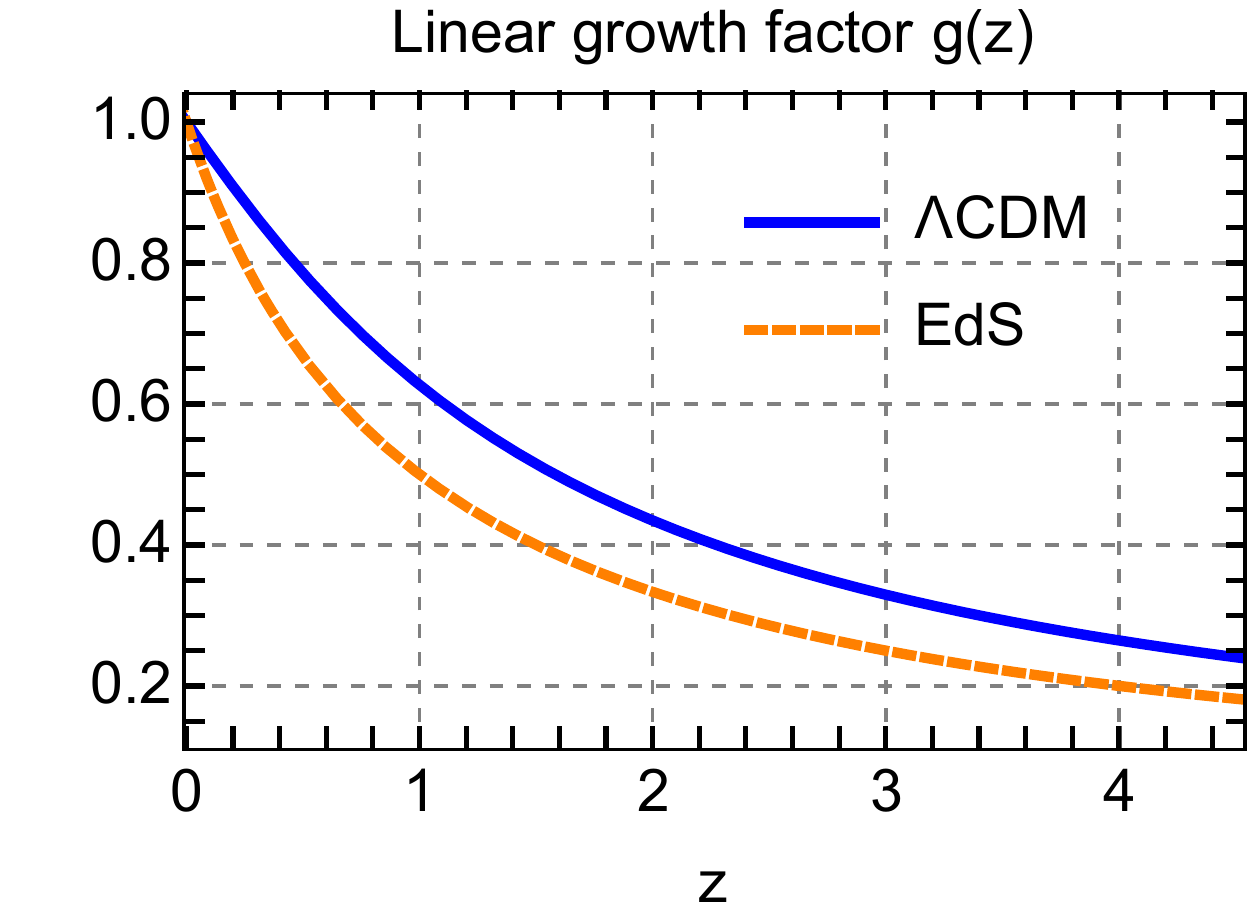}
\end{center}
\caption{\label{fig:sigmasq_g}
Left panel: the saddle point linear density variance as a function of
the final density in the cell at $z=0$ 
for comoving cell radii $10$ Mpc/$h$ and $15$ Mpc/$h$.
Right panel: the dependence of the linear growth factor on redhsift in
$\L$CDM and EdS cosmologies. In the latter case, it is equal to
$(1+z)^{-1}$.  
}
\end{figure}

The second ingredient is the linear density variance at redshift zero
$\sigma^2_{R_*}$. In contrast to $F$, it is determined only by the
linear power spectrum and is independent of the non-linear
dynamics. As already pointed out, it depends on the argument
$\delta_*$ of the
PDF through the Lagrangian radius $R_*$. This dependence is shown in
the left panel of
Fig.~\ref{fig:sigmasq_g} for two different cell radii. By definition,
$\sigma^2_{R_*}$ is independent of the redshift. The redshift
dependence of the PDF comes through the linear growth factor $g$,
shown as a function of $z$ in the right panel of
Fig.~\ref{fig:sigmasq_g}. From the way $g^2$ and $\sigma^2_{R_*}$
enter the leading exponent (\ref{leadingexp}) it is clear that
the physical expansion parameter controlling the validity of the
saddle-point approximation is the $z$-dependent linear variance
$g^2(z)\sigma^2_{R_*}$. One expects the semiclassical expansion to
work as long as $g^2\sigma^2_{R_*}\lesssim 1$. 
The numerical values of
the linear density variance for $\delta_*=0$   
are given in Table~\ref{tab:sigmas}.

\begin{table}[h!]
\begin{center}
\begin{tabular}{|l|c|c|}
\hline
 & $r_*=10$ Mpc$/h$ & $r_*=15$ Mpc$/h$ \\\hline
z=0 & 0.464 & 0.254 \\\hline
z=0.7 & 0.238 & 0.130\\\hline
z=4 & 0.0325 & 0.0177\\\hline
\end{tabular}
\caption{
The filtered density variance $g^2\sigma^2_{r_*}$ 
for various redshifts and 
cell radii.
}
\label{tab:sigmas}
\end{center}
\end{table} 

\begin{figure}[t]
\begin{center}
\includegraphics[width=0.50\textwidth]{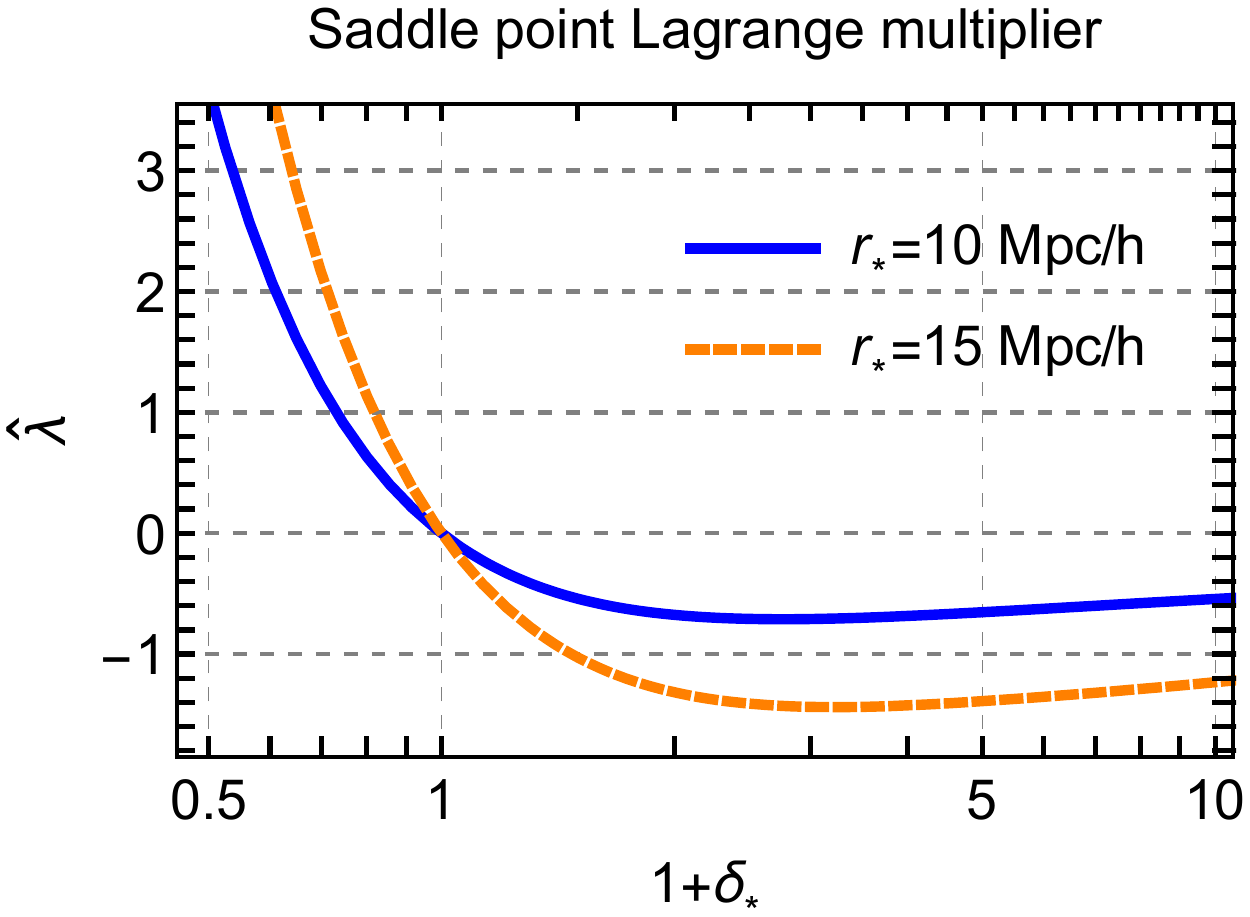}
\end{center}
\caption{\label{fig:lambda}
The saddle-point Lagrange multiplier, Eq.~(\ref{eq:lambda}), as a function
of $\delta_*$. The computation is performed in the EdS approximation.
}
\end{figure}

The Lagrange multiplier $\hat\l$ does not appear in the leading
exponent of the PDF. However, we will see below that it enters the
prefactor. So, it is instructive to plot its dependence on $\delta_*$,
see Fig.~\ref{fig:lambda}. Note that it is positive (negative) for
under- (over-) densities. It quickly grows at $\delta_*<0$.

\begin{figure}[t]
\begin{center}
\includegraphics[width=0.49\textwidth]{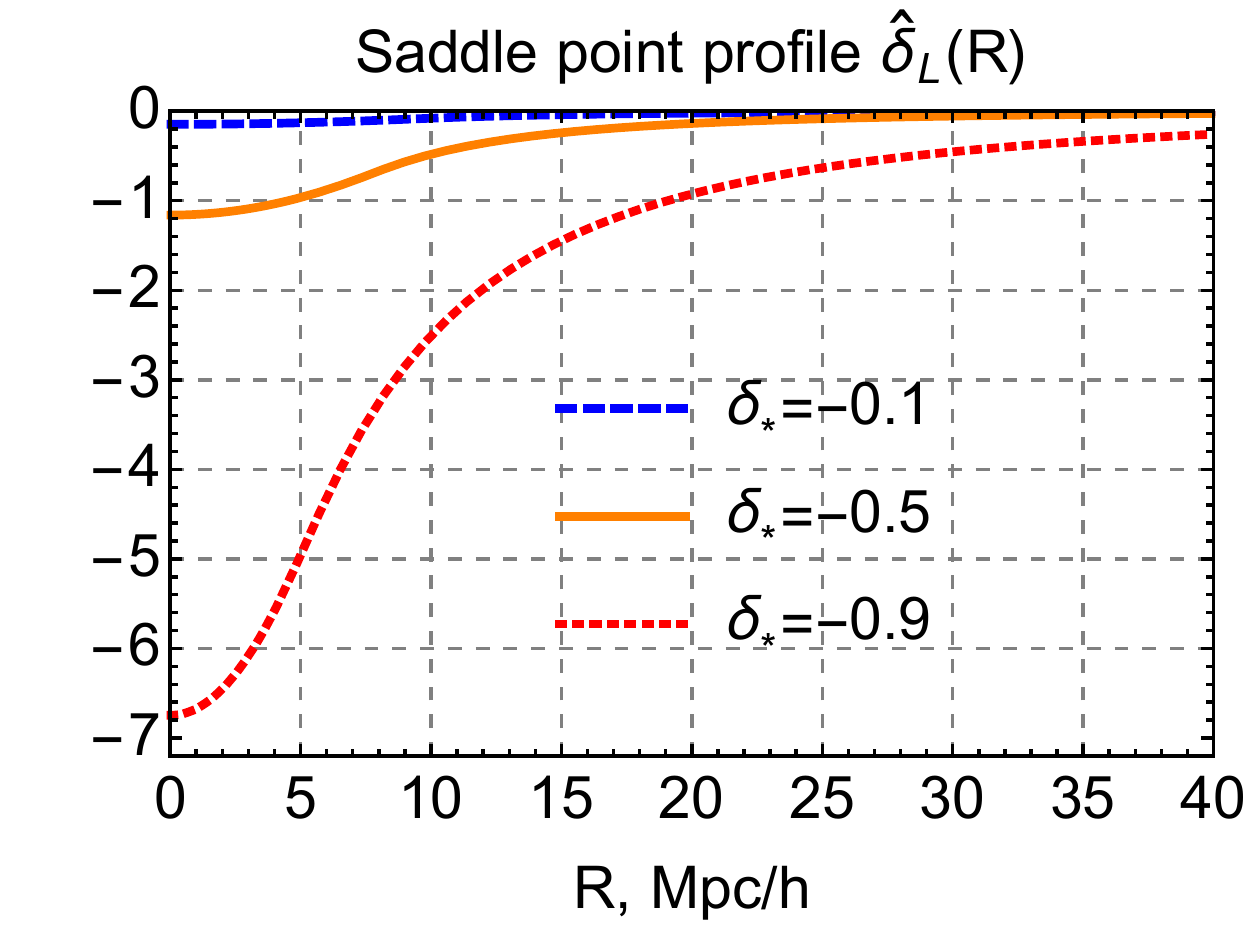}
\includegraphics[width=0.49\textwidth]{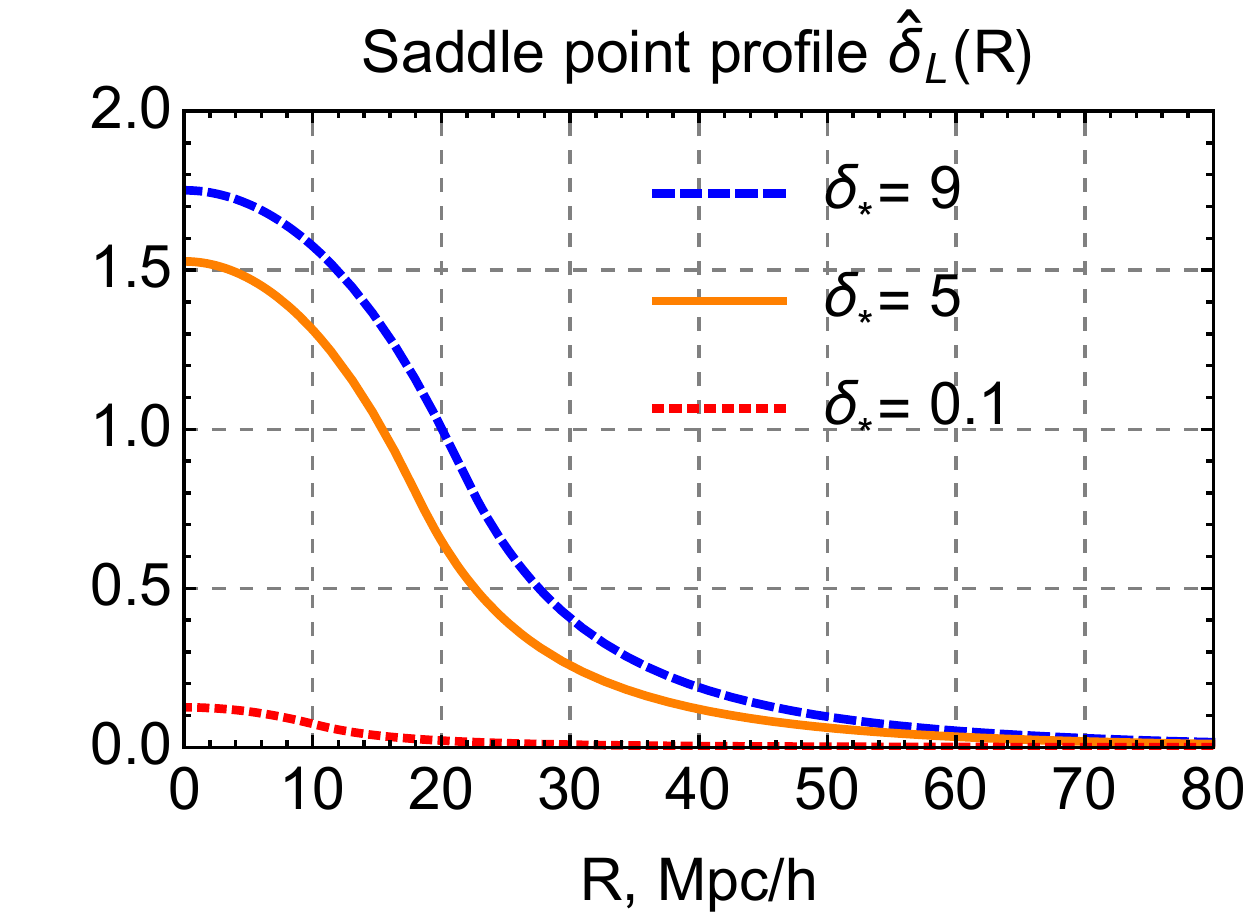}
\end{center}
\caption{\label{fig:profiles}
The saddle point linear density profiles in Lagrangian position
space for several values of $\delta_*$ corresponding to underdensities 
(left panel) and overdensities (right panel).
The results are shown for the cell radius $r_*=10$ Mpc$/h$.
}
\end{figure}

For completeness, we also present in Fig.~\ref{fig:profiles} the
saddle-point linear density profiles for several values of
$\delta_*$. 
For $\delta_* \gtrsim 7$ the density profile in the central region
exceeds the critical value\footnote{We give the critical value at $z=0$
for our reference $\L$CDM cosmology. It is somewhat lower than the
well-known EdS value  
$\delta_c=1.686$.
} 
$1.674$, and therefore the innermost part of the profile experiences 
shell-crossing. 
Conservatively, one would expect a breakdown of our saddle-point
expansion for such large overdensities. 
However, we will see shortly that the available data are consistent
with the semiclassical 
scaling even for $\delta_* \gtrsim 7$. 
This robustness of the semiclassical approach 
may be explained by the fact that 
the averaged density at $R_*$ 
is still less than the critical value even when the central regions 
undergo shell-crossing. 
Since the velocities of matter particles are rather low, it takes a
significant amount of time for the information about shell-crossing to
propagate to the boundary $R_*$. Until this happens, the dynamics of
the boundary remain the same as if no shell-crossing occurred, so that
the spherical collapse mapping used in the derivation of
(\ref{leadingexp}) still applies.

It should be stressed that having a spherical collapse saddle point does not 
mean that an exact spherical collapse happens 
inside each cell.
Recall that in the case of tunneling in quantum mechanics 
the saddle-point solution, by itself, has measure zero in the space of
all possible trajectories in the path integral, and thus is never realized precisely
(see e.g.~\cite{Andreassen:2016cvx,Kleinert:2002me}).  
What makes the tunneling amplitude finite are 
small perturbations around the saddle point solution that add up 
coherently and eventually contribute to the prefactor. 
From this argument it is clear that fluctuations around the saddle point 
are crucial for the consistency of 
our path integral calculation. 
If the saddle-point approximation works, the actual dynamics 
of the density field inside each cell
is spherical collapse perturbed by aspherical fluctuations.

\subsection{Prefactor from fluctuations}
\label{sec:aspfluc}

We now consider small fluctuations around the spherical collapse
saddle point found in the previous subsection. 
To leading order in
$g^2$, the path integral over these fluctuations is Gaussian and
produces the prefactor in front of the leading exponent
(\ref{leadingexp}), 
as was pointed out in Refs.~\cite{
Valageas:2001zr,Valageas:2001td}.
It is natural to expand the fluctuations of the
linear density field in spherical harmonics. We write,
\bseq
\begin{align}
\label{fluct1}
&\delta_L(\k)=\hat\delta_L(k)+\delta_{L,0}^{(1)}(k)
+\sum_{\ell >0}\sum_{m=-\ell}^\ell (-i)^\ell\, \delta_{L,\ell m}^{(1)}(k)\,
Y_{\ell m}(\k/k)\;,\\
&\l=\hat\l+\l^{(1)}\;,
\end{align}
\eseq
where we have singled out the monopole fluctuation
$\delta_{L,0}^{(1)}$. Note that due to our convention for the
spherical harmonics (see Appendix~\ref{sec:notations}), the reality
condition $\big(\delta_L(\k)\big)^*=\delta_L(-\k)$ translates into the
conditions 
\be
\label{reality}
\big(\delta_{L,0}^{(1)}(k)\big)^*=\delta_{L,0}^{(1)}(k)~,~~~~
\big(\delta_{L,\ell m}^{(1)}(k)\big)^*=\delta_{L,\ell ,-m}^{(1)}(k)\;.
\ee
Fluctuations give rise to a perturbation of the averaged density
contrast which up to second order can be written as,
\be
\label{perturbQl}
 \begin{split}
 \bar \delta_W=&\delta_*+\int [dk]\, 4\pi S(k)\,\delta_{L,0}^{(1)}(k)
+\int[dk]^2 \,4\pi Q_{0}(k_1,k_2)\,
\delta_{L,0}^{(1)}(k_1)\delta_{L,0}^{(1)}(k_2)\\
& +\sum_{\ell>0,m}
 \int [dk]^2 \,Q_{\ell}(k_1,k_2)\,
\delta_{L,\ell m}^{(1)}(k_1)\delta^{(1)}_{L,\ell,-m}(k_2)\;,
 \end{split}
\ee 
where we introduced the notation,
\be
\label{kmeasure}
[dk]^n\equiv \prod_{i=1}^n \frac{k^2_idk_i}{(2\pi)^3}\;,
\ee
and $S$, $Q_0$, $Q_\ell$ are some kernels. Below we will refer to
$Q_0$, $Q_\ell$ as {\em response matrices}. Note the factor $4\pi$
that we included in the definition of $S$ and $Q_0$; it reflects the
difference in our
normalization of spherical harmonics in the monopole
and higher multipole sectors, see Eq.~(\ref{harmnorm}).  
In the expression (\ref{perturbQl}) we have used the fact that
non-monopole fluctuations can contribute
only at quadratic order due to spherical symmetry. 
For the same reason, the kernels $Q_\ell$ do
not depend on the azimuthal number $m$. 

Substituting (\ref{fluct1}) and (\ref{perturbQl}) into the path
integral (\ref{eq:pdfLaplace}), after a straightforward calculation, we
find that the Gaussian integrals over fluctuations with different
multipole numbers $\ell$ factorize. This leads to the following
representation for the PDF,
\be
\label{eq:pdffull}
\P(\delta_*)= {\cal A}_0\cdot\prod_{\ell>0}{\cal A}_\ell(\delta_*) \cdot
\exp\left\{-\frac{F^2(\delta_*)}{2 g^2\sigma^2_{R_*}} \right\}\,,
\ee
where
\begin{align}
{\cal A}_0=&{\cal N}_0^{-1}
\int_{-i\infty}^{i\infty}\frac{d\l^{(1)}}{2\pi ig^2}
\int \mathscr{ D}\delta_{L,0}^{(1)}\,\exp\bigg\{-\frac{4\pi}{g^2}
\bigg[\int\frac{[dk]}{2P(k)}\big(\delta_{L,0}^{(1)}(k)\big)^2\notag\\
&\qquad+\l^{(1)}\int [dk]\,S(k)\,\delta_{L,0}^{(1)}(k)
+\hat\l\int[dk]^2 \,Q_{0}(k_1,k_2)\,
\delta_{L,0}^{(1)}(k_1)\delta_{L,0}^{(1)}(k_2)\bigg]\bigg\}\;,
\label{A01}\\
{\cal A}_\ell=&{\cal N}_\ell^{-1}\int [\mathscr{
  D}\delta_{L,lm}^{(1)}]\,
\exp\bigg\{-\frac{1}{g^2}\sum_m
\bigg[\int\frac{[dk]}{2P(k)}\delta_{L,\ell
  m}^{(1)}(k)\delta_{L,\ell,-m}^{(1)}(k)\notag\\
&\qquad\qquad\qquad\qquad\qquad+\hat\l\int [dk]^2 \,Q_{\ell}(k_1,k_2)\,
\delta_{L,\ell
  m}^{(1)}(k_1)\delta^{(1)}_{L,\ell,-m}(k_2)\bigg]\bigg\}\;.
\label{Al1}
\end{align}
The integration measure in the last expression is $[\mathscr{
  D}\delta_{L,lm}^{(1)}]=\prod_{m=-l}^l\mathscr{
  D}\delta_{L,lm}^{(1)}$, whereas the normalization factors are,
\begin{align}
\label{N01}
&{\cal N}_0=
\int \mathscr{ D}\delta_{L,0}\,\exp\bigg\{-\frac{4\pi}{g^2}
\int\frac{[dk]}{2P(k)}\big(\delta_{L,0}(k)\big)^2\bigg\}\;,\\
\label{Nl1}
&{\cal N}_\ell=\int [\mathscr{
  D}\delta_{L,lm}]\,
\exp\bigg\{-\frac{1}{g^2}\sum_m
\int\frac{[dk]}{2P(k)}\delta_{L,\ell m}(k)\delta_{L,\ell,-m}(k)\bigg\}\;.
\end{align}
Despite appearing more complicated, the monopole prefactor ${\cal
  A}_0$ can be evaluated analytically. This is not surprising, since
the dynamics in the monopole sector is known exactly. We postpone this
analysis to the next section and focus here on the prefactor stemming
from higher multipoles. 

The quadratic form in the exponent of Eq.~(\ref{Al1}) is a convolution
of the vector $\delta_{L,\ell m}^{(1)}$ with the matrix
\[
\frac{1}{g^2}\bigg(\mathbb{1}\cdot \frac{1}{P(k)}+2\hat\l Q_\ell\bigg)
\,\delta_{m,-m}\;, 
\]
where $\mathbb{1}$ is the unit operator in $k$-space whose kernel with
respect to the measure (\ref{kmeasure}) is,
\be
\label{kunity}
\mathbb{1}(k,k')=(2\pi)^3k^{-2}\delta_{\rm D}^{(1)}(k-k')\;,
\ee
and $\delta_{m,-m}$ is the Kronecker symbol. The Gaussian integral
over $\delta_{L,\ell m}^{(1)}$ is inversely proportional to the square
root of the determinant of this matrix. To get ${\cal A}_\ell$, this
determinant must be divided by the determinant of the corresponding
matrix in the normalization factor (\ref{Nl1}) which is simply
\[
\frac{1}{g^2}\bigg(\mathbb{1}\cdot\frac{1}{P(k)}\bigg)\,\delta_{m,-m}\;.
\]
In this way we obtain
\be
\label{AlDl}
{\cal A}_\ell=\D_\ell^{-(\ell+1/2)}\,,
\ee
where
\be
\label{Dl}
\D_{\ell} = \det\left(\mathbb{1}+2\hat\lambda \sqrt{P}Q_{\ell}\sqrt{P}\right)\,,
\ee
is the $\ell$th \textit{aspherical fluctuation determinant}. The
second term in $\D_\ell$ denotes an operator with the kernel 
$\sqrt{P(k)}Q_\ell(k,k')\sqrt{P(k')}$.
It is convenient to introduce the {\it aspherical prefactor} that
aggregates contributions of all multipoles with strictly positive
$\ell$,
\be
\label{prefl>0}
\mathcal{A}_{\text{ASP}}\equiv 
\prod_{\ell>0}\mathcal{A}_{\ell} = \prod_{\ell>0}\D_\ell^{-(\ell+1/2)} \,.
\ee
We see that its computation requires knowledge of the aspherical
response matrices~$Q_\ell$.

Let us make an important remark. The growth factor $g$ has dropped out
of the expression for the fluctuation determinants (\ref{Dl}). Also,
it can be shown that the response matrices $Q_\ell$ do not depend on
the redshift\footnote{Strictly speaking, this statement is true only
  in the EdS universe. However, the response matrices computed in the
  exact $\L$CDM cosmology coincide with the EdS approximation better
  than at a per
cent level. Another source of a weak $z$-dependence is a UV
counterterm in the prefactor, required to renormalize the
short-distance contributions, see Sec.~\ref{sec:counterterm}.} (see
Sec.~\ref{sec:mainEq}). This implies that the 
aspherical prefactor is redshift-independent. We are going to see in
the next section that this theoretical expectation is confirmed by the
N-body data. 

The redshift-independence of ${\cal A}_{\rm ASP}$ may be somewhat
puzzling. Indeed, being a non-trivial function of $\delta_*$, the
aspherical prefactor affects the shape of PDF even at early times,
when the distribution must be Gaussian. To resolve this apparent
paradox,  
we notice that 
at high redshifts (in the limit $g^2 \to 0$) the distribution
\eqref{eq:pdffull} 
approaches the delta-function centered at $\delta_*=0$.
On the other hand, recall that $\hat\l$ vanishes at $\delta_*$ (see
Fig.~\ref{fig:lambda}) and hence $\D_\ell(\delta_*=0)=1$ for all
$\ell$. This implies 
${\cal A}_{\text{ASP}}(\delta_*=0)=1$ and 
in the limit $g^2\to 0$ the whole aspherical prefactor
reduces to unity.
One concludes that 
the role of the aspherical prefactor decreases as the distribution becomes sharper
towards high redshifts.

\section{Closer look at the prefactor}
\label{sec:prefactor}

In this section we explicitly compute the monopole prefactor ${\cal
  A}_0$ from the spherical collapse dynamics.
We then use N-body data to extract 
the aspherical prefactor ${\cal A}_{\rm ASP}$ and discuss its main properties.

\subsection{Monopole}
\label{sec:saddle}

The factorization property (\ref{eq:pdffull}) implies that in the
computation of the monopole prefactor all aspherical perturbations can
be set to zero. Thereby it is convenient to consider the path integral
over the spherically symmetric sector as a whole, without splitting
the density field into the saddle-point configuration and
fluctuations. In this way we arrive at what can be called `spherical
PDF', 
\be 
\label{PSP1}
\mathcal{P}_{\text{SP}}(\delta_*)=
\mathcal{N}_0^{-1}\int \mathscr{ D}\delta_{L,0}
\exp\bigg\{-\frac{4\pi}{g^2}
\int \frac{[dk]}{2P(k)}\big(\delta_{L,0}(k)\big)^2\bigg\}
\,\delta_D^{(1)}\big(\delta_* - \bar \delta_W[\delta_{L,0}]\big)\,,
\ee
with the normalization factor given in Eq.~(\ref{N01}). We stress that
${\cal P}_{\rm SP}$ is {\em not} equal to the true PDF, as it
restricts the original path integral (\ref{eq:pdf2}) to spherically
symmetric configurations only, and thus misses the contribution of
aspherical modes.

\begin{figure}[t]
\begin{center}
\includegraphics[width=0.6\textwidth]{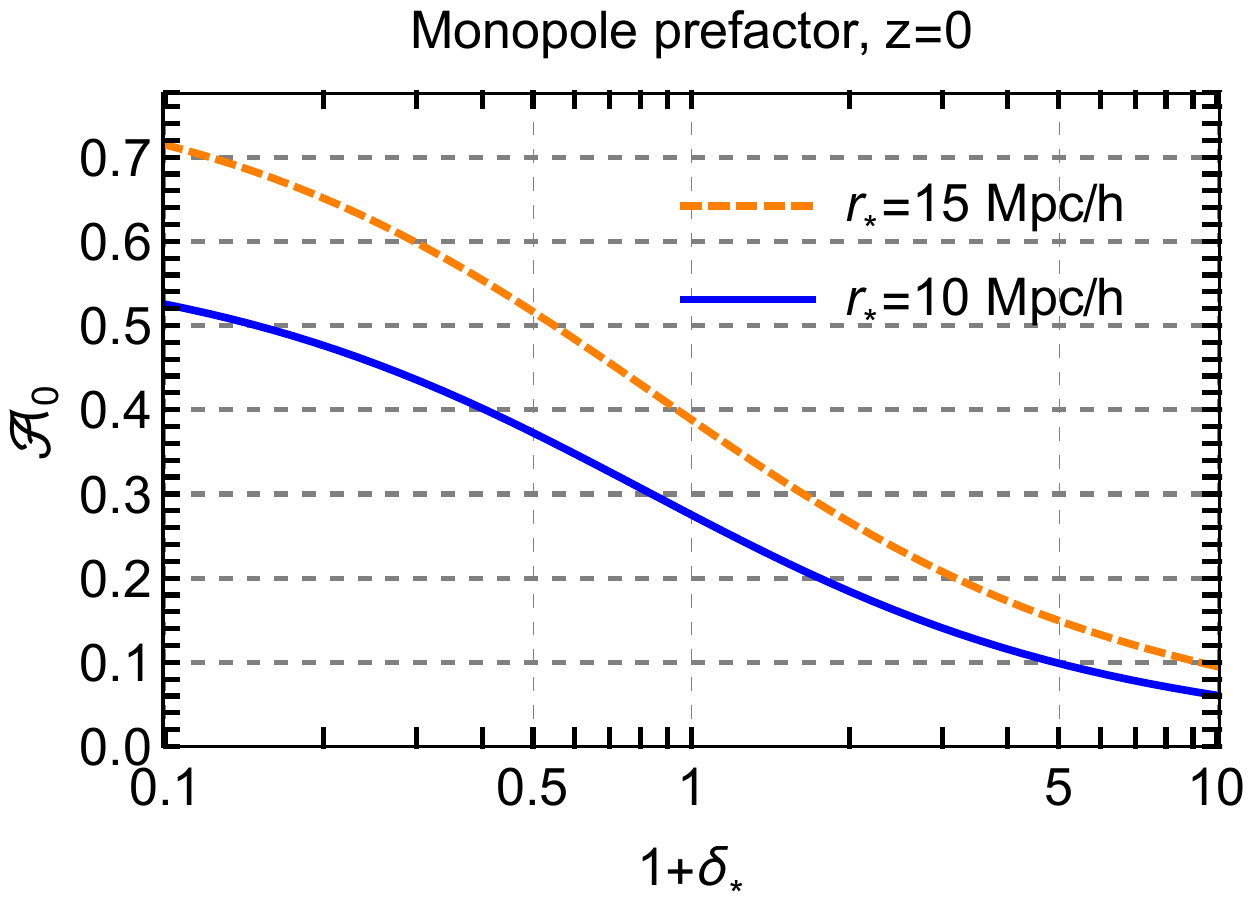}
\end{center}
\caption{\label{fig:sph}
The monopole prefactor at $z=0$.
}
\end{figure}

Due to the existence of the spherical collapse mapping
(\ref{eq:scmap}), the condition $\delta_*=\bar\delta_W[\delta_{L,0}]$
is equivalent to the condition
$F(\delta_*)=\bar\delta_{L,0}(R_*)$. Thus, the delta-function in
(\ref{PSP1}) is proportional to the delta-function of the argument 
$F(\delta_*)-\bar\delta_{L,0}(R_*)$,  
\be 
\label{eq:deltaf}
\delta^{(1)}_D\big(\delta_* - \bar\delta_{W}[\delta_{L,0}]\big)=
C[\delta_{L,0}]\cdot\delta^{(1)}_D\big(F(\delta_*) - \bar \delta_{L,0}(R_*)\big)\,.
\ee
The proportionality coefficient $C$ is given in Eq.~(\ref{eq:G1}); it
is fixed by the requirement that the integral of both sides of
(\ref{eq:deltaf}) over $\delta_*$ produces unity. Substituting this
relation into Eq.~(\ref{PSP1}) and using the integral representation
for the delta-function we obtain,
\be 
\label{eq:pdf3}
\begin{split}
\mathcal{P}_{\text{SP}}(\delta_*)=&\mathcal{N}_{0}^{-1}
\int_{-i\infty}^{i\infty} \frac{d\l}{2\pi i g^2}\,\e^{\l F/g^2}
\int \mathscr{ D}\delta_{L,0} \,
C[\delta_{L,0}]\\
&\times\exp\bigg\{-\frac{4\pi}{g^2}\bigg[
\int \frac{[dk]}{2P(k)} \big(\delta_{L,0}(k)\big)^2 
+ \lambda 
\int [dk] W_{\rm th}(kR_*)\delta_{L,0}(k)
\bigg]\bigg\} 
\,.
\end{split}
\ee
It is now straightforward to evaluate this integral by the saddle
point method, which yields\footnote{This result is actually 
    exact as $C[\delta_{L,0}]$ is a linear functional of
  $\delta_{L,0}$, and for this type of integrals there are no
  corrections to the saddle-point approximation.},
\be 
\label{eq:pdf-spher}
\mathcal{P}_{\text{SP}}(\delta_*)
=\frac{\hat C(\delta_*)}{\sqrt{2\pi g^2\sigma^2_{R_*}}}
\exp \left(-\frac{F^2(\delta_*)}{2g^2\sigma^2_{R_*}}\right)\,,
\ee
where $\hat C$ is defined in (\ref{eq:lambda}). From this expression
we infer the monopole prefactor,
\be
\label{Amono}
{\cal A}_0(\delta_*)=\frac{\hat C(\delta_*)}{\sqrt{2\pi g^2\sigma^2_{R_*}}}\;.
\ee
We plot its dependence on the density contrast in Fig.~\ref{fig:sph}.
It varies roughly by an order of magnitude in the range $\delta_*=[-0.9,9]$.
Since it is inversely proportional to the r.m.s density contrast
$g\sigma_{R_*}$, it significantly varies with the window function
radius and redshift.  
For illustration purposes we show the results for $z=0$. The curves 
for other redshifts are qualitatively similar and 
can be obtained upon rescaling 
by an appropriate growth factor (shown in the right panel of
Fig.~\ref{fig:sigmasq_g}). 

By construction, the spherical PDF (\ref{PSP1}) is normalized to
unity, 
\be
\int_{-1}^\infty d\delta_* \,\P_{\text{SP}}(\delta_*)=1\,.
\ee
However, it does not reproduce the correct zero mean value of the density
contrast,
\be
\label{meandeltaSP}
\langle \delta_*\rangle_{\rm SP}\equiv\int_{-1}^\infty d\delta_*\,
{\cal P}_{\rm SP}(\delta_*)\,\delta_*\neq 0\;.
\ee
To see this, we define the variable $\nu=F/\sigma_{R_*}$ and rewrite
(\ref{eq:pdf-spher}) as
\be
\label{PSPnu}
{\cal P}_{\rm SP}=\frac{1}{\sqrt{2\pi
    g^2}}\,\frac{d\nu}{d\delta_*}\,\e^{-\frac{\nu^2}{2g^2}}\;. 
\ee 
The expectation value (\ref{meandeltaSP}) becomes,
\be
\label{meandeltaSP2}
\langle \delta_*\rangle_{\rm SP}=\int_{-\infty}^\infty
\frac{d\nu}{\sqrt{2\pi g^2}}\,\delta_*(\nu)\,\e^{-\frac{\nu^2}{2g^2}}
=\frac{g^2}{2}\frac{d^2\delta_*}{d\nu^2}\bigg|_{\nu=0}\;,
\ee
where we have evaluated the integral at leading order in $g^2$. It
is straightforward to compute the second derivative appearing in the
above equation. One finds,
\be
\label{eq:d2deltanu}
\frac{d^2\delta_*}{d\nu^2}\bigg|_{\nu=0}=-\sigma_{r_*}^2\bigg[F''(0)
+2\bigg(1-\frac{\xi_{r_*}}{\sigma^2_{r_*}}\bigg)\bigg]\;.
\ee
Using also the Taylor expansion (\ref{scpert1}) for the function $F$ one
obtains, 
\be
\label{meandeltaSP3}
\langle \delta_*\rangle_{\rm SP}=-g^2\sigma_{r_*}^2 a_1\,,~~~~~~
\text{where}~~a_1=\frac{4}{21}-\frac{\xi_{r_*}}{\sigma^2_{r_*}}\;.
\ee
The numerical values of $a_1$ for different cell radii are given in
Table~\ref{tab:a123} in the next subsection.

At first sight, the fact that the spherical PDF fails to reproduce the
zero mean value 
of $\delta_*$ may seem surprising. However, it becomes less so once we
realize that vanishing of
$\langle\delta_*\rangle$ is related to translational
invariance. Indeed, it is implied
  by the vanishing of $\langle\delta(\x)\rangle$, the mean density
  contrast at each space point. The latter, in turn, involves two
  ingredients: ({\sf i}) the constraint $\int d^3x\,\delta(\x)=0$ which
  follows trivially from the definition of the density contrast, and
  ({\sf ii}) the fact that, due to translational invariance,
  $\langle\delta(\x)\rangle$ is the same at all points. But the
  translational invariance
has been explicitly broken by the reduction
of the path integral to the spherically symmetric sector that 
singles out the origin as a preferred point in space. The
correct identity $\langle\delta_*\rangle=0$ will be restored once we
take into account the aspherical prefactor generated by 
fluctuations beyond the monopole sector.

\subsection{Aspherical prefactor from N-body data}
\label{sec:asppref}

\begin{figure}[t]
\begin{center}
\includegraphics[width=0.49\textwidth]{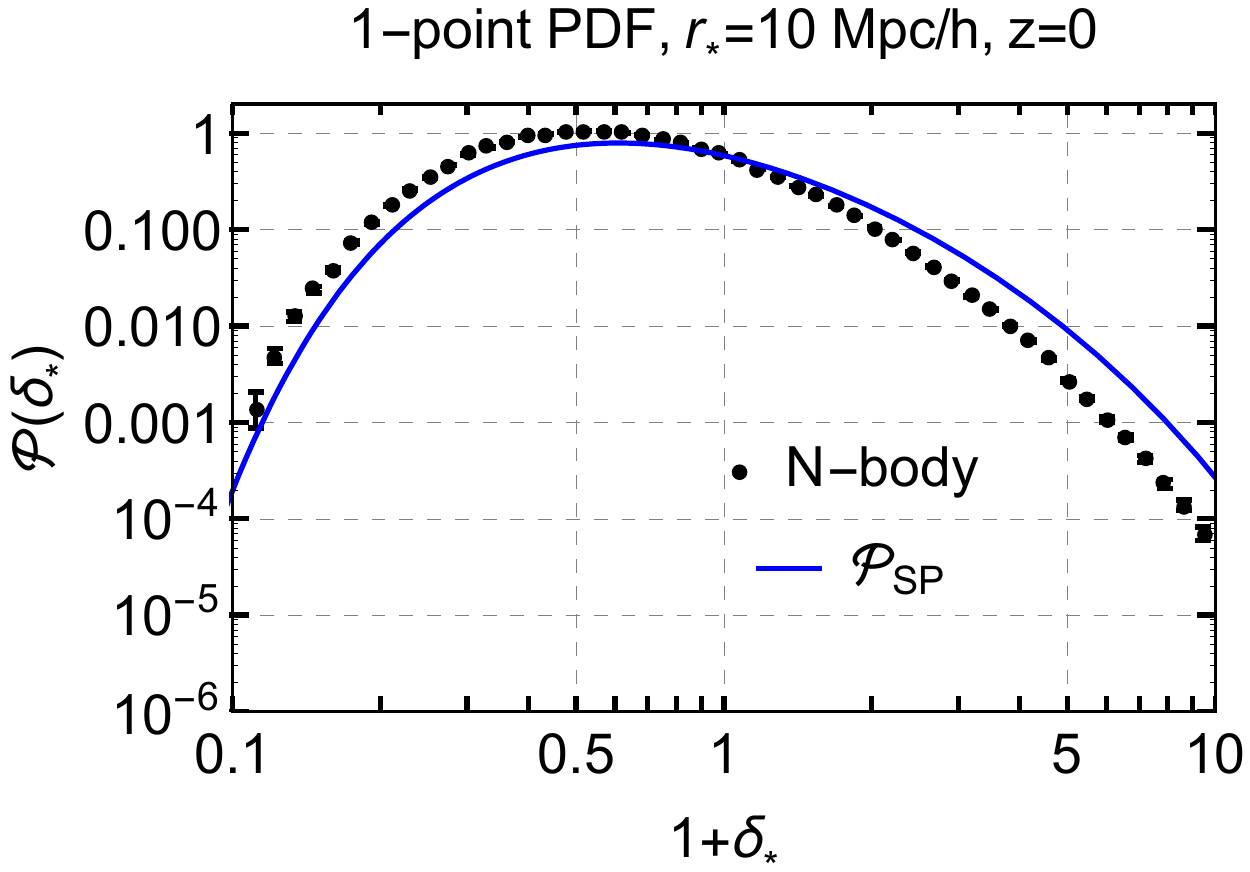}
\includegraphics[width=0.49\textwidth]{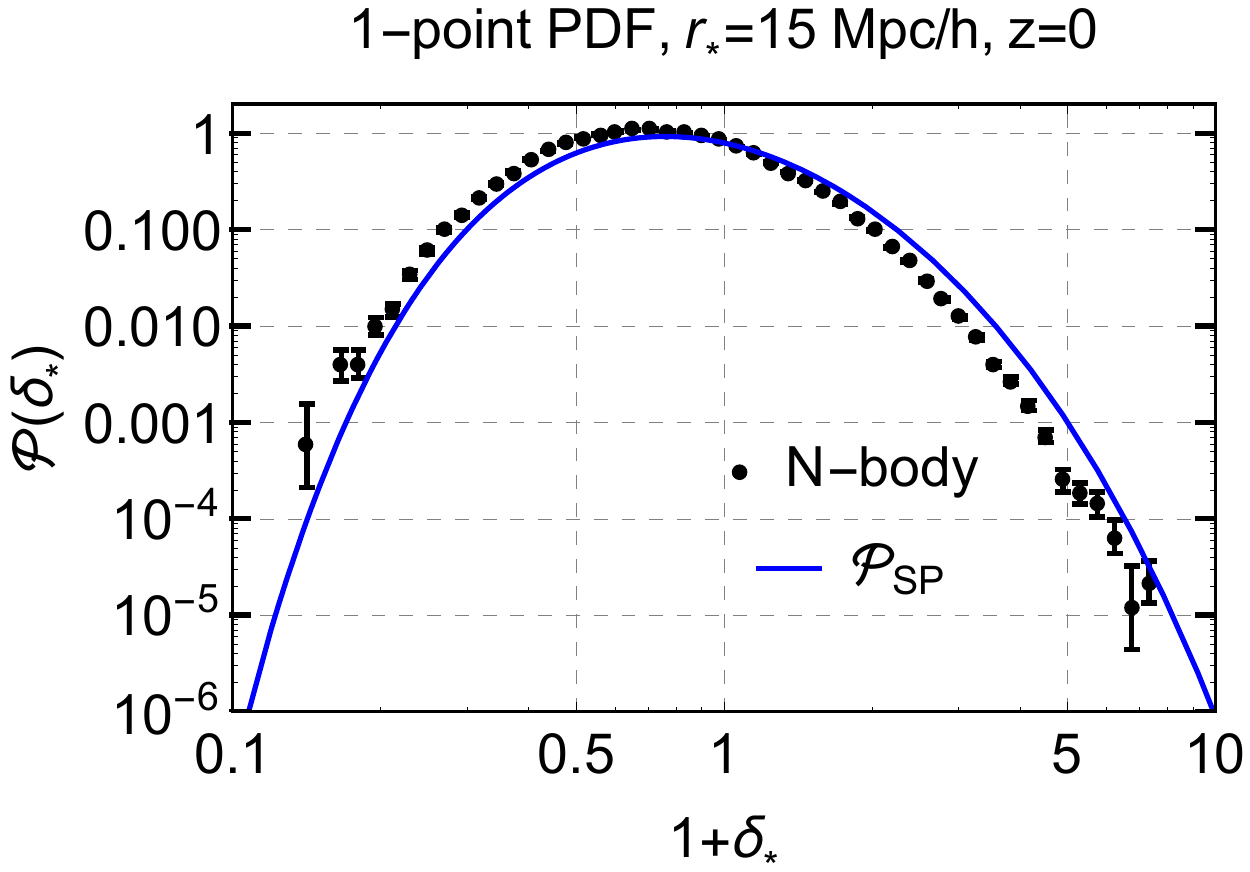}
\end{center}
\caption{\label{fig:spherpdfs}
1-point PDF of the smoothed density field at redshift
$z=0$ for $r_*=10$ Mpc/$h$ (left panel) 
and $r_*=15$ Mpc/$h$ (right panel): the spherical PDF given by 
Eq.~\eqref{eq:pdf-spher} (blue line) against the N-body data (black
dots). Error-bars on the data points show the statistical uncertainty.  
}
\end{figure}

\begin{figure}[tb]
\begin{center}
\includegraphics[width=0.49\textwidth]{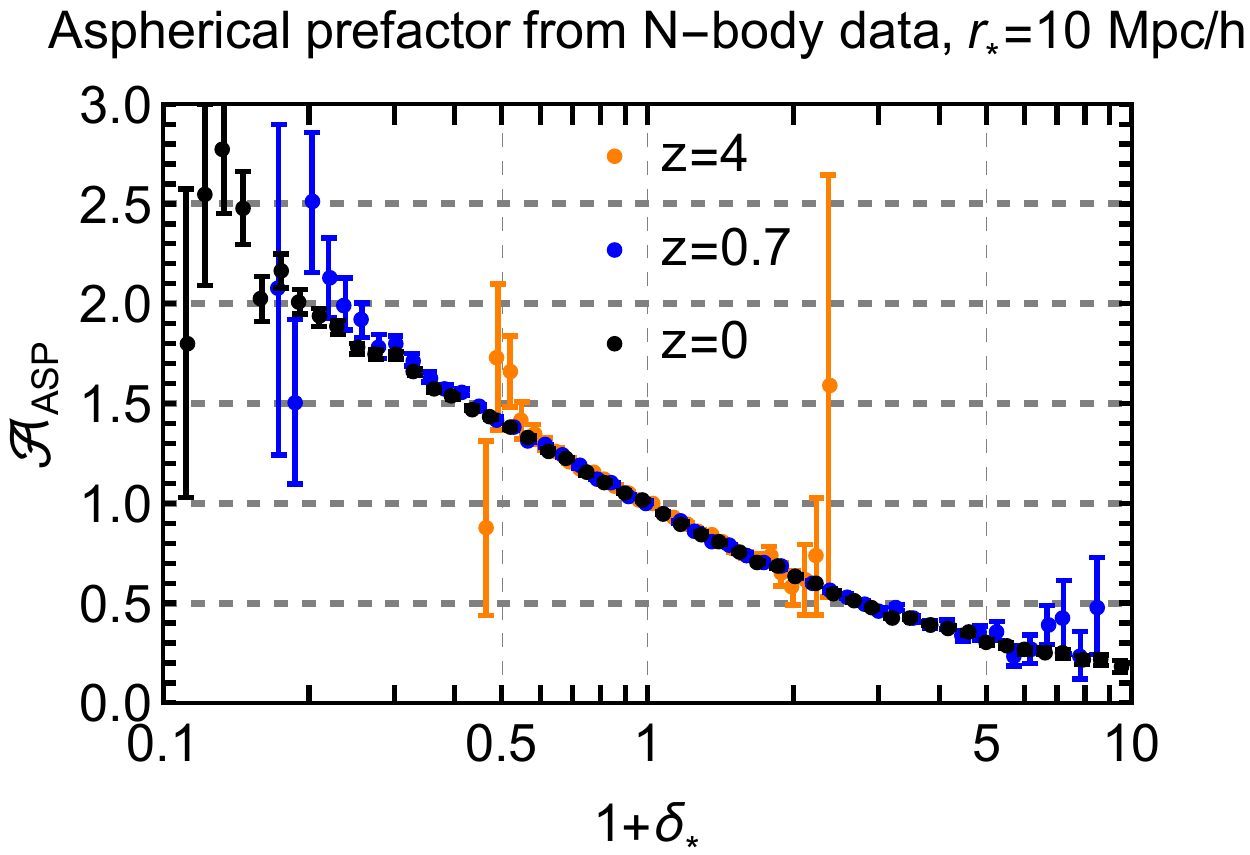}
\includegraphics[width=0.49\textwidth]{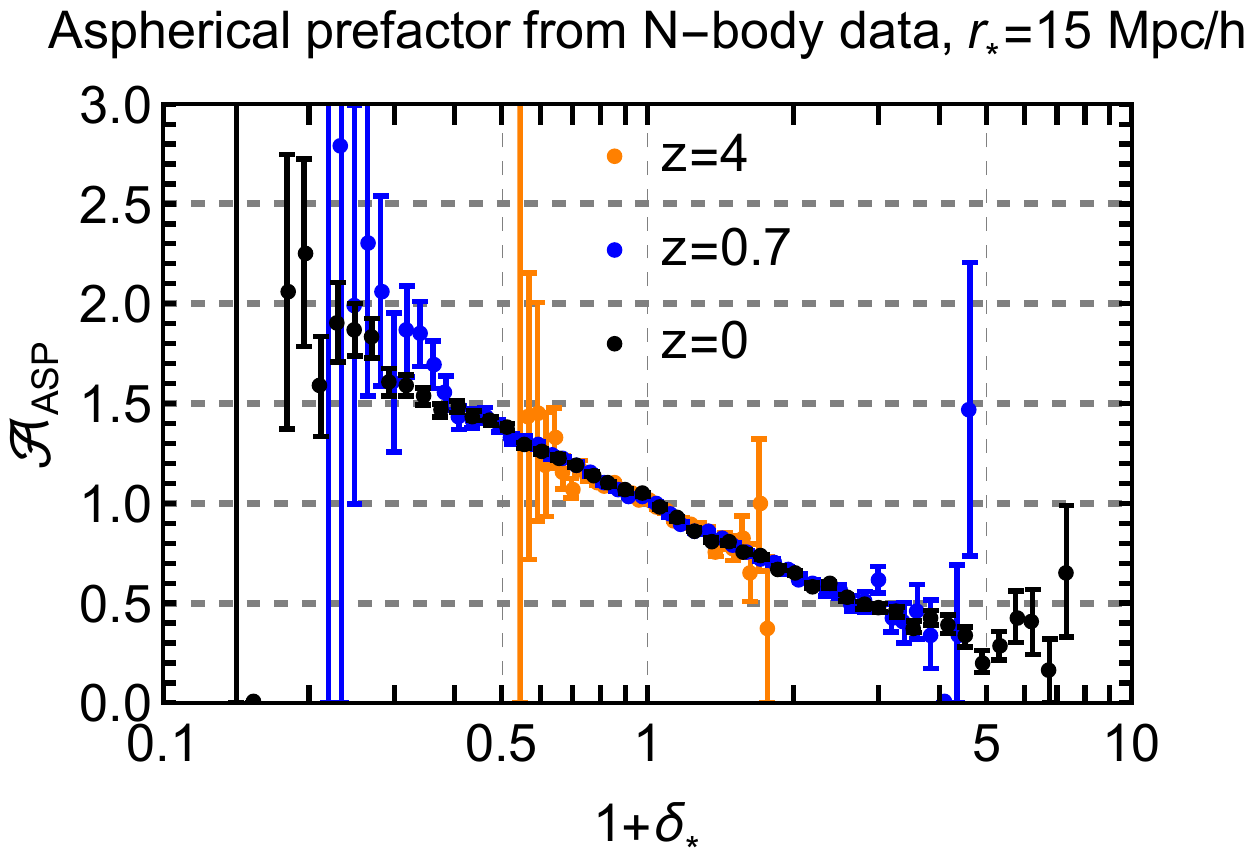}
\end{center}
\caption{\label{fig:aspfrom data}
The aspherical prefactor ${\cal A}_{\text{ASP}}=\P_{\rm data}/\P_{\text{SP}}$
extracted from the simulations. 
The results are shown for the cell radii 10 Mpc/$h$ (left panel)
and 15 Mpc/$h$ (right panel).
}
\end{figure}

Before delving into the calculation of the aspherical prefactor, let us
verify the semiclassical factorization formula (\ref{eq:pdffull})
against the N-body data. To this end, we have run a suite of N-body
simulations using the \texttt{FastPM} code
\cite{Feng:2016yqz} and obtained the counts-in-cells statistics for a
total of $518400$ cells with radius $r_*=10$~Mpc$/h$ and $153600$
cells with $r_*=15$~Mpc$/h$. The details of our simulations are
presented in Appendix~\ref{app:Nbody}.
Figure~\ref{fig:spherpdfs} shows the data points together
with the spherical PDF ${\cal P}_{\rm SP}$. The results are shown
for redshift $z=0$. 
The PDFs for other redshifts are qualitatively similar and will be
discussed shortly.  
From Fig.~\ref{fig:spherpdfs}
we see that although the spherical PDF correctly captures the
exponential falloff of the data points at large over-/under-densities,
it is clearly off-set from the data even at $\delta_*=0$.
According to (\ref{eq:pdffull}), this off-set should be compensated by
the aspherical prefactor ${\cal A}_{\text{ASP}}$.
Using the full PDF $\P_{\rm data}(\delta_*)$ measured from the data, 
we can extract the aspherical prefactor as
\be
\label{AASPdata}
{\cal A}_{\rm ASP}(\delta_*)=\frac{{\cal P}_{\rm
    data}(\delta_*)}{{\cal P}_{\rm SP}(\delta_*)}\;.
\ee
The result is shown in Fig.~\ref{fig:aspfrom data} for various redshifts and cell radii.
At higher redshifts the distribution becomes sharper, which increases 
the measurement errors away from the origin. This is especially
visible in the case $z=4$ 
where the available $\delta_*$-range in the data significantly
shrinks compared to $z=0$.
The errorbars shown in the plots represent the statistical uncertainty
of our data. It is worth noting that the bins at the tails of the
distribution are expected to contain also a systematic error
comparable to the statistical one, see the discussion in  
Appendix~\ref{app:Nbody}.

The spherical PDF has an exponential sensitivity to the density
variance, which changes by  
an order of magnitude across the considered redshifts, see
Tab.~\ref{tab:sigmas}.  
Similarly, the measured PDF's at different redshifts and
cell radii are exponentially different. Nevertheless, 
we observe that
the results of their division by the spherical PDF's depend very
weakly on the redshift and the size of the window function.  
This is a strong confirmation of the validity of the semiclassical
scaling \eqref{eq:pdffull}.
In particular, we conclude that the spherical collapse saddle point
indeed dominates the probability: if it were not the case, one would
expect exponentially large difference between ${\cal P}_{\rm data}$
and ${\cal P}_{\rm SP}$. Moreover, the data are clearly consistent
with the redshift-independence of ${\cal A}_{\rm ASP}$, as predicted by
the theory (see Sec.~\ref{sec:aspfluc}). 
Note that the aspherical prefactor is a very smooth function that
varies only by an order of  
magnitude within 
the density range where the whole PDF varies by six-seven orders of
magnitude. 

In complete agreement with the theoretical expectation (recall the discussion 
at the end of Sec.~\ref{sec:aspfluc}), 
we see that
${\cal A}_{\rm ASP}\big|_{\delta_*=0}=1$. 
Note that this ensures the correct
normalization of the full PDF ${\cal P}={\cal A}_{\rm ASP}\P_{\rm SP}$ 
in the leading semiclassical
approximation. 
Indeed, in this approximation the 
PDF is concentrated
around $\delta_*=0$ and
we have,
\[
\int d\delta_*\,{\cal A}_{\rm ASP}(\delta_*)\P_{\rm SP}(\delta_*)
={\cal A}_{\rm ASP}\big|_{\delta_*=0}\int d\delta_*\,\P_{\rm SP}(\delta_*)
={\cal A}_{\rm ASP}\big|_{\delta_*=0}\;.
\]
Let us now see how inclusion
of the aspherical prefactor restores the zero expectation value of the
density contrast. To this end, we introduce the variable $\nu$ as in
(\ref{PSPnu}) and write, 
\be
\begin{split}
\langle\delta_*\rangle & =\int_{-1}^\infty d\delta_*\,{\cal A}_{\rm
  ASP}(\delta_*)\P_{\rm SP}(\delta_*)\,\delta_*
=\int_{-\infty}^\infty \frac{d\nu}{\sqrt{2\pi g^2}}\,{\cal
  A}_{\text{ASP}}(\nu)\,\delta_*(\nu)\,\e^{-\frac{\nu^2}{2g^2}}\\
&=g^2\bigg(\frac{d{\cal A}_{\rm ASP}}{d\nu}\cdot
\frac{d\delta_*}{d\nu}+\frac{1}{2}\frac{d^2\delta_*}{d\nu^2}\bigg)
\bigg|_{\nu=0}\;,
\end{split}
\ee
where in the last equality we evaluated the integral at leading
order in $g^2$. For $\langle\delta_*\rangle$ to vanish, the first
derivative of ${\cal A}_{\rm ASP}$ at $\delta_*=0$ must satisfy,
\[
\frac{d{\cal A}_{\rm ASP}}{d\delta_*}\bigg|_{\delta_*=0}
=-\frac{1}{2}\bigg(\frac{d\delta_*}{d\nu}\bigg)^{-2}
\frac{d^2\delta_*}{d\nu^2}\bigg|_{\nu=0}\;.
\]
Comparing with Eq.~(\ref{eq:d2deltanu}) we obtain the condition 
\be
\label{AASPderiv}
\frac{d{\cal A}_{\rm ASP}}{d\delta_*}\bigg|_{\delta_*=0}=a_1\;,
\ee
where $a_1$ has been defined in (\ref{meandeltaSP3}).

\begin{figure}[t]
\begin{center}
\includegraphics[width=0.49\textwidth]{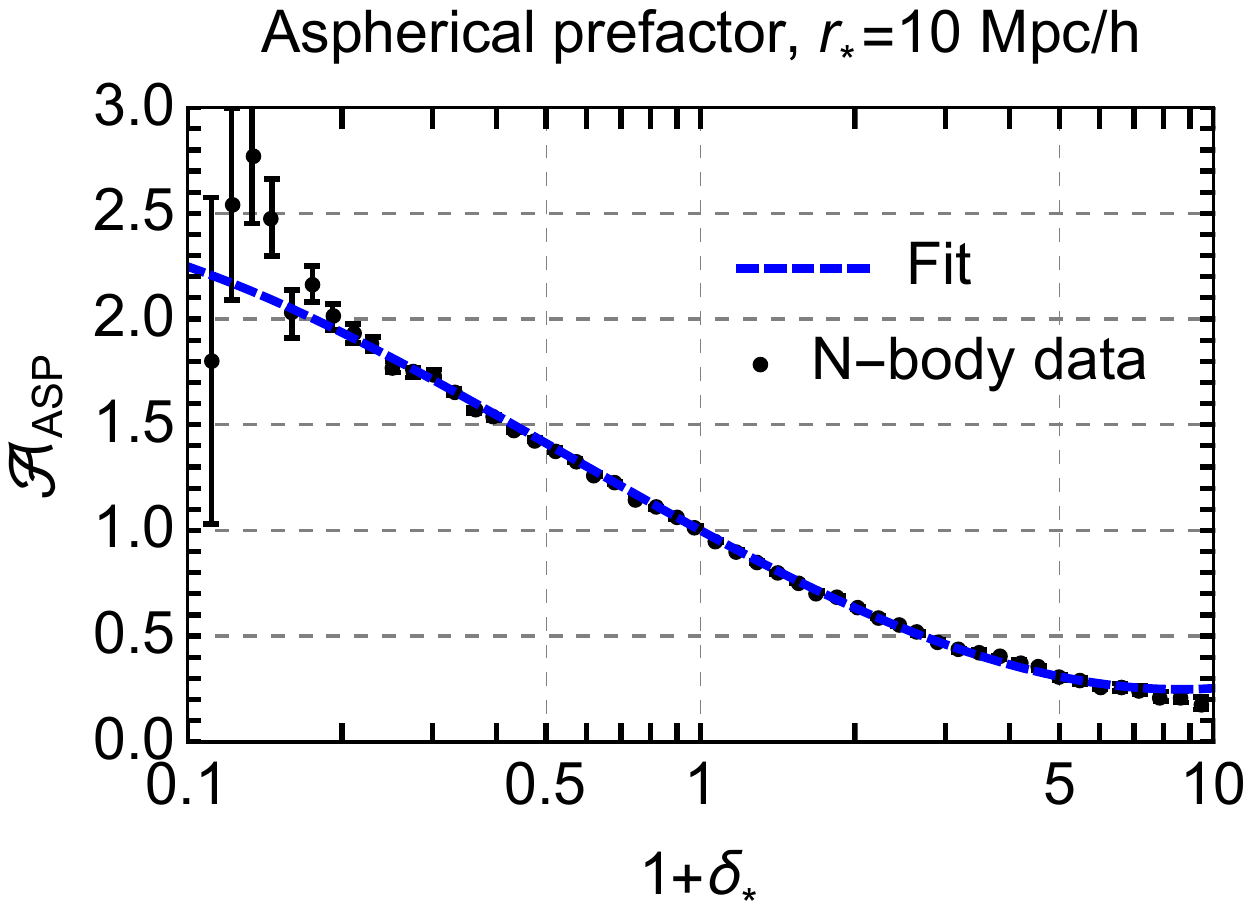}
\includegraphics[width=0.49\textwidth]{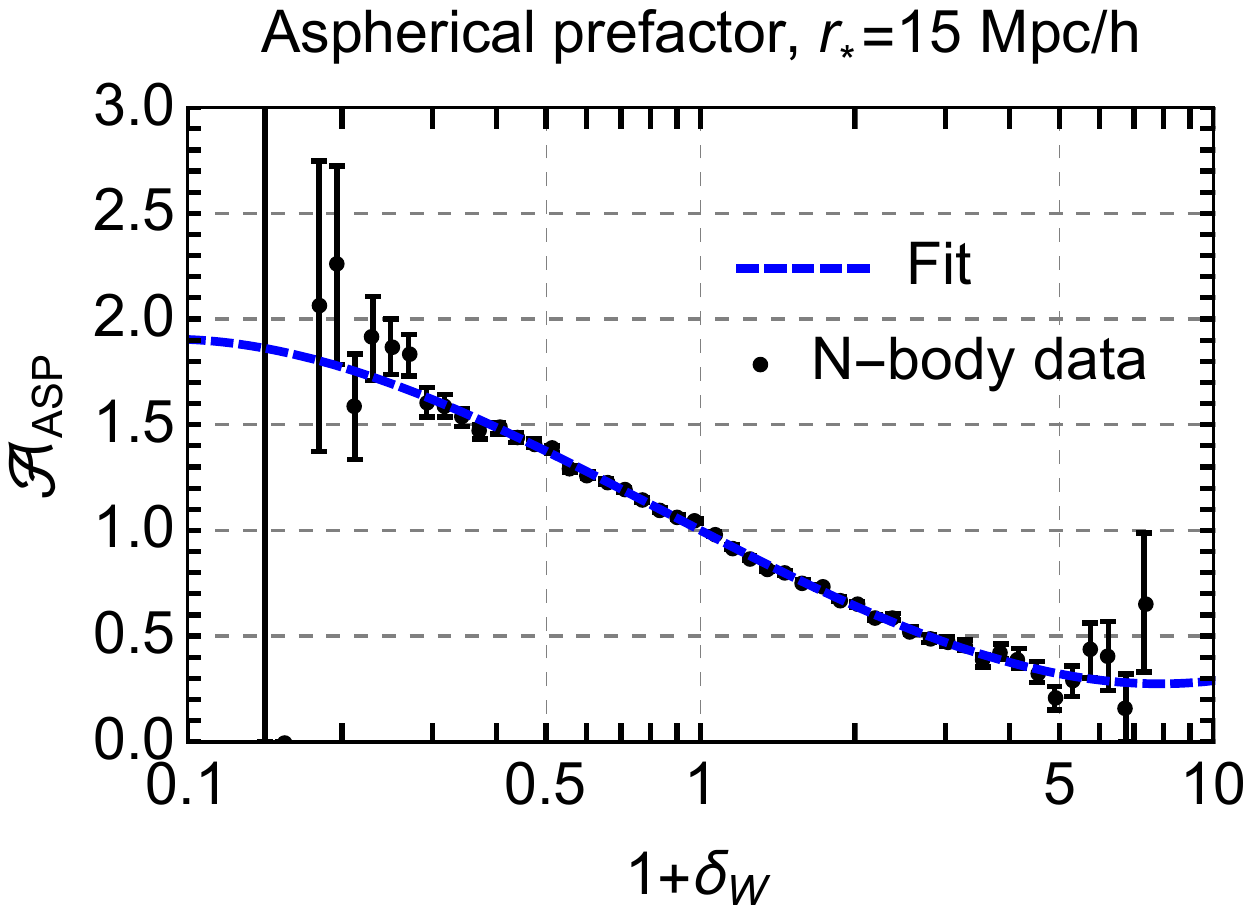}
\end{center}
\caption{\label{fig:fit}
The fitting formula for the aspherical prefactor 
\eqref{aspfit1} against the N-body data 
for $r_* = 10$ Mpc$/h$ (left panel) and $r_* = 15$ Mpc$/h$ (right panel).
All results are shown for $z=0$.
}
\end{figure}

We have checked that the N-body data are fully consistent with this
requirement. Namely, we fit the dependence ${\cal A}_{\rm
  ASP}(\delta_*)$ extracted from the data with the formula
\be 
\label{aspfit1}
\begin{split}
{\cal A}_{\text{ASP}}&=1+a_1\ln(1+\delta_*)+a_2\ln^2(1+\delta_*)
+a_3\ln^3(1+\delta_*)\,,
\end{split}
\ee
where we fix $a_1$ to the numerical values predicted by
Eq.~(\ref{meandeltaSP3}), whereas $a_2$ and $a_3$ are treated as free
parameters of the fit. The results of the fit are shown in
Fig.~\ref{fig:fit} and the parameters are summarized in
Table~\ref{tab:a123}. 
We observe that the expression (\ref{aspfit1}) accurately describes
the data throughout the whole available range of densities. In
particular, there is a perfect match between the slopes of the fitting
curve and the data at the origin. Note that the precise values of the
coefficients $a_2$, $a_3$ listed in Table~\ref{tab:a123}
should be taken with a grain of salt as
they are determined by the tails of the measured distribution, which are
subject to systematic errors. 

\begin{table}[h!]
\begin{center}
\begin{tabular}{|l|c|c|c|}
\hline
 & $a_1$ & $a_2$ & $a_3$ \\\hline
$r_*=10$ Mpc$/h$ & $-0.575$ & $~0.047~ $ & $~0.027~ $\\\hline
$r_*=15$ Mpc$/h$ & $-0.546$ & $~0.018~ $ & $~0.037~$ \\\hline
\end{tabular}
\caption{Parameters of the fitting formula (\ref{aspfit1}) for the
  aspherical prefactor for two different cell radii. The parameter
  $a_1$ is computed from Eq.~(\ref{meandeltaSP3}), and {\em is
    not} fitted from the data.
}
\label{tab:a123}
\end{center}
\end{table} 

We have seen that the aspherical prefactor is independent of the
linear growth factor. 
We also observe that the prefactor depends rather weakly on the size of the
window function. 
The leading response of the PDF to a
change in the cosmological model (such as
e.g. variation of the cosmological parameters or beyond-$\L$CDM
physics) will clearly enter through the exponent of the spherical part ${\cal
  P}_{\rm SP}$. 
The modification of the PDF due to the change of ${\cal A}_{\rm ASP}$
is expected to be subdominant.
Hence, for practical applications of the 1-point PDF to
constraining the cosmological parameters or exploring new physics one
can, in principle, proceed with the simple fitting formula
(\ref{aspfit1}) with the parameters extracted from N-body simulations
of a fiducial $\L$CDM cosmology.

Nevertheless, from the theoretical perspective, it is highly
instructive to perform the full first-principle calculation of the
aspherical prefactor. The rest of the paper is devoted to this task.  
In the four subsequent sections we derive and analyze the relevant
equations. A reader interested in the final results can jump directly to
Sec.~\ref{sec:results}.

\section{Perturbative calculation at small density contrast }
\label{sec:aspt}

In this section we compute the 
aspherical prefactor treating the saddle point configuration 
perturbatively. This approximation is valid at
small contrasts $|\delta_*|\ll 1$.
We will work at quadratic order in $\delta_*$ which, as we will see shortly, 
corresponds to the 1-loop order of standard perturbation theory.
We first consider standard cosmological perturbation theory (SPT)
\cite{Bernardeau:2001qr} 
and then discuss its extension,
the effective field theory (EFT) of large scale structure
\cite{Baumann:2010tm,Carrasco:2012cv}. 
Eventually, we are interested in large averaged density contrasts 
$|\delta_*|\sim 1$ where perturbation theory does not apply. Still, it
will serve us to grasp
important features of a fully non-linear calculation. 

It is convenient  to introduce an alternative representation of 
the aspherical prefactor. Let us multiply and divide the expression
(\ref{prefl>0}) by the square root of the monopole fluctuation
determinant 
\be 
\D_{0}= \text{det}\left[\mathbb{1}+2\hat\l \sqrt{P}Q_{0}\sqrt{P}\right]\,, 
\label{Dl0}
\ee
where $Q_0$ is the monopole response matrix introduced in
(\ref{perturbQl}). Next we observe that
\be
\label{Dlprod}
\begin{split}
\prod_{\ell =0}^\infty {\cal D}_\ell^{-(\ell+1/2)}
={\cal N}^{-1}\int&\mathscr{D} \delta_L^{(1)} 
\exp\bigg\{-\frac{1}{g^2}\bigg[\int_{\k}
\frac{\big(\delta_L^{(1)}(\k)\big)^2}{2P(k)}\\
&+\hat\lambda\int_{\k_1}\int_{\k_2}
Q_{tot}(\k_1,\k_2)\,\delta_L^{(1)}(\k_1)\,
\delta_L^{(1)}(\k_2)\bigg]\bigg\}\,,
\end{split}
\ee
where 
\be
\label{eq:defQtot}
 Q_{tot}(\k_1,\k_2)= 
\frac{1}{2}\frac{\d^2 \bar\delta_W}{\d \delta_L(\k_1) \d \delta_L(\k_2)}\,
\ee
is the {\it total quadratic response operator}. Note that it is
defined in the space of functions depending on the full 3-dimensional
wavevectors $\k$, unlike the partial multipole operators $Q_\ell$ defined
in the space of functions of the radial wavenumber $k$. The
expression on the r.h.s. of (\ref{Dlprod}) is the inverse square root
of the {\it total fluctuation determinant},
\be
\label{eq:Dtotal}
\D_{tot}=\text{det}[\mathbb{1}+2\hat\l \sqrt{P}Q_{tot}\sqrt{P}]\,.
\ee
In this way we obtain the following formula for the aspherical
prefactor, 
\be
\label{eq:AaspthruDtot}
\mathcal{A}_{\text{ASP}}=\sqrt{ \frac{\D_{0}}{{\D_{tot}}} }\,.
\ee
The monopole determinant ${\cal D}_0$ can be computed analytically for
any value of $\delta_*$, see Appendix~\ref{sec:spher}.
Note that, by itself, it does not have any physical meaning 
as the quadratic monopole fluctuations are already taken into account
in the monopole prefactor ${\cal A}_0$.  
The introduction of the monopole determinant is just a useful trick to
simplify the calculation, ${\cal D}_{tot}$ being more convenient to treat in
perturbation theory than the determinants in separate multipole
sectors.

\subsection{Fluctuation determinant in standard perturbation theory}
\label{sec:pt}

In order to find the response matrix we use the SPT solution
\cite{Bernardeau:2001qr} for the mildly non-linear density field, 
\be
\label{eq:spt}
\delta(\k)=\delta_L(\k)+\sum_{n=2}^{\infty}\int_{\k_1}...\int_{\k_n} (2\pi)^3\delta^{(3)}_D\Big(\k-\sum_i \k_i\Big)F_n(\k_1,...,\k_n)\prod_{i=1}^{n}\delta_L(\k_i)\,.
\ee
We work in the EdS approximation, where the SPT kernels $F_n$ are
redshift-independent, e.g.
\be
\label{eq:F2}
F_2(\k_1,\k_2)=\frac{17}{21}+(\k_1\cdot \k_2)\left(\frac{1}{2k_1^2}+ 
\frac{1}{2k_2^2} \right)+ \frac{2}{7}\left(\frac{(\k_1\cdot \k_2)^2}{k_1^2k_2^2}-\frac{1}{3}\right)\,.
\ee
We will discuss the EFT corrections later on.
Using \eqref{eq:spt} we obtain
\be
\label{qtotpt}
\begin{split}
 Q_{tot}(\k_1,\k_2)
 =\sum_{n=2}^{\infty}\frac{n(n-1)}{2}\int_{\q_1}...\int_{\q_{n-2}}
&F_n(\k_1,\k_2,\q_1,...,\q_{n-2})\\
&\times W_{\rm th}(|\k_{12}+\q_{1...n-2}|r_*)\prod_{i=1}^{n-2}\hat\delta_L(\q_i)\,,
\end{split}
\ee
where $\q_{1...m}\equiv \q_1+...+\q_{m}$.
We will keep only the first two terms in the expansion \eqref{qtotpt}:
\be
\label{eq:Qpt}
 Q_{tot}(\k_1,\k_2)=F_2(\k_1,\k_2)\,W_{\rm th}(|\k_{12}|r_*)
 +3\int_{\q}F_3(\k_1,\k_2,\q)\,
W_{\rm th}(|\k_{12}+\q|r_*)\,\hat\delta_L(\q) 
 \,.
\ee
An important comment is in order. The SPT kernels $F_n(\k_1,...,\k_n)$
are known to contain poles when one or several momenta vanish, see
e.g. the second term in (\ref{eq:F2}). These lead to the
so-called\footnote{For the realistic power spectrum there are no true
  divergences, but rather spurious enhanced contributions of soft
  modes.}
 `IR divergence' in the individual SPT loop integrals that cancel in the
final results for the correlation functions
\cite{Scoccimarro:1995if}. Equation (\ref{eq:Qpt}) implies that the
response matrix has IR poles when $\k_1$ or $\k_2$ (or both) tend to
zero. Nevertheless, we are going to see that the IR divergences
associated with these poles cancel in the determinant ${\cal
  D}_{tot}$. In other words, the aspherical prefactor, and hence the
full 1-point PDF, is IR safe.
In Sec.~\ref{sec:noIR} this property will be related to the
equivalence principle. 

To compute the determinant ${\cal D}_{tot}$, we make use of the trace formula,
\be 
\label{eq:dexp}
\begin{split}
\D_{tot} & =\exp\left\{\text{Tr}\ln \left(\mathbb{1}+2\hat\l
    \sqrt{P}Q_{tot}\sqrt{P}\right)\right\}\\
& \approx \exp\left\{
\left[-2\frac{\delta_*}{\s^2_{r_*}}+
6\frac{\delta_*^2}{\sigma^2_{r_*}}
\left(-\frac{4}{21}+\frac{\xi_{r_*}}{\sigma^2_{r_*}}\right)\right] 
\text{Tr}(PQ_{tot})-2\frac{\delta_*^2}{\s_{r_*}^4}
\text{Tr}(PQ_{tot}PQ_{tot})\right\}
\,,
\end{split}
\ee
where in the second line we perturbatively expanded the Lagrange
multiplier $\hat\l$ and kept only the terms that can contribute at order
$\delta_*^2$.
Let us first compute the leading-order contribution $O(\delta_*)$. 
From Eq.~\eqref{eq:Qpt} it is proportional to
\be 
\label{eq:qtotf2}
\begin{split}
\text{Tr}(Q_{tot}P)_{\text{LO}}=W_{\rm th}(0)\int_{\k}\,F_2(\k,-\k)P(k)\,.
\end{split}
\ee
But this vanishes due to $F_2(\k,-\k)=0$. Note that this property can
be traced back to the translational invariance. Indeed, the latter
implies conservation of momentum, so that at quadratic order of
SPT around homogeneous background one has,
\[
\delta(\k)=\delta_{L}(\k)+\int_\q F_2(\k-\q,\q)\,\delta_L(\k-\q)\delta_L(\q)\;.
\]
Averaging over the Gaussian initial conditions and recalling that
$\langle\delta(\k)\rangle=\langle\delta_L(\k)\rangle=0$ by
construction, one obtains that the integral entering (\ref{eq:qtotf2})
must vanish. As this should be true for any power spectrum, one
further infers vanishing of $F_2(\k,-\k)$.

At next-to-leading order one has,
\be 
\label{eq:p13}
\text{Tr}(Q_{tot}P)_{\text{NLO}}=3\frac{\delta_*}{\sigma^2_{r_*}}
\int_{\k}\int_{\q}F_3(\q,-\q,\k)P(k)P(q)|W_{\rm th}(kr_*)|^2\,.
\ee
This term is similar to the $P_{13}$-contribution 
to the filtered density variance in SPT. 
It is known to contain a spurious IR-enhancement, which cancels upon
adding the $P_{22}$ 
contribution, whose counterpart in our calculation is the rightmost
term in \eqref{eq:dexp},  
\be
 \text{Tr}(Q_{tot}P Q_{tot}P)=\int_{\k_1}\int_{\k_2}F_2^2(\k_1,\k_2)P(k_1)
 P(k_2)\big|W_{\rm th}(|\k_1+\k_2|r_*)\big|^2\,.
\ee
The net expression for the prefactor generated by total fluctuations reads:
\be 
\label{eq:Dtot}
\mathcal{A}_{tot}
\equiv 
\D_{tot}^{-1/2}
\approx 
\exp\left\{\frac{\delta_*^2}{2}\frac{\sigma^2_{\text{1-loop}}}{\sigma^4_{r_*}}
\right\}\,,
\ee
where we defined the filtered 1-loop density variance:
\bseq
\begin{align}
&\sigma^2_{\text{1-loop}}=\int_\k P_{\text{1-loop}}(k) |W_{\rm th}(kr_*)|^2\;,
\label{eq:1lvariance}\\
&P_{\text{1-loop}}(k)=
\int_\q \Big(2F_2^2(\k-\q,\q)P(q) P(|\k-\q|)+
6 F_3(\k,-\q,\q)P(q)P(k)\Big)\,.\label{eq:1lps}
\end{align}
\eseq
This result has an intuitive interpretation.
One can get expression~\eqref{eq:Dtot}
by replacing the linear matter power spectrum in
the density variance of the saddle-point
exponent \eqref{leadingexp} by its 1-loop version,
\be 
\label{eq:pert}
\exp\left\{-\frac{\delta_*^2}{2g^2(\sigma^2_{r_*}
+g^2\sigma^2_{\text{1-loop}})}\right\}
\approx 
\exp\left\{-\frac{\delta_*^2}{2g^2\sigma^2_{r_*}}
+\frac{\delta_*^2}{2}\frac{\sigma_{\text{1-loop}}^2}{\sigma^4_{r_*}}
\right\}
\,.
\ee
The replacement of the linear variance by the 1-loop expression in
\eqref{eq:pert}  
is reminiscent of the coupling constant renormalization 
due to radiative corrections in instanton calculations in QFT
(see e.g. \cite{tHooft:1976snw}).

\subsection{Effective field theory corrections}
\label{sec:EFT}

SPT does not capture correctly the effect of very short modes that
become deeply non-linear by $z=0$. This problem is addressed in EFT of
LSS. The latter augments the pressureless hydrodynamics equations
solved in SPT by the effective stress tensor, which is treated within
a gradient expansion
\cite{Baumann:2010tm,Carrasco:2012cv,Baldauf:2014qfa}. At the leading
(1-loop) order it produces the following correction (counterterm) to
the density contrast,
\be
\label{eq:dctr}
\delta_{\text{ctr}}(\k) = -\gamma(z) k^2\delta_L(\k)\,,
\ee
which must be added to the SPT expression (\ref{eq:spt}). Here
$\gamma(z)$ is a $z$-dependent coefficient with the dimension of
(length)$^2$
whose value and scaling
with $g(z)$ will be discussed below. 
Note that this contribution is linear in $\delta_L$.
However, it has the same order of magnitude as the one-loop correction
because the combination $\gamma k^2$ 
is assumed to be small according to the rules of gradient expansion.

Addition of the term (\ref{eq:dctr}) to the relation between linear
and non-linear density contrasts slightly modifies the saddle-point
solution. To find this correction we observe that, at the order we are
working, the final smoothed density contrast is related to the linear density
field as,
\be
\label{deltaWctr}
\bar\delta_W=\int_\k W_{\rm th}(kr_*)\,\delta_L(\k)\, (1-\gamma k^2).
\ee
Substituting this into the saddle-point equations (\ref{eq:sp}) we obtain,
\be
\label{deltahatctr}
\hat\delta_L=\frac{\delta_*}{\s^2_{r_*}}
\bigg(1+\frac{2\gamma\Sigma^2_{r_*}}{\s_{r_*}^2}\bigg)
P(k)W_{\rm th}(kr_*)(1-\g k^2)\;,
\ee
where
\be
\label{Sigmar*}
\Sigma^2_{r_*}=\int_\k |W_{\rm th}(kr_*)|^2\,P(k)\,k^2\;.
\ee
The modification of the saddle point produces a shift in the leading
exponent of the PDF and results in 
the following counterterm prefactor:
\be
\label{eq:ActrPT}
{\cal A}_{\text{ctr}}
 =\exp\left(-\delta_*^2\frac{\g(z)}{g^2(z)}\frac{\Sigma_{r_*}^2}{\sigma^4_{r_*}}
 \right)\,.
\ee

It is instructive to derive this result in an alternative way. One
recalls that the 1-loop SPT correction to the power spectrum
(\ref{eq:1lps}) receives a large contribution from short modes that
has the form (see e.g. \cite{Baldauf:2014qfa}),
\be
\label{eq:1lpsUV}
g^2 P_{\text{1-loop, UV}}(k)=\bigg(-\frac{61}{630\pi^2}\int_{q\gg k}
dq P(q)\bigg) g^2 k^2 P(k)\;.
\ee
This contributons would be divergent for a universe where the spectrum
$P(q)$ falls slower than $q^{-1}$ at $q\to \infty$. In EFT of LSS it
is renormalized by the counterterm $-2\gamma k^2P(k)$ coming from the
correction (\ref{eq:dctr}). Performing the renormalization inside the
filtered 1-loop density variance we obtain the expression,
\be
\label{eq:1lvarianceren}
\s_{\text{1-loop, ren}}^2=\s_{\text{1-loop}}^2-\frac{2\g}{g^2}\Sigma^2_{r_*}\;,
\ee 
which translates into the multiplication of the 1-loop
prefactor ${\cal A}_{tot}$ by the counterterm
(\ref{eq:ActrPT}). 

We obtain the value of the EFT coefficient $\g(z=0)$
by fitting the dark matter power spectrum of the
simulations\footnote{For the fit we use the power spectrum of the
  Horizon Run 2 \cite{Kim:2011ab} that has the same cosmology as
  assumed in this paper. This gives a better precision than our own
  simulations performed in relatively small boxes and contaminated by
  systematic errors at large scales.} 
at $z=0$
to the 1-loop IR-resummed theoretical template of \cite{Blas:2016sfa}. 
We follow Ref.~\cite{Baldauf:2016sjb} to include the theoretical error
in our analysis, which 
yields the following result:
\be 
\label{cs}
\g_0\equiv \g\big|_{z=0} =1.51\pm 0.07 \quad (\text{Mpc}/h)^2\,. 
\ee
In general, the redshift dependence of $\g$ should be also fitted from
the power spectrum in different redshift bins. In our analysis we use
a simplified model of a scaling universe \cite{Pajer:2013jj}.
In the range of wavenumbers $k\sim 0.1\,h/\text{Mpc}$ relevant for the
EFT considerations the broad-band part of the power spectrum can be
approximated as a power law \cite{Angulo:2014tfa,Assassi:2015jqa},
\be
\label{eq:psn}
P(k)\sim \frac{2\pi^2}{k_{NL}^3}\left(\frac{k}{k_{NL}}\right)^{n}\,,
\ee
where $k_{NL}$ is the non-linear scale and the spectral index is
estimated in the range $n\simeq -(1.5\div 1.7)$. In a universe with
such spectrum, the EFT coefficient is expected to scale as $\g\propto
k_{NL}^{-2}$, whereas $k_{NL}$ depends on the growth factor as 
$k_{NL}\propto \big(g(z)\big)^{\frac{n+3}{2}}$. This gives the
dependence, 
\be
\label{eq:gzed}
\g(z)=\g_0 \big(g(z)\big)^{\frac{4}{n+3}}\;.
\ee
It has been found consistent with the results of N-body
simulations \cite{Baldauf:2014qfa,Foreman:2015uva}. For numerical
estimates we will adopt the value $n=-3/2$ and the corresponding scaling
$\g(z)=\g_0 \big(g(z)\big)^{8/3}$.

\begin{figure}[t]
\begin{center}
\includegraphics[width=.49\textwidth]{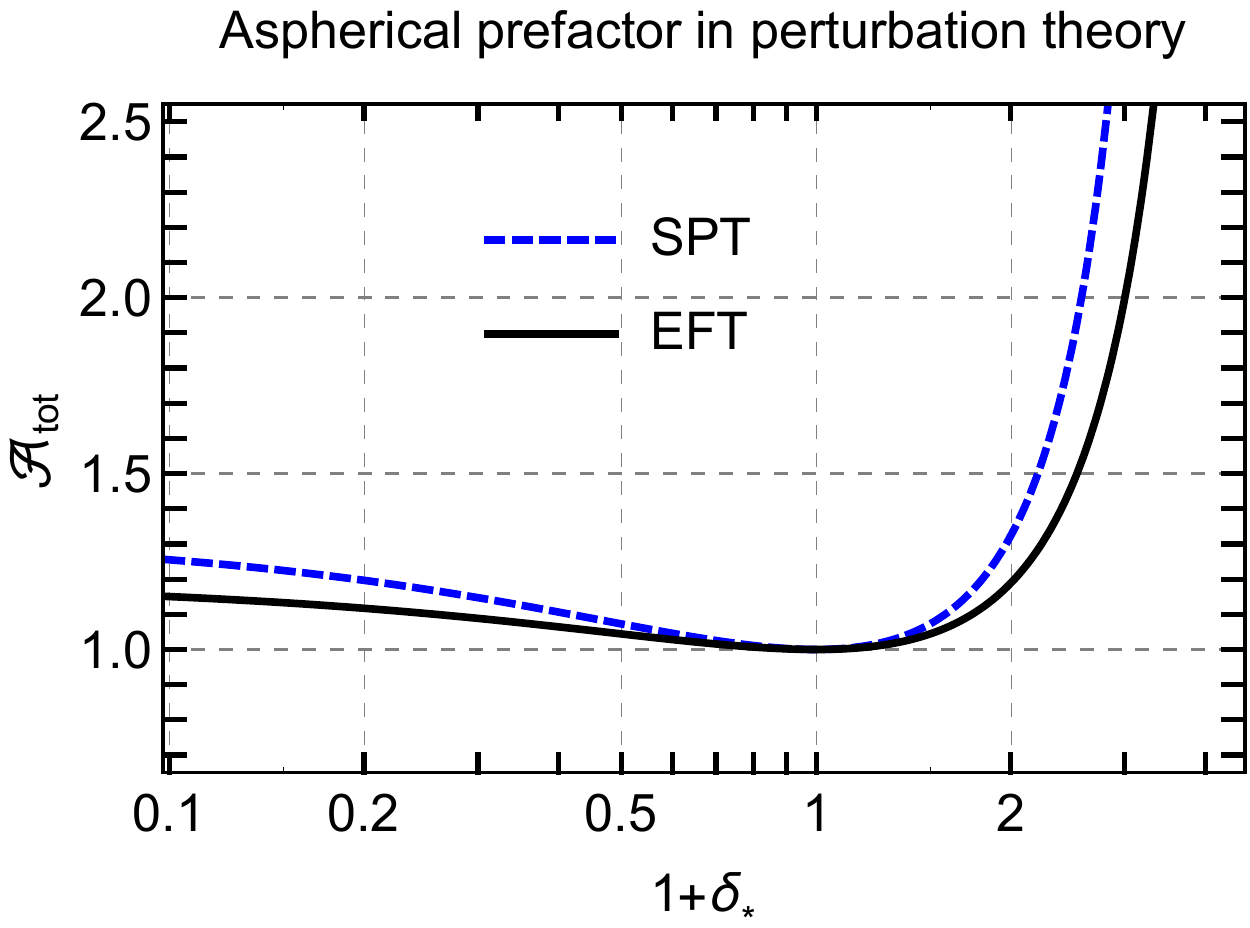}
\caption{\label{fig:aspPTspt} 
The prefactor ${\cal A}_{tot}$ due to quadratic fluctuations in perturbation theory
computed at 1-loop order in SPT and EFT. The results are shown at
$z=0$. Perturbation theory is strictly applicable in the neighborhood
of $\delta_*=0$.
}
\end{center}
\end{figure}

In Fig.~\ref{fig:aspPTspt} we compare the numerical results for
$\mathcal{A}_{tot}$ at $z=0$ computed in SPT and upon inclusion of the
EFT correction (we use the value $\g_0=1.5\,(\text{Mpc}/h)^2$). 
We see that the EFT correction 
has a sizable effect on the prefactor, and somewhat reduces its value.

\subsection{Aspherical prefactor at second order in background density}

In order to compute the full aspherical prefactor we have to combine
the total determinant  
with the spherical one, see Eq.~\eqref{eq:detl0}.
Unlike the total determinant, the spherical determinant differs from unity
at leading order 
in $\delta_*$ and yields
\be 
\label{eq:PTLO}
\begin{split}
\mathcal{A}^{\text{LO}}_{\text{ASP}}=\D_{0}^{1/2}
=
\exp\Bigg\{
\delta_*\left(\frac{4}{21}-\frac{\xi_{r_*}}{\s^2_{r_*}}\right)\Bigg\}\,.
\end{split} 
\ee
Remarkably, the aspherical prefactor at order $O(\delta_*)$ 
is fully controlled by translational invariance
which forces the corresponding terms in ${\cal A}_{tot}$
to vanish. 
Thus, the slope of the aspherical prefactor at the origin 
is encoded in the spherical collapse dynamics.
Note that this slope has precisely the value necessary to restore the zero
mean of the density contrast, Eqs.~(\ref{AASPderiv}),
\eqref{meandeltaSP3}.
This is an important consistency check of our approach.

\begin{figure}[t]
\begin{center}
\includegraphics[width=.49\textwidth]{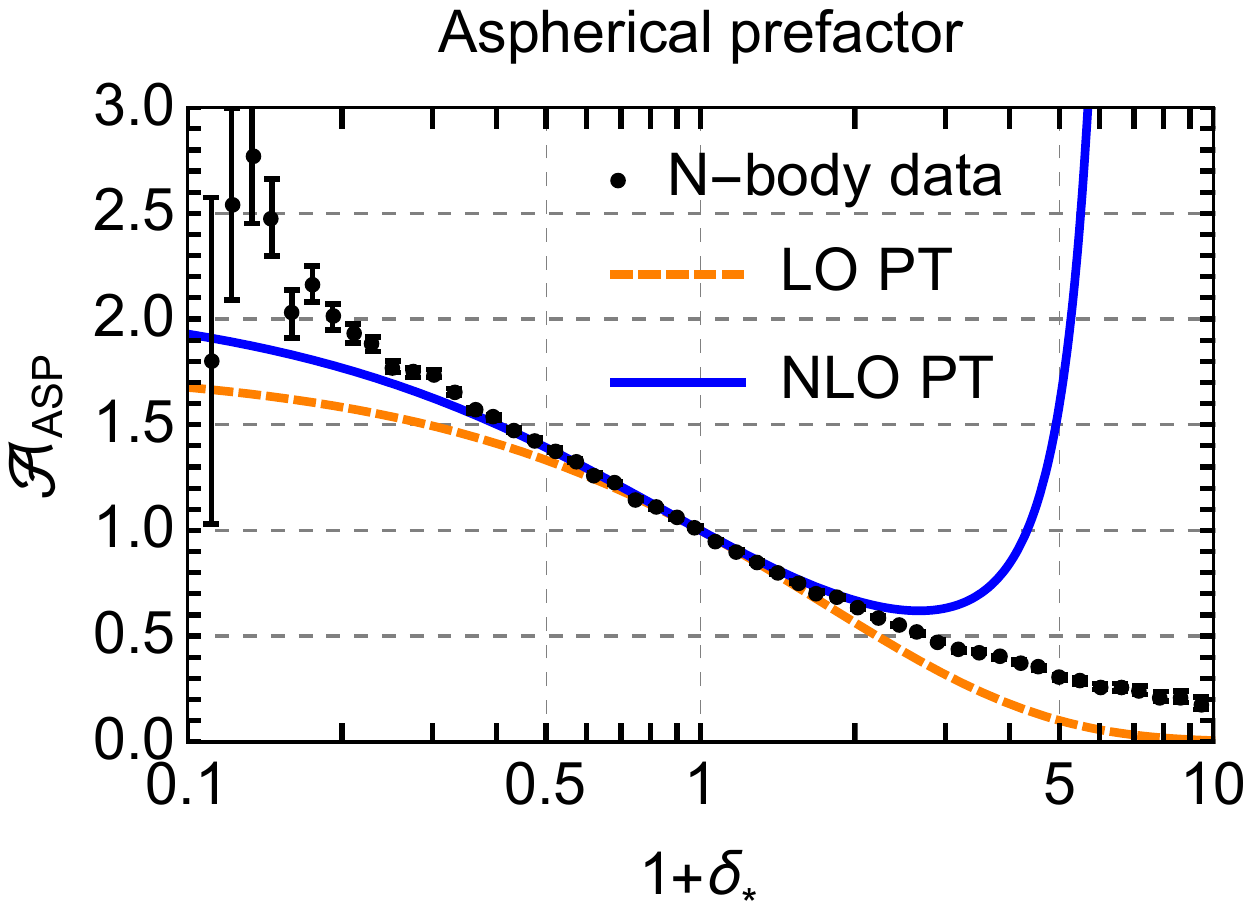}
\includegraphics[width=.5\textwidth]{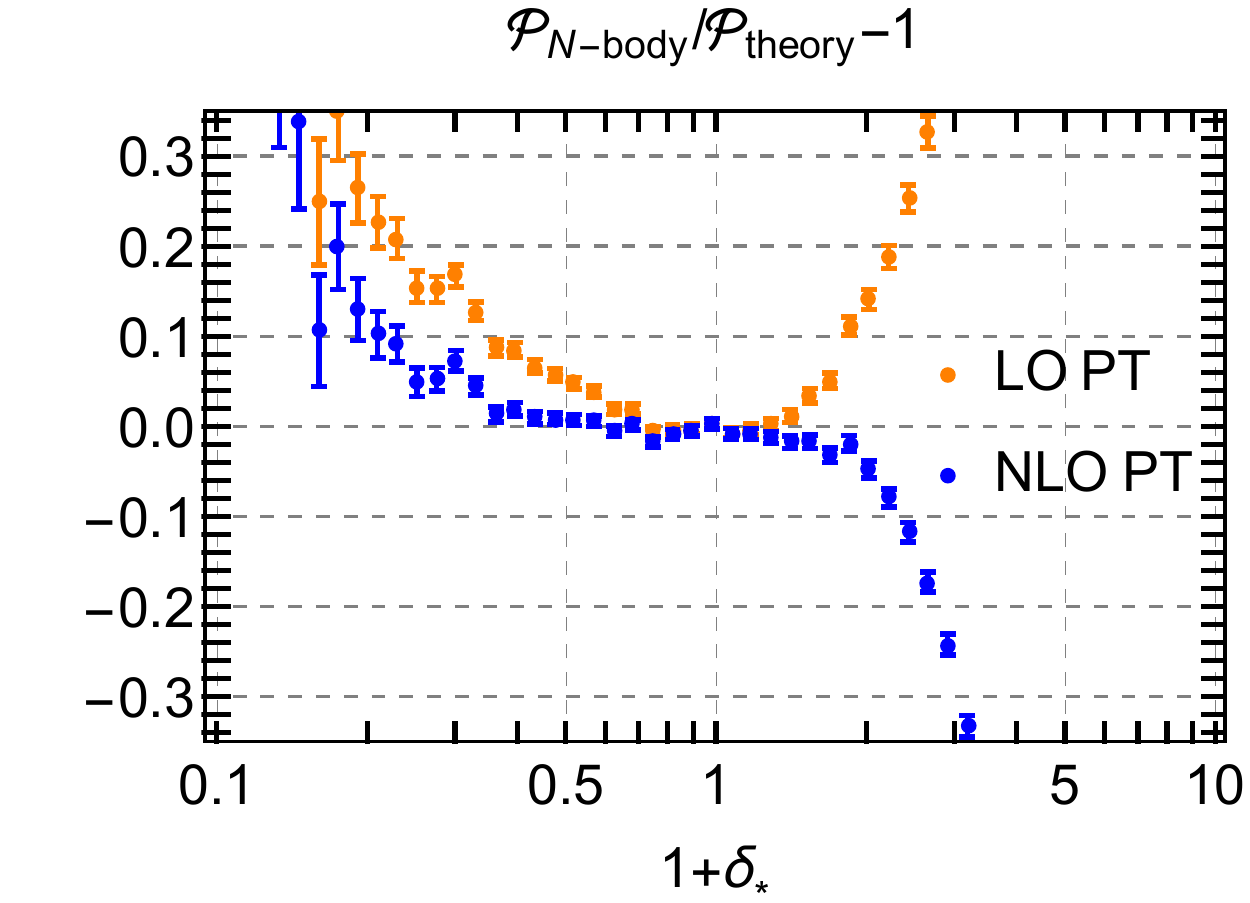}
\caption{\label{fig:aspPT} Left panel: the aspherical prefactor in
  perturbation theory at leading (LO) 
and next-to-leading (NLO) orders shown against the N-body data for cell radius \mbox{$r_*=10$~Mpc$/h$} at $z=0$.
Right panel: the corresponding residuals.
}
\end{center}
\end{figure}

Expanding the monopole determinant, one finds
at the next-to-leading order:
\be
\label{eq:PTaspfull}
\begin{split}
\mathcal{A}^{\text{NLO}}_{\text{ASP}}=\exp\Bigg\{&
\delta_*\left(\frac{4}{21}-\frac{\xi_{r_*}}{\s^2_{r_*}}\right)
+\frac{\delta_*^2}{2}\frac{\sigma^2_{\text{1-loop,
      ren}}}{\sigma^4_{r_*}}\\
&+\frac{\delta_*^2}{2}\Bigg(
-\frac{1180}{1323}
+\frac{40\xi_{r_*}}{21\sigma^2_{r_*}}
+\frac{r_*^2\Sigma^2_{r_*}}{3\s^2_{r_*}}
+\frac{\xi_{r_*}^2}{\s^4_{r_*}}
-\frac{3\s_{1\,r_*}^2}{\s_{r_*}^2}
\Bigg)
\Bigg\} \,,
\end{split}
\ee
where $\Sigma_{r_*}^2$ is defined in (\ref{Sigmar*}) and
\be
\s^2_{1\,r_*} = \int_\k
\left(\frac{\sin(kr_*)}{kr_*}\right)^2 P(k)\,.
\ee
In the left panel of Fig.~\ref{fig:aspPT} we show the aspherical
prefactor evaluated at leading and next-to-leading orders in
perturbation theory. 
We observe that the LO result works surprisingly well and 
does not deviate from the data by more than $10\%$ in the range 
$\delta_*\approx [-0.5,1]$,
while the NLO results extends the agreement up to $\delta_*\approx [-0.8,1.5]$.
In the right panel of Fig.~\ref{fig:aspPT} we show the 
residuals for the perturbation theory PDF.
One sees that the NLO corrections reduce the residuals close to the
origin, 
but quickly blow up towards large overdensities.  

One takes four main lessons from the perturbative calculation:
\begin{enumerate}
\item The response matrix contains spurious IR enhanced terms
that cancel in the determinant.
\item Including the aspherical corrections amounts, in part, to
  replacing the linear density 
variance by its non-linear version. 
\item The short-scale contributions should be renormalized by appropriate EFT 
counterterms. 
\item The slope of the aspherical prefactor at the origin is dictated
  by translational invariance and is such that the mean value 
  $\langle\delta_*\rangle$ vanishes.
\end{enumerate}

\section{Aspherical prefactor at large density contrasts: main
  equations} 
\label{sec:mainEq}

In Sec.~\ref{sec:aspfluc} we expressed the aspherical prefactor as the
product of fluctuation determinants in different multipole
sectors. Calculation of these determinants requires knowledge of the
aspherical response matrices $Q_\ell$. In this and the subsequent
section we set up the
equations for the determination of $Q_\ell$ that we will solve
numerically afterwards. For simplicity, we work in the EdS
approximation. The equations for $\L$CDM cosmology are summarized in
Appendix~\ref{sec:LCDM}. We have checked that the difference in the
final answers for the prefactor in $\L$CDM and in EdS does not exceed
1\%. Thus, the EdS approximation is vastly sufficient for our
purposes.

\subsection{Linearized fluctuations with $\ell > 0$}
\label{sec:5.1}

We first derive the evolution equations for linearized aspherical
perturbations in the background of the saddle-point solution. 
We start from the standard pressureless Euler--Poisson equations for
the density,  
peculiar velocity, and the Newtonian gravitational potential in an EdS
universe,
\bseq
\label{EP}
\begin{align}
\label{EP1}
&\frac{\d\delta}{\d t}+\d_i\big((1+\delta)u_i\big)=0\;,\\
\label{EP2}
&\frac{\d u_i}{\d t}+{\cal H}u_i+(u_j\d_j) u_i=-\d_i \Phi\;,\\
\label{EP3}
&\Delta \Phi=\frac{3{\cal H}^2}{2}\delta\;,
\end{align}
\eseq
where $t$ is conformal time, $\HH=\d_{t}a/a=2/t$ is the conformal
Hubble parameter and $a$ is the scale factor. 
We expand all quantities into background and first-order perturbations,
$\delta=\hat\delta+\delta^{(1)}$, etc. 
Next, we take the divergence of
(\ref{EP2}) and introduce the velocity potential $\Psi$:
\be
\label{PsiThet}
u_i^{(1)}=-{\cal H}\d_i \Psi^{(1)}~,~~~~
\d_i u_i^{(1)}=-{\cal H}\Theta^{(1)}\;.
\ee
From now on we also switch to a new time variable 
\be
\eta \equiv \ln a(t)\,,
\ee
To linear order in perturbations, the system (\ref{EP}) takes the
form,
\bseq
\label{EPp}
\begin{align}
\label{EPp1}
&\dot{\delta}^{(1)}-\Theta^{(1)}+{\cal H}^{-1}\hat u_i\,\d_i\delta^{(1)}
+{\cal H}^{-1}\d_i\hat u_i\,\delta^{(1)}-\d_i\hat\delta\,\d_i\Psi^{(1)}
-\hat\delta\,\Theta^{(1)}=0\;,\\
\label{EPp2}
&\dot{\Theta}^{(1)}+\frac{1}{2}\Theta^{(1)}-\frac{3}{2}\delta^{(1)}
+{\cal H}^{-1}\hat u_i\,\d_i\Theta^{(1)}
+{\cal H}^{-1}\d_i\d_j\hat u_j\,\d_i\Psi^{(1)}
+2{\cal H}^{-1}\d_i\hat u_j\,\d_i\d_j\Psi^{(1)}=0\;,\\
\label{EPp3}
&\Delta\Psi^{(1)}=\Theta^{(1)}\;,
\end{align}
\eseq
where dot denotes the derivative with respect to $\eta$.
Note that the background quantities have only radial
dependence and the velocity $\hat u_i$ has only the
radial component, so that
\be
{\cal H}^{-1}\d_i\hat u_j=-\d_i\d_j\hat\Psi=
\frac{x_ix_j}{r^2}\d_r^2\hat\Psi
+\bigg(\delta_{ij}-\frac{x_ix_j}{r^2}\bigg)\frac{\d_r\hat\Psi}{r}\;.
\ee 
We now expand the perturbations in spherical harmonics,
\be
\label{sphexp}
\delta^{(1)}({\bf x})=\sum_{\ell>0}\sum_{m=-\ell}^\ell 
Y_{\ell m}({\bf x}/r)\,\delta_{\ell m}(r)\;,
\ee
and similarly for the other fields.
To simplify notations, we have omitted the superscript `$(1)$' on the
multipole components of the fluctuations. 
In what follows we will also omit the azimuthal quantum number
$m$ as it does not appear explicitly in the equations. 
Substituting the expansion into Eqs.~(\ref{EPp}) we obtain,
\bseq
\label{EPppp}
\begin{align}
\label{EPppp1}
&\dot{\delta}_\ell-\Theta_\ell-\d_r\hat\Psi\,\d_r\delta_\ell
-\hat\Theta\,\delta_\ell-\d_r\hat\delta\,\d_r\Psi_\ell
-\hat\delta\,\Theta_\ell=0\;,\\
\label{EPppp2}
&\dot{\Theta}_\ell+\frac{1}{2}\Theta_\ell-\frac{3}{2}\delta_\ell
-\d_r\hat\Psi\,\d_r\Theta_\ell
-\d_r\hat\Theta\,\d_r\Psi_\ell
-2\d_r^2\hat\Psi\,\Theta_\ell\notag\\
&\qquad\qquad\qquad+2\bigg(\d_r^2\hat\Psi-\frac{\d_r\hat\Psi}{r}\bigg)
\,\bigg(\frac{2}{r}\d_r\Psi_\ell-\frac{\ell(\ell+1)}{r^2}\Psi_\ell\bigg)=0\;,\\
\label{EPppp3}
&\d_r^2\Psi_\ell+\frac{2}{r}\d_r\Psi_\ell-\frac{\ell(\ell+1)}{r^2}\Psi_\ell=\Theta_\ell\;.
\end{align}
\eseq
This is a system of (1+1)-dimensional partial differential equations
for the set of functions $(\delta_\ell,\T_\ell,\Psi_\ell)$.

To determine the initial conditions, we reason as follows.
At early times the saddle-point background vanishes and 
a solution to the previous system goes into 
\[
\delta_\ell(r)\to \e^\eta\, \delta_{L,\ell}(r)\;,
\]
where $\delta_{L,\ell}$ is a linear density field. 
Just like one decomposes linear perturbations over plane
waves in 3-dimensional space, we need to choose a basis of functions on the half-line which
are properly normalized w.r.t. to the radial integration measure,
\be
\label{normlmk}
\int_0^\infty dr\,r^2\,\delta_{L,\ell,k}^*(r)\delta_{L,\ell,k'}(r)
=(2\pi)^3k^{-2}\delta_D^{(1)}(k-k')\;.
\ee
The expression on the r.h.s. is the radial delta-function compatible
with the mo\-men\-tum-space measure $\int [dk]$, Eq.~(\ref{kintradial}). A
convenient basis with these 
properties is provided by the spherical Bessel functions (see
Appendix A),
\[
\delta_{L,\ell,k}(r)=4\pi\,j_\ell(kr)\;.
\] 
We conclude that the relevant initial conditions are,
\bseq
\label{linit}
\begin{align}
\label{linit1}
&\delta_{\ell,k}(r)=\Theta_{\ell,k}(r)=\e^{\eta}\cdot 4\pi j_\ell(kr)
\;,\\
\label{linit2}
&\Psi_{\ell,k}=-\e^\eta
\cdot \frac{4\pi}{k^2}\,j_\ell(kr)
~~~~~~~~~~~~\text{ at}~~\eta\to-\infty\;.
\end{align}
\eseq
In setting up the initial conditions for $\Psi$  we have used 
that Bessel functions are
eigenstates of the radial part of the Laplace operator, see
Eq.~(\ref{Besseleq}).

\subsection{Quadratic fluctuations in the monopole sector}
\label{2ndorder}

To find the response matrix, we need the second-order monopole perturbation
$\delta^{(2)}_0$ 
induced by a pair of first-order aspherical modes with a given
$\ell$. 
For simplicity, we will take the
latter in the form,
\be
\delta^{(1)}_k(r)=Y_{\ell ,m=0}(\x/r)\,\delta_{\ell,k}(r)\;,
\ee
so that, according to (\ref{deltalmprop}), $\delta_{\ell,k}$ is real. 
Let us first focus on the diagonal elements of the response matrix, i.e. consider 
the case when the fluctuation $\delta^{(2)}_0$ is sourced by two linear 
modes with the same wavenumber $k$. Generalization to a pair with
different wavenumbers will be discussed at the end of the subsection.
For compactness  
we will omit this wavenumber in the subscript of $\delta_\ell,
\Theta_\ell, \Psi_\ell$ in what follows. 

Expanding the Euler--Poisson
equations to the quadratic order and averaging over the angles we obtain,
\bseq
\label{EP2nd}
\begin{align}
\label{EP2nd1}
&\dot\delta_0^{(2)}-\Theta_0^{(2)}-\d_r\hat\Psi\,\d_r\delta_0^{(2)}
-\hat\Theta\,\delta_0^{(2)}-\d_r\hat\delta\,\d_r\Psi_0^{(2)}
-\hat\delta\,\Theta_0^{(2)}=\Xi_{\delta}\;,\\
\label{EP2nd2}
&\dot{\Theta}_0^{(2)}+\frac{1}{2}\Theta_0^{(2)}-\frac{3}{2}\delta_0^{(2)}
-\!\d_r\hat\Psi\,\d_r\Theta_0^{(2)}
-\!\d_r\hat\Theta\,\d_r\Psi_0^{(2)}\\
\nonumber
& ~~~~~~~~~~~~~~~~~~
-2\d_r^2\hat\Psi\,\Theta_0^{(2)}
+\frac{4}{r}\bigg(\d_r^2\hat\Psi-\frac{\d_r\hat\Psi}{r}\bigg)
\,\d_r\Psi_0^{(2)}=\Xi_\Theta\,,\\
\label{EP2nd3}
&\d_r^2\Psi_0^{(2)}+\frac{2}{r}\d_r\Psi_0^{(2)}=\Theta_0^{(2)}\;,
\end{align}
\eseq
where the sources on the r.h.s. are,
\be
\label{Xi}
\Xi_\delta=-\frac{1}{\cal H}\int \frac{d\Omega}{4\pi}\,
\,\d_i(\delta^{(1)}u_i^{(1)})~,\qquad
\Xi_\Theta=\frac{1}{{\cal H}^2}\int \frac{d\Omega}{4\pi}\,
\,\d_i(u_j^{(1)}\d_j u_i^{(1)})\;.
\ee
Performing the angular integration and using the Poisson equation
(\ref{EPppp3}) the sources can be cast in a suggestive form,
\be
\label{XiUps}
\Xi_\delta=\frac{1}{r^2}\d_r(r^2\Upsilon_\delta)~,\qquad
\Xi_\Theta=\frac{1}{r^2}\d_r(r^2\Upsilon_\Theta)\;,
\ee
where 
\bseq
\label{Ups}
\begin{align}
\label{Upsd}
\Upsilon_\delta&=\frac{1}{4\pi}\,\delta_\ell\d_r\Psi_\ell\;,\\
\Upsilon_\Theta&
=\frac{1}{4\pi}\,\bigg[\Theta_\ell\d_r\Psi_\ell-\frac{2}{r}(\d_r\Psi_\ell)^2
+\frac{2\ell(\ell+1)}{r^2}\Psi_\ell\d_r\Psi_\ell-\frac{\ell(\ell+1)}{r^3}\Psi_\ell^2\bigg]\;.
\label{UpsT}
\end{align}
\eseq

Let us introduce a second-order 
overdensity integrated over a sphere of radius\footnote{Note that we
  do not divide by the volume of the sphere, so $\mu^{(2)}$ differs
  from the spherically averaged density contrast by a factor
  $r_\eta^3/3$. }~$r_\eta$, 
\be
\label{mass*def}
\mu^{(2)}=\int_0^{r_\eta}dr\,r^2\,\delta_0^{(2)}(r)\;,
\ee
where we allow $r_\eta$ to be time dependent. We 
now show that if $r_\eta$ satisfies an appropriate evolution equation,
the system (\ref{EP2nd}) reduces to an ordinary differential equation
for $\mu^{(2)}$.  
It is convenient to work with the total quantities (background plus
second order perturbations), $\delta_0=\hat\delta+\delta_0^{(2)}$
etc. Then Eqs.~(\ref{EP2nd}) become,
\bseq
\label{0Xi}
\begin{align}
\label{0Xid}
&\dot\delta_0-\d_r\Psi_0\d_r\delta_0-(1+\delta_0)\Theta_0=\Xi_\delta\;,\\
\label{0XiT}
&\dot\Theta_0+\frac{1}{2}\Theta_0-\frac{3}{2}\delta_0
-\d_r\Psi_0\d_r\Theta_0-\Theta_0^2+\frac{2}{r^2}(\d_r\Psi_0)^2
+\frac{4}{r}\d_r^2\Psi_0\d_r\Psi_0=\Xi_\Theta\;,\\
\label{0XiP}
&\d_r(r^2\d_r\Psi_0)=r^2\Theta_0\;.
\end{align}
\eseq
Let us multiply the first equation by $r^2$ and integrate from $0$ to
$r_\eta$, 
\be
\int_0^{r_\eta}dr\;r^2\frac{\d\delta_0}{\d\eta}
-\int_0^{r_\eta}dr\;r^2\d_r\Psi_0\d_r\delta_0
-\int_0^{r_\eta}dr\;r^2(1+\delta_0)\Theta_0=
\int_0^{r_\eta}dr\;r^2\Xi_\delta\;.
\ee 
The last two terms on the l.h.s. combine into a total derivative due
to Eq.~(\ref{0XiP}). Also pulling the time derivative outside of the
integral in the first term we obtain,  
\be
\label{masschange}
\frac{d}{d\eta}\int_0^{r_\eta}dr\;r^2\big(1+\delta_0(r)\big)
-r_\eta^2 \big(1+\delta_0(r_\eta)\big) \big(\dot r_\eta+
\d_r\Psi_0(r_\eta)\big)
=r^2_\eta\Upsilon_\delta(r_\eta)\;.
\ee
The boundary terms on the l.h.s. cancel 
if we choose the time-dependence of $r_\eta$ in such a way
that 
\be
\label{RPsi}
\dot r_\eta=-\d_r\Psi_0(r_\eta)\;.
\ee
In other words, we shall choose the boundary to be moving with the
angular-averaged fluid velocity.
Then Eq.~(\ref{masschange}) simplifies,
\be
\label{masschange1}
\dot \mu
=r^2_\eta\Upsilon_\delta(r_\eta)\;,
\ee
where we introduced
\be
\label{massdef}
\mu=\int_0^{r_\eta}dr\;r^2\big(1+\delta_0(r)\big).
\ee
This equation has a clear physical
interpretation. 
It tells us that the mass inside a spherical region comoving with the
average spherical fluid flow changes due to the inflow through the
boundary generated by aspherical modes.

Equation (\ref{masschange1}) allows us to determine $\mu$ once the
time dependence of $r_\eta$ is known. However, we still need an
evolution equation for $r_\eta$ in terms of $r_\eta$ and $\mu$ to
close the system\footnote{Equation (\ref{RPsi}) is not sufficient as
  it involves the monopole velocity potential $\Psi_0$ which is
  unknown.}.
This is obtained from (\ref{0XiT}) by multiplying it 
with $r^2$
and integrating from $0$ to $r_\eta$. Using (\ref{XiUps}) and
(\ref{UpsT}) we obtain,
\be
r_\eta^2\bigg(\frac{\d}{\d\eta}\d_r\Psi_0
+\frac{1}{2}\d_r\Psi_0
-\d_r\Psi_0\d_r^2\Psi_0\bigg)\bigg|_{r_\eta}
-\frac{3}{2}\bigg(\mu-\frac{r^3_\eta}{3}\bigg)
=r_\eta^2\Upsilon_\Theta(r_\eta)\;.
\ee
It follows from (\ref{RPsi}) that 
\be
\ddot r_\eta=
\bigg(-\frac{\d}{\d\eta}\d_r\Psi_0+\d^2_r\Psi_0\d_r\Psi_0\bigg)\bigg|_{r_\eta}\;.
\ee
Thus, we arrive at
\be
\label{Rchange}
\ddot r_\eta+\frac{\dot r_\eta}{2}
-\frac{r_\eta}{2}+\frac{3\mu}{2r_\eta^2}=-\Upsilon_\Theta(r_\eta)\;.
\ee
There is again a transparent physical interpretation: the aspherical
perturbations exert an effective force on the spherical flow that
modifies its
acceleration. 

As a final step, we decompose $\mu$ and $r_\eta$ into background
values and second-order perturbations,
\bseq
\begin{align}
\label{R2def}
&r_\eta=\hat r_\eta+r^{(2)}_\eta\;,\\
\label{massdecomp}
&\mu=\hat\mu+\mu^{(2)}
+r_\eta^{(2)}\hat r_\eta^2\big(1+\hat\delta(\hat r_\eta)\big)\;,
\end{align}
\eseq
where $\hat r_\eta$, $\hat \mu$ satisfy source-free
Eqs.~(\ref{masschange1}), (\ref{Rchange}). Subtracting the background
contributions from the evolution equations we obtain,
\bseq
\label{mu2r2}
\begin{align}
\label{mass*change}
&\dot \mu^{(2)}+\dot r_\eta^{(2)}
\hat r_\eta^2\big(1+\hat\delta(\hat r_\eta)\big)
+r_\eta^{(2)} \frac{d}{d\eta}\Big(
\hat r_\eta^2\big(1+\hat\delta(\hat r_\eta)\big)\Big)
=\hat r_\eta^2 \Upsilon_\delta(\hat r_\eta)\;\\
\label{dR2}
&\ddot r^{(2)}_\eta+\frac{\dot r_\eta^{(2)}}{2}
+\bigg(1+\frac{3}{2}\hat\delta(\hat r_\eta)
-\frac{R_*^3}{\hat r_\eta^3}\bigg) r^{(2)}_\eta
+\frac{3}{2\hat r_\eta^2}\mu^{(2)}
=-\Upsilon_\Theta(\hat r_\eta)\;,
\end{align}
\eseq
where we have used the asymptotics $\hat r_\eta\to R_*$, $\hat\mu\to R^3_*/3$ at
$\eta\to -\infty$. Equations (\ref{mu2r2}) provide a closed system of
linear ordinary differential equations for the variables 
$\mu_*^{(2)}$, $r^{(2)}_\eta$ once the sources $\Upsilon_{\delta,\T}$
are known.

We must supplement (\ref{mu2r2}) by three boundary conditions. One
of them is set at the final time and expresses the fact that we are
interested in the overdensity within the fixed radius $r_*$, so that
the final radius is not perturbed,
\bseq
\label{init}
\be
\label{0init}
r^{(2)}_\eta\big|_{\eta=0}=0\;.
\ee
The conditions at the initial time $\eta\to-\infty$ are more
subtle. The source-free Eqs.~(\ref{mu2r2}) admit solutions
corresponding to first order monopole fluctuations, that can also
change the mass within the cell. We need to eliminate such
solutions. For this purpose, we observe that for the spurious solutions
the fields $\delta_0$ etc. behave as $\e^\eta$ at early times, whereas
the second-order perturbations that we are interested in are
proportional to $\e^{2\eta}$. We conclude that we must require,
\be
\label{1init}
\mu^{(2)}\propto \e^{2\eta}~,\quad \text{at}~\eta\to-\infty\;. 
\ee 
As for $r_\eta^{(2)}$, it need not vanish in the beginning. Rather, it
should approach a constant value in a specific way. Indeed, from
Eqs.~(\ref{masschange1}), (\ref{massdecomp}) and the fact that
$\dot{\hat\mu}$ vanishes we conclude that the derivative of the
combination $r_\eta^{(2)}\hat r_\eta^2\big(1+\hat\delta(\hat r_\eta)\big)$
must fall off as $\e^{2\eta}$. Thus, we obtain the third condition,
\be
\label{2init}
\dot r_\eta^{(2)}
+r_\eta^{(2)}\frac{d}{d\eta}\ln\big[\hat r^2_\eta\big(1+\hat\delta(\hat
r_\eta)\big)\big] \propto \e^{2\eta}~,\quad \text{at}~\eta\to-\infty\;.
\ee
\eseq

It is straightforward to generalize the above analysis to the case
when the second-order perturbation is sourced by a pair of aspherical
modes with different radial wavenumbers $k$ and $k'$ (but, of course,
the same angular numbers $\ell$ and $m$). In that case the sources
(\ref{Ups}) are replaced by symmetric combinations constructed from
the two modes, 
\bseq
\label{Upsnond}
\begin{align}
\label{Upsdnond}
\Upsilon_{\delta,kk'}&=\frac{1}{8\pi}\,\delta_{\ell, k}\d_r\Psi_{\ell,k'}+ (k\longleftrightarrow k')\;,\\
\nonumber
\Upsilon_{\Theta,kk'}&
=\frac{1}{8\pi}\,\bigg[\Theta_{\ell,k}\d_r\Psi_{\ell,k'}-\frac{2}{r}\d_r\Psi_{\ell,k}
\d_r\Psi_{\ell,k'}
+\frac{2\ell(\ell+1)}{r^2}\Psi_{\ell,k}\d_r\Psi_{\ell,k'}
\\
&~~~~~~~~~~-\frac{\ell(\ell+1)}{r^3}\Psi_{\ell,k}\Psi_{\ell,k'}\bigg]
+ (k\longleftrightarrow k')
\;.
\label{UpsTnond}
\end{align}
\eseq
The rest of the derivation goes exactly the same as above, leading to
Eqs.~(\ref{mu2r2}) with the new sources.

\subsection{Summary of the algorithm}
\label{sec:algorithm}

Summarizing the results of this section, one obtains the following
algorithm to find the response matrix $Q_\ell(k,k')$ and the
fluctuation determinant ${\cal D}_\ell$:
\begin{enumerate}
\item One solves Eqs.~(\ref{EPppp}) with the initial
conditions (\ref{linit}) 
and finds the mode functions $\delta_\ell,\Theta_\ell,\Psi_\ell$
for each basis function from a set of $N+1$ momenta $\{k_0,...,k_{N}\}$.

\item One uses these solutions to construct the
sources (\ref{Upsnond}) 
for a pair of wavevectors $k_i$ and $k_j$.

\item 
One solves (\ref{mu2r2})
with
the initial
conditions \eqref{init}\,.
The final variation in the 
averaged overdensity
gives the element of $Q_\ell$,
\be
\label{Qdiag}
Q_\ell(k_i,k_j)=\frac{3}{r_*^3}\mu^{(2)}(\eta=0)\;. 
\ee

\item 
One repeats the above procedure for all different pairs of wavenumbers
$(k_i,k_j)$, construct the operator
$\mathbb{1}+2\hat\l\sqrt{P}Q_\ell\sqrt{P}$ and evaluates its determinant.
\end{enumerate}

The implementation of this algorithm on a discrete grid is
described in Appendix~\ref{sec:aspy}.

The algorithm requires a modification in the dipole sector
($\ell=1$) due to the IR sensitivity of the matrix $Q_1$. 
We now focus on this issue.

\section{Removing IR divergences in the dipole contribution}
\label{sec:dipole}

A complication
arises in the dipole sector ($\ell=1$). The initial conditions
(\ref{linit2}) imply that the velocity potential 
$\Psi_{1,k}\propto \e^\eta\cdot  r/k$ has an $1/k$ pole\footnote{This
  problem does not arise for higher multipoles. The Bessel functions
behave at the origin as $(kr)^\ell$, and hence the corresponding velocity potential $\Psi_{\ell,k}$ is regular
at $k\to 0$ for $\ell>1$.}
 at $kr\sim
kr_*\ll 1$. 
Substitution of this
expressions into equations of motion (\ref{EPppp1}), (\ref{EPppp2}) 
leads to further $1/k$ contributions in $\delta_{1,k}$ and $\T_{1,k}$ 
proportional to the derivatives
of the background 
configuration $\d_r\hat\delta$, $\d_r\hat\Theta$. Thus, the linear solution
$(\delta_1,\Theta_1,\Psi_1)$ contains singular $1/k$ terms
which translate into first- and second-order poles 
in the matrix 
$Q_{1}(k,k')$ at $k,k'\ll 1/r_*$. As we discuss below, 
these infrared (IR) enhanced contribution
must cancel in the determinant ${\cal D}_1$
entering the prefactor (\ref{prefl>0}), which is IR-safe\footnote{We
  have already seen this cancellation in the perturbative calculation
  in Sec.~\ref{sec:pt}.}.
However, the presence of
the `IR-divergent'\footnote{Here the term `divergence' is used in the
  sense adopted in the perturbation theory literature, where it
refers to the fact that 
loop integrals would be divergent in IR 
for power-law spectra $P(k)\propto k^n$ with
$n \leq -1$.  
The $\L$CDM power spectrum vanishes quickly at small $k$, so the loop 
integrals are actually convergent, albeit strongly enhanced.} 
terms makes a straightforward 
numerical evaluation of the determinant unfeasible. 
The purpose of this section is to show that the
IR-enhanced contributions can be isolated and the IR-divergences can 
be removed, whereby reducing the task to numerical
evaluation of IR-safe quantities only. 

\subsection{IR safety of the prefactor}
\label{sec:noIR}

We start by showing that the aspherical prefactor (\ref{prefl>0}) is IR
safe. We first give a heuristic argument and then a more direct proof.
Let us assume that the mapping from the linear to non-linear density
fields is invertible\footnote{This would be true in the absence of
  shell-crossing, but in general is not correct.}. Then the
counts-in-cells PDF can be written in the schematic form,
\be
\label{pdf}
{\cal P}(\delta_*)={\cal N}^{-1}\int \mathscr{ D}\delta\,\int \frac{d\l}{2\pi i g^2}\,
\exp\left\{-\frac{\Gamma[\delta]}{g^2}+\frac{\l}{g^2}
\left(\delta_*-\bar\delta(r_*)\right)\right\}\;,
\ee
where 
the path integral runs over all density configurations at the final
moment of time and $\Gamma[\delta]$ is a weighting functional obtained
from the Gaussian weight using the map $\delta\mapsto \delta_L$. A
perturbative expansion for the functional $\Gamma[\delta]$ was derived
in \cite{Blas:2015qsi} and it was shown that all coefficients in this
expansion are IR-safe. Extrapolating this property to the
non-perturbative level, one concludes that the matrix of second
variational derivatives around the saddle-point solution 
\be
\frac{\d^2
  \Gamma}{\d\delta^{(1)}(\x)\,\d\delta^{(1)}(\x')}\bigg|_{\delta=\hat\delta(\x)}\; 
\ee 
is also IR-safe. The prefactor of the PDF is given by the determinant
of this matrix, hence it is IR-safe as well. 

We now give a more rigorous argument that does not require
invertibility of the density mapping. We split the integration
variables in the path integral (\ref{eq:pdfLaplace}) into soft ($k\ll
1/r_*$) and hard ($k\gtrsim 1/r_*$) modes. Omitting for clarity the
normalization factors we obtain,
\be
{\cal P}(\delta_*)
=\int \mathscr{ D}\delta^{\rm soft}_L 
\exp\bigg\{-\int_\k \frac{|\delta_L^{\rm soft}(\k)|^2}{2g^2P(k)}\bigg\}
\; {\cal P}[\delta_*;\delta_L^{\rm soft}]\;,
\ee
where 
\be
\label{pdfsoft}
{\cal P}[\delta_*;\delta_L^{\rm soft}]\equiv 
\int  \mathscr{ D}\delta_L^{\rm hard}\, d\l
\,\exp\left\{-\frac{1}{g^2}\bigg[\int_\k\frac{|\delta_L^{\rm hard}(\k)|^2}{2P(k)}
-\l\big(\delta_*-\bar\delta_W[\delta_L^{\rm hard}+\delta_L^{\rm soft}]
\big)\bigg]\right\}\;,
\ee
has the physical meaning of the PDF for short-scale 
overdensities in the background of soft modes. 

Now, the addition of a soft mode corresponds to immersion of the
system into a large-scale flow. Due to the equivalence principle, 
the main effect of such flow is
an overall translation of the hard modes by a distance 
proportional to the gradient
of the Newtonian potential 
\cite{Kehagias:2013yd,Peloso:2013zw,Blas:2013bpa,
Creminelli:2013mca,Horn:2014rta}. 
In other words,
\be
\label{deltashift}
\delta[\delta_L^{\rm hard}+\delta_L^{\rm soft}](\x,\eta)
=\delta[\delta_L^{\rm hard}]\Big(\x-\frac{\nabla}{\Delta}
\delta_L^{\rm soft}(0,\eta),\eta\Big) 
+\ldots\;.
\ee 
The shift is enhanced for long-wavelength perturbations
leading to $1/k_{\rm soft}$ poles in the perturbative expansion of the expression
(\ref{deltashift}) in $\delta_L^{\rm soft}$. 
On the other hand, the remaining terms represented by 
dots in (\ref{deltashift}) contain more derivatives acting on the
Newtonian potential, and thus 
are regular in the limit when the soft momentum $k_{\rm soft}$
goes to zero.

The PDF (\ref{pdfsoft}) can be evaluated in the
saddle-point approximation. 
The saddle-point solution is
\be
\label{saddleshift}
\hat\delta_L[\delta_L^{\rm soft}](\x)
=\hat\delta_L\Big(\x+\frac{\nabla}{\Delta}\delta_L^{\rm soft}|_{\x=0}\Big)\;,
\ee
where $\hat\delta_L$ is the saddle-point configuration in the absence
of soft modes. Likewise, the fluctuations around the
solution (\ref{saddleshift}) are obtained from those around
$\hat\delta_L$ by the same translation, so that the integral over them
does not contain any poles. We
conclude that ${\cal P}[\delta_*;\delta_L^{\rm soft}]$ is IR-safe which
implies the IR safety of the original PDF~${\cal P}(\delta_*)$.

\subsection{Factorization of IR divergences}
\label{sec:IRisol}
At $\ell=1$ the equations (\ref{EPppp}) admit an exact solution 
\be
\label{shift}
\delta_1=\d_r\hat\delta\cdot \e^\eta~,~~~~
\Theta_1=\d_r\hat\Theta\cdot \e^\eta~,~~~~
\Psi_1=\d_r\hat\Psi\cdot \e^\eta+r\cdot\e^\eta\;.
\ee
Notice that in the far past all contributions here vanish faster than
$\e^\eta$ (actually, as $O(\e^{2\eta})$), except for the last term in
$\Psi_1$. The latter corresponds
to a uniform motion of all fluid elements\footnote{Recall that
the gradient of $\Psi$ is proportional to the fluid velocity, see
eq.~(\ref{PsiThet}).}, i.e. to a large bulk flow.
Existence of the solution (\ref{shift}) follows from the equivalence principle 
obeyed by the Euler-Poisson equations. Indeed, we can
impose on any solution an infinitely large bulk flow that 
will sweep
the original solution as a whole. The dipolar solution
(\ref{shift}) precisely corresponds to imposing such a large bulk flow
on the saddle-point configuration $(\hat\delta,\hat\Theta,\hat\Psi)$. 

The solution (\ref{shift}) can be added with an arbitrary
coefficient
to any other solution of
eqs.~(\ref{EPppp}). In particular, the perturbation  
with the initial conditions
(\ref{linit}) for $\ell=1$ can be written as,
\bseq
\label{solut}
\begin{align}
\label{solut1}
&\delta_{1,k}=\breve\delta_k-\frac{4\pi}{3k}\d_r\hat\delta\,\e^\eta\;,\\
\label{solut2}
&\Theta_{1,k}=\breve\Theta_k-\frac{4\pi}{3k}\d_r\hat\Theta\,\e^\eta\;,\\
\label{solut3}
&\Psi_{1,k}=\breve\Psi_k-\frac{4\pi}{3k}\d_r\hat\Psi\,\e^\eta-\frac{4\pi
r}{3k}\,\e^\eta\;,
\end{align}
\eseq
where the triple
$\big(\breve\delta_k,\breve\Theta_k,\breve\Psi_k\big)$
is also a solution of eqs.~(\ref{EPppp}) satisfying the same initial
conditions (\ref{linit1}) for $\breve\delta_k$, $\breve\Theta_k$, but
with modified
initial condition for $\breve\Psi_k$,
\be
\label{brevePsiinit}
\breve\Psi_k=\bigg[-\frac{4\pi}{k^2}j_{1}(kr)
+\frac{4\pi r}{3k}\bigg]\cdot \e^\eta\;.
\ee
Importantly, this modification eliminates the dangerous $1/k$ pole, so
that the initial conditions for all functions
$(\breve\delta_k,\breve\Theta_k,\breve\Psi_k)$ are regular at $k\to
0$. In fact,
\be
\label{brevesmallk}
\breve\delta_k,\breve\Theta_k,\breve\Psi_k=O(k)\;. 
\ee
Then, by linearity of eqs.~(\ref{EPppp}), this property holds at all
times.  

The next step is to isolate the IR divergences in the sources
$\Upsilon_\delta$, $\Upsilon_\Theta$. Substituting (\ref{solut}) into
(\ref{Upsnond}), we obtain
\bseq
\label{Upsdec}
\begin{align}
\label{Uspdec1}
&\Upsilon_{\delta,kk'}=\frac{A_\delta}{kk'}+\frac{B_{\delta,k'}}{k}+
\frac{B_{\delta,k}}{k'}+\breve\Upsilon_{\delta,kk'}\;,\\
\label{Upsdec2}
&\Upsilon_{\Theta,kk'}=\frac{A_\Theta}{kk'}+\frac{B_{\Theta,k'}}{k}+
\frac{B_{\Theta,k}}{k'}+\breve\Upsilon_{\Theta,kk'}\;,
\end{align}
\eseq
where
\bseq
\label{Upscoeff}
\begin{align}
\label{Upscoeff1}
&A_\delta=\frac{4\pi}{9}\d_r\hat\delta(\d^2_r\hat\Psi
+1)\,\e^{2\eta}\;,\\
\label{Upscoeff2}
&B_{\delta,k}=-\frac{1}{6}\big[(\d_r^2\hat\Psi+1)\breve\delta_k+
\d_r\hat\delta\d_r\breve\Psi_k\big]
\,\e^{\eta}\;,\\
\label{Upscoeff3}
&A_\Theta= \frac{4\pi}{9}
\bigg[\d_r\hat\Theta(\d^2_r\hat\Psi+1)
-\frac{2}{r}(\d_r^2\hat\Psi)^2
+\frac{4}{r^2}\d_r\hat\Psi\d_r^2\hat\Psi
-\frac{2}{r^3}(\d_r\hat\Psi)^2\bigg]\,\e^{2\eta}
\;,\\
\label{Upscoeff4}
&B_{\Theta,k}\!=\!-\frac{1}{6}\bigg[(\d_r^2\hat\Psi\!+\!1)\breve\Theta_k
\!+\!\bigg(\!\d_r\hat\Theta\!-\!\frac{4}{r}\d_r^2\hat\Psi
\!+\!\frac{4}{r^2}\d_r\hat\Psi\!\bigg)\d_r\breve\Psi_k
\!+\!\bigg(\frac{4}{r^2}\d_r^2\hat\Psi\!-\!\frac{4}{r^3}\d_r\hat\Psi\!\bigg)
\breve\Psi_k\bigg]\,\e^\eta 
\,,
\end{align}
\eseq
and $\breve\Upsilon_{\delta,kk'}$, $\breve\Upsilon_{\Theta,kk'}$ are
computed using the regular solutions 
$(\breve\delta_{k},\breve\Theta_{k},\breve\Psi_{k})$, 
$(\breve\delta_{k'},\breve\Theta_{k'},\breve\Psi_{k'})$. 
Due to linearity of Eqs.~(\ref{mu2r2}), the pole structure of the
sources (\ref{Upsdec}) propagates into the 
pole structure of the matrix
\be
Q_1(k,k')=\frac{A}{kk'}+\frac{B(k')}{k}+\frac{B(k)}{k'}+\breve Q(k,k')\;,
\ee
where $A$, $B$, $\breve Q$ are found by solving Eqs.~(\ref{mu2r2})
with the sources 
$(A_\delta,A_\Theta)$, $(B_\delta,B_\Theta)$ and $(\breve\Upsilon_\delta,\breve
\Upsilon_\Theta)$ respectively.
Due to the property (\ref{brevesmallk}) we have
\be
\label{BbQsmallk}
B(k)=O(k)~,~~~~ \breve Q(k,k')=O(kk')~~~\text{at}~k,k'\to 0\;. 
\ee

We now observe that the sought-for determinant has the form,
\be
\label{D1-1}
\D_{1}=
\det\Big(\mathbb{1}+2\hat\l\sqrt{P}\breve Q\sqrt{P}+a(k)b(k')+b(k)a(k')\Big)\;,
\ee
with
\be
\label{ab}
a(k)=\hat\l\frac{\sqrt{P(k)}}{k}~,~~~~~
b(k)=\bigg(\frac{A}{k}+2B(k)\bigg)\sqrt{P(k)}\;.
\ee
Denoting
\be
\label{Mkk}
M(k,k')=(2\pi)^3k^{-2}\delta_{\rm D}^{(1)}(k-k')
+2\hat\l\sqrt{P(k)}\breve Q(k,k')\sqrt{P(k')}
\ee
we write,
\be
\label{D1decomp}
\D_{1}=\det M\cdot \D^{IR}\;,
\ee
where all IR-sensitive contributions have been collected into
\be
\label{D1IR1}
\D^{IR}=\det\big(\mathbb{1}+\tilde a\otimes \tilde b+\tilde b\otimes
\tilde a\big)\;.
\ee
We have 
introduced $\tilde a=M^{-1/2}a$, $\tilde b=M^{-1/2}b$ and
used the fact that the matrix $M$ is symmetric.

The determinant \eqref{D1IR1} can be easily
evaluated using Eq.~(\ref{detvec}) from Appendix~\ref{app:determinant},
\be
\label{D1IR2}
\D^{IR}=1+2(\tilde a\cdot \tilde b)+(\tilde a\cdot \tilde
b)^2-\tilde a^2\,\tilde b^2\;.
\ee 
Here dot denotes the scalar product, 
$$
\tilde a\cdot\tilde b=\int [dk]\, \tilde a(k)\tilde b(k)
=\int [dk][dk']\, a(k) M^{-1}(k,k')b(k')
\;,
$$
and similarly for $\tilde a^2$ and $\tilde b^2$. The inverse matrix
$M^{-1}$ has the form, 
\be
\label{Minv}
M^{-1}=\mathbb{1}-2\hat{\l}\sqrt{P}\QQ \sqrt{P}
\ee
with
\be
\label{Qring}
\QQ(k,k')=\breve Q(k,k')+\ldots=O(kk')~~~\text{at small}~k,k'\;.
\ee
Using this property one isolates the `IR-divergences' in the different
terms\footnote{
As in perturbation theory,
for the realistic power spectrum these terms are
finite, but still dangerously enhanced. They would be actually 
divergent if the power spectrum behaved as $P(k)\propto
k^{n}$ with $n \leq -1$ at small $k$.} in (\ref{D1IR2}), 
\bseq
\begin{align}
2(\tilde a\cdot \tilde b) & \ni 2\hat\l\int [dk] \frac{P(k)}{k^2}A
\label{det1}\\
(\tilde a\cdot\tilde b)^2&-\tilde a^2\tilde b^2
\ni -4\hat{\l}^2\int[dk]^2\frac{P(k_1)}{k_1^2}P(k_2)
\big(B(k_2)\big)^2\notag\\
&\qquad\qquad\;+8\hat{\l}^3\int[dk]^3\frac{P(k_1)}{k_1^2}P(k_2)P(k_3)\QQ(k_2,k_3)
B(k_2)B(k_3)\;.
\label{det2}
\end{align}
\eseq
A necessary and sufficient condition for their cancellation in the 
determinant (\ref{D1IR2})~is, 
\be
\label{detcancel}
A=2\hat{\l}\int[dk]^2\sqrt{P(k_1)P(k_2)}B(k_1)M^{-1}(k_1,k_2)B(k_2)\;.
\ee 
While we do not have a direct proof of this identity,  
the arguments of the previous subsection imply that it
must be satisfied. We also checked it numerically and found that it is
fulfilled in our computations within the accuracy of the numerical
procedure. 

Using (\ref{detcancel}) we can simplify the expression (\ref{D1IR2}).
A straightforward calculation yields,
\be
\label{D1IR3}
\D^{IR}=\bigg[1+2\hat{\l}\int [dk]^2\sqrt{P(k_1)P(k_2)}
\;\frac{1}{k_1}M^{-1}(k_1,k_2)B(k_2)\bigg]^2
\;.
\ee
This is the final expression to be used in numerical evaluation. The
algorithm for the computation of ${\cal D}_1$ consists of the
following steps: 
\begin{enumerate}
\item Solve the linear equations (\ref{EPppp}) with initial conditions
  (\ref{linit1}), (\ref{brevePsiinit}) to find the functions $\breve
  \delta_k$, $\breve\Theta_k$, $\breve\Psi_k$;
\item Find the matrix $\breve Q(k,k')$ by solving
  Eqs.~(\ref{mu2r2}) with the sources
  $\breve\Upsilon_{\delta,kk'}$, $\breve\Upsilon_{\Theta,kk'}$;
\item Find the vector $B(k)$ by solving
  Eqs.~(\ref{mu2r2}) with the sources
  $B_{\delta,k}$,  $B_{\Theta,k}$,\\ Eqs.~(\ref{Upscoeff2}),
  (\ref{Upscoeff4});
\item Construct the matrix $M(k,k')$, Eq.~(\ref{Mkk}), compute its
  determinant and inverse;
\item Use the inverse matrix $M^{-1}(k,k')$ and the vector 
  $B(k)$ to compute the IR
  contribution (\ref{D1IR3});
\item Compute the full determinant in the dipole sector as a product
  of $\det M$ and $\D^{IR}$.
\end{enumerate}

\section{WKB approximation for high multipoles}
\label{sec:WKB}

In general the computation of the aspherical fluctuation determinant 
requires solving the system of linear partial differential equations
(\ref{EPppp}) on a grid. However, in the sectors with large orbital
numbers 
$\ell \gg 1$ one can
use the Wentzel-Kramers-Brillouin (WKB) technique to simplify the
problem and gain insights 
into the structure of the aspherical response matrix. 
Remarkably, in the WKB regime the system \eqref{EPppp} reduces to a system of
ordinary differential equations and can be easily solved, e.g. in
\texttt{Mathematica}. 
The WKB analysis serves
both to cross check the results of the full numerical integration of
Eqs.~(\ref{EPppp}) and to study the UV sensitivity of
the aspherical prefactor in Sec.~\ref{sec:counterterm}.

We start by noticing that the basis functions (\ref{linit}) are
suppressed at $kr<\ell$ due to the centrifugal barrier. Indeed, at these
values of $r$
we obtain using Eq.~(10.19.3) from \cite{DLMF},
\be
\label{Jsmallr}
j_{\ell}\big((\ell +1/2)\vk r\big)\sim 
\frac{e^{-(\ell+1/2)\big({\rm arcch}\frac{1}{\vk r}-\sqrt{1-(\vk
      r)^2}\big)}}{(2\ell+1)\sqrt{\vk r}[1-(\vk r)^2]^{1/4}}
\;, 
\ee
where we have introduced the ratio
\be
\label{vk}
\vk\equiv\frac{k}{\ell +1/2}
\ee
which will be kept fixed in the limit $\ell \to \infty$. We see that
$j_{\ell}(kr)$ is exponentially suppressed at $\vk r<1$. Thus, if
$k r_*\ll \ell$ the perturbation has support outside of the
window function and does not contribute into the variation of the
overdensity: $Q_\ell(k,k')\approx 0$ whenever $k$ or $k'$ is much
smaller than $\ell/r_*$. We conclude
that the dominant contribution into the response matrix comes from the modes
with  
\be
\label{klarge}
k\gtrsim \ell/r_* \gg 1/r_*\;.
\ee
These modes oscillate much faster than the background, so we can use
the WKB technique to find their evolution. 

We will see that we have to
go up to the second order in the WKB expansion, hence we write
the following Ansatz:
\be
\label{WKBform}
\begin{split}
&\delta_\ell=(\delta_{\ell 1}+k^{-1}\delta_{\ell 2})\e^{ikS_\ell}+{\rm h.c.}~,~~\\
&\Theta_\ell=(\Theta_{\ell 1}+k^{-1}\Theta_{\ell 2})\e^{ikS_\ell}+{\rm h.c.}~,~~\\
&\Psi_\ell=\!(k^{-2}\Psi_{\ell 1}+k^{-3}\Psi_{\ell 2})\e^{ikS_\ell}+{\rm h.c.}
\end{split}
\ee
where $\delta_{\ell 1},\delta_{\ell 2}$ etc. are slowly varying functions.
Note that
$\Psi_\ell$ is suppressed by two powers of $k$ compared to
$\delta_\ell$ and $\Theta_\ell$.
From the Poisson equation (\ref{EPppp3}) we find at leading order
\bseq
\be
\label{PsiLO}
\Psi_{\ell 1}=-\frac{\Theta_{\ell 1}}{(S_\ell')^2+(\vk r)^{-2}}\;.
\ee
The next-to-leading expansion yields,
\be
\label{PsiNLO}
\Psi_{\ell 2}=\frac{-\Theta_{\ell 2}+iS_\ell''\Psi_{\ell 1}+2iS_\ell'\Psi_{\ell 1}'
+\frac{2iS_\ell'}{r}\Psi_{\ell 1}}{(S_\ell')^2+(\vk r)^{-2}}\;.
\ee
\eseq
Further, we susbstitute the form (\ref{WKBform}) into the dynamical
Eqs.~(\ref{EPppp1}), (\ref{EPppp2}). At leading order $O(k)$ both
equations reduce to
\be
\label{Seq}
\dot S_\ell -\d_r\hat\Psi S_\ell'=0\;.
\ee
In the combination on the l.h.s. one recognizes the time-derivative
along the background flow, so one concludes that $S_\ell$ is conserved along
the flow,
\be
\label{flowcons}
\frac{d S_\ell}{d\eta}\bigg|_{\rm flow}=0\;.
\ee
In other words, $S_\ell(r,\eta)=S_\ell\big(R(r,\eta)\big)$, where $R$ is
the Lagrangian coordinate of the spherical shell in the background
solution. 
It
is related to the Eulerian coordinate $r$ and $\eta$ by
Eq.~(\ref{eq:LagR}) where for the density contrast one takes the
saddle-point profile $\hat\delta(r,\eta)$. At $\eta\to-\infty$ the
Lagrangian and Eulerian coordinates coincide, $R=r$.

From the orders $O(1)$ and $O(1/k)$ of Eqs.~(\ref{EPppp1}),
(\ref{EPppp2}) we obtain the equations for the coefficient functions
in the WKB Ansatz:\\
Next-to Leading Order,
\bseq
\label{WKBNLO}
\begin{align}
\label{WKBNLO1}
&\frac{d\delta_{\ell 1}}{d\eta}\bigg|_{\rm flow}-\hat\Theta\delta_{\ell 1}
-(1+\hat\delta)\Theta_{\ell 1}=0\;,
\\
\label{WKBNLO2}
&\frac{d\Theta_{\ell 1}}{d\eta}\bigg|_{\rm flow}-\frac{3}{2}\delta_{\ell 1}
+\bigg[\frac{1}{2}
-\frac{2(\vk r S_\ell')^2\d_r^2\hat\Psi}{1+(\vk r S_\ell')^2}
-\frac{2\d_r\hat\Psi}{r(1+(\vk r S_\ell')^2)}\bigg]\Theta_{\ell 1}=0\;.
\end{align}
\eseq
Next-to-Next-to-Leading Order,
\bseq
\label{WKBNNLO}
\begin{align}
\label{WKBNNLO1}
\frac{d\delta_{\ell 2}}{d\eta}\bigg|_{\rm flow}-&\hat\Theta\delta_{\ell 2}
-(1+\hat\delta)\Theta_{\ell 2}=i\d_r\hat\delta\,S_\ell'\Psi_{\ell 1}\;,
\\
\frac{d\Theta_{\ell 2}}{d\eta}\bigg|_{\rm flow}-&\frac{3}{2}\delta_{\ell 2}
+\bigg[\frac{1}{2}
-\frac{2(\vk r S_\ell')^2\d_r^2\hat\Psi}{1+(\vk r S_\ell')^2}
-\frac{2\d_r\hat\Psi}{r(1+(\vk r S_\ell')^2)}\bigg]\Theta_{\ell 2}\notag\\
=&\frac{4iS_\ell'}{1+(\vk rS_\ell')^2}
\bigg(\d_r^2\hat\Psi-\frac{\d_r\hat\Psi}{r}\bigg)
\Psi_{\ell 1}'\notag\\
&+\bigg[i\d_r\hat\Theta S_\ell'+\frac{2i}{1+(\vk rS_\ell')^2}
\bigg(S_\ell''-\frac{2S_\ell'(\vk rS_\ell')^2}{r}\bigg)
\bigg(\d_r^2\hat\Psi-\frac{\d_r\hat\Psi}{r}\bigg)\bigg]\Psi_{\ell 1}
\;.
\label{WKBNNLO2}
\end{align}
\eseq
We notice that Eqs.~(\ref{WKBNLO}) do not contain
spatial derivatives of $\delta_{\ell 1},\Theta_{\ell 1}$, so that they form a
system of ordinary differential equations for these functions. The
same is true for Eqs.~(\ref{WKBNNLO}) with respect to the
functions $\delta_{\ell 2},\Theta_{\ell 2}$.

To set up the initial conditions we use the asymptotic expansion for
the Bessel function at large order (Eq.~(10.19.6) from \cite{DLMF}),
\be
\label{asymplargel}
j_\ell\bigg(\!\frac{\ell\!+\!1/2}{\cos \beta}\!\bigg)
=\frac{\cos\b}{(\ell\!+\!1/2)\sqrt{\sin \b}}
\bigg(\!\!\cos\xi+\frac{1}{8(\ell\!+\!1/2)}
\bigg(\!\ctg\beta+\frac{5}{3}(\ctg\beta)^3\bigg)\sin\xi
+O(\ell^{-2})\!\bigg)\;,
\ee
where
\be
\xi=(\ell+1/2)(\tg \b-\b)-\pi/4\;.
\ee
Substituting this into (\ref{linit}) and comparing with the WKB Ansatz
(\ref{WKBform}) we find the initial conditions at $\eta\to-\infty$,
\bseq
\label{WKBinit}
\begin{align}
\label{Slinit}
&S_\ell=\frac{1}{\vk}\bigg[\sqrt{(\vk r)^2-1}-\arccos\frac{1}{\vk
  r}\bigg]\;,\\
\label{dl1init}
&\delta_{\ell 1}=\Theta_{\ell 1}=-\Psi_{\ell 1}=\frac{\e^\eta}{\ell+1/2}
\cdot\frac{2\pi}{\sqrt{\vk r}\,[(\vk r)^2-1]^{1/4}}\cdot e^{-i\pi/4}\;,\\
\label{dl2init}
&\delta_{\ell 2}=\Theta_{\ell 2}=-\Psi_{\ell 2}=\frac{\e^\eta}{\ell+1/2}
\cdot \frac{\pi\vk}{4\sqrt{\vk r}}\bigg(\frac{1}{[(\vk r)^2-1]^{3/4}}
+\frac{5}{3[(\vk r)^2-1]^{7/4}}\bigg)\cdot e^{-i3\pi/4}\;.
\end{align}
\eseq
Equation (\ref{flowcons}) with the initial conditions (\ref{Slinit})
is readily solved giving
\be
\label{Ssol}
S_\ell=\frac{1}{\vk}\bigg[\sqrt{(\vk R)^2-1}-\arccos\frac{1}{\vk R}\bigg]\;.
\ee
We observe that in the large-$\ell$ limit the function $S_\ell$ becomes
universal ($\ell$-independent).
The WKB approximation is valid as long as
\be
\label{WKBval}
S_\ell''/(S_\ell')^2\ll k
\ee
which is equivalent to
\be
\label{WKBval1}
|\vk R-1|\gg (\ell+1/2)^{-2/3}\;.
\ee
Next, Eqs.~(\ref{WKBNLO}) for the first-order WKB coefficients can be
integrated numerically along the flow lines (i.e. at fixed $R$)
starting from the initial conditions
(\ref{dl1init}). We will see
shortly that the functions $\delta_{\ell 1}$, $\T_{\ell 1}$ need to be evaluated
only in the vicinity of the flow line $R=R_*$ corresponding to the
boundary of the spherical region that collapses to the cell of radius
$r_*$ at the final time. Knowing $\T_{\ell 1}$, one finds $\Psi_{\ell 1}$ by
Eq.~(\ref{PsiLO}) and inserts it in the r.h.s. of
(\ref{WKBNNLO}). Finally, Eqs.~(\ref{WKBNNLO}) are integrated at fixed
$R$ starting from the initial configuration (\ref{dl2init}). Again, we
will need $\delta_{\ell 2}$, $\T_{\ell 2}$ only at $R_*$. Notice that
the r.h.s. of (\ref{WKBNNLO}) involves the radial derivative
$\Psi_{\ell 1}'$. Thus, evaluating the first-order functions precisely
at $R_*$ would be insufficient: one needs to know them in a small vicinity
of this point\footnote{Alternatively, one can take radial derivatives
  of Eqs.~(\ref{WKBNLO}) and in this way obtain a system of ordinary
  differential equations for $\delta_{\ell 1}'$, $\T'_{\ell 1}$. Then 
$\Psi_{\ell 1}'$ is computed from $\T_{\ell 1}'$ by using the radial
derivative of Eq.~(\ref{PsiLO}).}. 

The factor $(\ell+1/2)^{-1}$ in the initial 
conditions (\ref{dl1init}), (\ref{dl2init}) implies that the 
WKB solution is suppressed in the limit $\ell\to \infty$.
This leads to a suppression of the sources $\Upsilon_{\delta,\T}$
appearing on the r.h.s. of Eqs.~(\ref{mu2r2}) and hence a suppression
of the response matrix $Q_\ell$. Then the fluctuation determinant can
be approximated using the trace formula,
\be
{\cal D}_\ell\approx \exp\left(2\hat{\l}\,\text{Tr}PQ_\ell\right)\,,
\ee 
and for its calculation it suffices to focus on the diagonal elements
of the response matrix $Q_\ell(k,k)$.
The sources for these elements are obtained by substituting the WKB
solution with a single wavenumber $k$ into (\ref{Ups}). For the first
source this yields,
\be
\begin{split}
\Upsilon_\delta=&\frac{1}{4\pi}
\bigg[\frac{iS_\ell'}{k}\delta_{\ell 1}\Psi_{\ell 1}
+\frac{1}{k^2}\Big(\delta_{\ell 1}\Psi_{\ell 1}'+iS_\ell'\delta_{\ell 1}\Psi_{\ell 2}
+iS_\ell'\delta_{\ell 2}\Psi_{\ell 1}\Big)\bigg]\,\e^{i2kS_\ell}\\
&+\frac{1}{4\pi k^2}\Big(\delta_{\ell 1}^*\Psi_{\ell 1}'+iS_\ell'\delta_{\ell 1}^*\Psi_{\ell 2}
+iS_\ell'\delta_{\ell 2}^*\Psi_{\ell 1}\Big)+{\rm h.c.}
\end{split}
\ee
The term in the first line is quickly oscillating. In the eventual
integral over $k$ that appears in 
the $Q_\ell$-trace 
it will average
to zero. Neglecting it we get,
\bseq
\label{UpsWKB}
\be
\label{UpsdWKB}
\overline{\Upsilon_\delta(\hat r_\eta)}=
\frac{1}{2\pi k^2}\bigg[\tilde\delta_{\ell 1}\frac{\d\tilde\Psi_{\ell 1}}{\d
  R}
+\frac{\d S_{\ell}}{\d R}
(\tilde\delta_{\ell 1}\tilde\Psi_{\ell 2}-\tilde\delta_{\ell 2}\tilde\Psi_{\ell 1})\bigg]
\frac{\d R}{\d r}\bigg|_{R_*}\;,
\ee
where the overline means averaging over the oscillations. 
Here we denoted by tildes the functions with the
complex phases stripped off\footnote{In other words, $\tilde\delta_{\ell 1}\equiv
  \delta_{\ell 1}\e^{i\pi/4}$, $\tilde\delta_{\ell 2}\equiv
  \delta_{\ell 2}\e^{i3\pi/4}$ and so on.} and switched from the Eulerian
to the Lagrangian
radial coordinate $R$. Similarly, for the source $\Upsilon_\Theta$ we
have,
\be
\label{UpsTWKB}
\begin{split}
\overline{
\Upsilon_\Theta(\hat r_\eta)}=
\frac{1}{2\pi k^2}\bigg\{&\bigg[\tilde\Theta_{\ell 1}\frac{\d\tilde\Psi_{\ell 1}}{\d
  R}
+\frac{\d S_{\ell}}{\d R}
(\tilde\Theta_{\ell 1}\tilde\Psi_{\ell 2}-\tilde\Theta_{\ell 2}\tilde\Psi_{\ell 1})
+\frac{2}{(\vk\hat r_\eta)^2}
\tilde\Psi_{\ell 1}\frac{\d\tilde\Psi_{\ell 1}}{\d R}
\bigg]
\frac{\d R}{\d r}\\
&-\frac{2}{\hat r_\eta}\bigg(\frac{\d S_\ell}{\d R}\bigg)^2 
\tilde\Psi_{\ell 1}^2\bigg(\frac{\d R}{\d r}\bigg)^2 
-\frac{1}{\vk^2\hat r_\eta^3}\tilde\Psi_{\ell 1}^2\bigg\}
\bigg|_{R_*}\;.
\end{split}
\ee
\eseq

These relations allow us to extract the
asymptotic dependence of the 
response matrix on $\ell$ and $k$.
We
first observe that $k$ and $\ell$ appear in the dynamical equations
(\ref{WKBNLO}), (\ref{WKBNNLO}) only in the combination
$\vk$. Together with the form (\ref{WKBinit}) of the initial
conditions this implies that the coefficient function 
$\delta_{\ell 1}$, $\delta_{\ell 2}$ etc. have a universal dependence
on $\vk$, up to an overall factor $(\ell+1/2)^{-1}$. This, in turn,
implies that the sources (\ref{UpsWKB}) are functions of $\vk$ times
an overall factor $k^{-2}(\ell+1/2)^{-2}$. 
On general grounds, the matrix elements of $Q_\ell$ are linear
functionals of the sources,
\be
\label{linsource}
Q_\ell(k,k)=\int_{-\infty}^0
d\eta\,\Big(
K_\delta(\eta)\Upsilon_\delta(\hat r_\eta,\eta; k,\ell)+
K_\T(\eta)\Upsilon_\Theta(\hat r_\eta, \eta; k,\ell)\Big)\;,
\ee
with some kernels $K_{1,2}$ that do not depend on $\ell$ and $k$.
This leads to the expression,
\be
\label{Qq}
\overline{Q_\ell(k,k)}= k^{-2}(\ell+1/2)^{-2} q(\vk)\,,
\ee
where the function $q$ depends only on the ratio (\ref{vk}).

We can now collect the contributions of all high-$\ell$
multipoles to the prefactor,
\be
\label{Ahighl}
\begin{split}
{\cal A}_{{\rm high}-\ell}&=\exp\Big[-2\hat\l \sum_\ell (\ell+1/2)\Tr Q_\ell P\Big]\\
&=\exp\bigg[-2\hat\l \int d\vk\, q(\vk) \sum_\ell\,(2\pi)^{-3} P\big(\vk(\ell+1/2)\big)\bigg]\;.
\end{split}
\ee
The sum over $\ell$ converges as long as the power spectrum
falls down faster than $k^{-1}$ in the UV, 
which coincides with the condition for the 
convergence of the 1-loop corrections in the standard cosmological
perturbation theory. 
One can show that $q(\vk)\propto \vk^{-2}$ at large values of $\vk$
(see below), so the integral over $\vk$ will converge as well. 
Still, the expression (\ref{Ahighl}) receives large contributions from
unphysical 
UV modes and must be renormalized just like the 1-loop
correction to the power spectrum is renormalized in EFT of LSS. We
will return to this issue in Sec.~\ref{sec:counterterm}.

Let us discuss the lower limit of integration in (\ref{Ahighl}).  
From the arguments of the beginning of this section we know that
$q(\vk)=0$ for $\vk R_*<1$, so the integral in (\ref{Ahighl}) should be
taken from $\vk=R_{*}^{-1}$ to infinity. The WKB result for the
function $q(\vk)$ and hence for the integral is valid at
\be
\label{WKBvk}
\vk > (1+\epsilon)/R_*~,~~~~ \epsilon\gg \ell^{-2/3}\;.
\ee
One would like to extend the WKB expression for the integral down to 
$\vk=R_*^{-1}$ hoping that the error made in the region 
$1<\vk R_{*}<1+\epsilon$ is small. However, here we encounter a
problem. The expressions (\ref{dl1init}),
(\ref{dl2init}) imply that the functions $\delta_{\ell 1}$, $\delta_{\ell 2}$,
etc. have a singular behavior at $\vk \to 
R_{*}^{-1}$. Due to
the locality of Eqs.~(\ref{WKBNLO}),
(\ref{WKBNNLO}) this singularity survives the time evolution and gives
rise to singular terms in the sources 
(\ref{UpsWKB}) behaving as 
$[(\vk R_*)^2-1]^{-3/2}$. 
Further, the representation (\ref{linsource}) implies that the
singularity is inherited by the function 
$q(\vk)$, so its  integral
actually diverges at the lower limit as $\epsilon^{-1/2}$. 
As shown in Appendix~\ref{app:BT}, this is an artifact of
the WKB approximation and the divergence is canceled by a boundary
term produced by the integral over the interval 
$(1-\epsilon)/R_*<\vk<(1+\epsilon)/R_*$ which is not
captured by the WKB method. The net result is 
that an integral of $q(\vk)$ with a smooth function $\varphi(\vk)$
should be understood as
\be
\label{qint}
\dashint d\vk\, q(\vk) \varphi(\vk)=\lim_{\epsilon\to 0}
\bigg(\int_{(1+\epsilon)/R_*}^\infty d\vk\, q(\vk) \varphi(\vk)
-\frac{2C}{\sqrt{\epsilon}}\bigg)\;,
\ee
with
\be
\label{CWKB}
C=R_*^{-1}\varphi(R_*^{-1})\lim_{\vk\to 1/R_*}
[\vk R_*-1]^{3/2}\, q(\vk)\;.
\ee
A numerically efficient way to evaluate this integral is described in
Appendix~\ref{app:integral}.
In the next section we will see that the WKB approximation becomes accurate 
for orbital numbers $\ell \geq 9$.

Before closing this section, let us discuss the limit $\vk\to \infty$,
corresponding to $k\gg (\ell+1/2)/r_*$.
In this limit all the above formulas greatly simplify. From
Eq.~(\ref{Ssol}) we get $S_\ell=R$. Also $\vk$ drops off the equations
(\ref{WKBNLO}), (\ref{WKBNNLO}) for the coefficient functions. In the
initial conditions (\ref{dl1init}), (\ref{dl2init}) $\vk$ factors out,
so that all coefficient functions become simply proportional to
$1/\vk$. This translates into the following asymptotics of the
function $q(\vk)$,
\be
\label{qasymp}
q(\vk)= \frac{q_\infty}{\vk^2}\;,\qquad \text{at}~~\vk\gg 1/R_*\;.
\ee  
Alternatively, for the diagonal elements of the response matrix we obtain
\be
\label{Qluniv}
\overline{Q_\ell(k,k)}= \frac{q_{\infty}}{k^4}~,~~~~ \text{at}~~ 
k\gg (\ell+1/2)/R_*\;.
\ee
Note that this high-$k$ asymptotics is $\ell$-independent. Although it
has been derived under the assumption of large $\ell$, one can show
that in fact it holds for any\footnote{To obtain (\ref{Qluniv}) 
  at arbitrary fixed $\ell$ and $k\to \infty$, one can use a slightly modified
  version of the WKB expansion based on the asymptotics of Bessel
  functions at large arguments.} $\ell$, including $\ell=0$. 
Thus, we can determine
$q_\infty$ using the exact expression for the response matrix in the
monopole sector. Comparing (\ref{Qluniv}) to (\ref{Q0}) we get,
\be
\label{qinfty}
q_\infty=\frac{6\pi}{R_*^4}\bigg(-\frac{3\hat E}{\hat C^3}
+\frac{1}{\hat C^2(1+\delta_*)}\bigg)\;,
\ee
where $\hat C$, $\hat E$ are defined in (\ref{eq:lambda}),
(\ref{Edef}) respectively. We have verified that the numerically
computed function $q(\vk)$ satisfies the asymptotics (\ref{qasymp})
with $q_\infty$ from (\ref{qinfty}) with very high precision.

\section{Aspherical prefactor: results}
\label{sec:results}

\subsection{Evaluation of fluctuation determinants}
\label{sec:numerics}

In this section we present the results obtained by a 
fully-nonlinear numerical calculation of the aspherical prefactor.
We follow the algorithm discussed in Sec.~\ref{sec:algorithm}: compute
the linear aspherical fluctuations on the grid, 
use them to build the sources $\Upsilon_{\T,\delta}$,  
solve the ODE's governing the time evolution  
of the response matrix, and finally compute the fluctuation determinants.  
For the dipole sector we have implemented the IR safe algorithm discussed in
Sec.~\ref{sec:IRisol}.
The details of our numerical procedure are presented in
Appendix~\ref{sec:aspy}. 
We have evaluated the aspherical prefactor both in the EdS
approximation and for the  
exact $\L$CDM cosmology and found that the results agree within one
percent accuracy. 
This is consistent with the fact that the departures from EdS 
appear only at late times. However, at this stage the coupling of the
fluctuations to the local spherical collapse background already
dominates the effect of the cosmological expansion, so the effect
of the cosmological constant is suppressed. In what follows we display
the results obtained within the EdS approximation.  

\begin{figure}[t]
\begin{center}
\includegraphics[width=.47\textwidth]{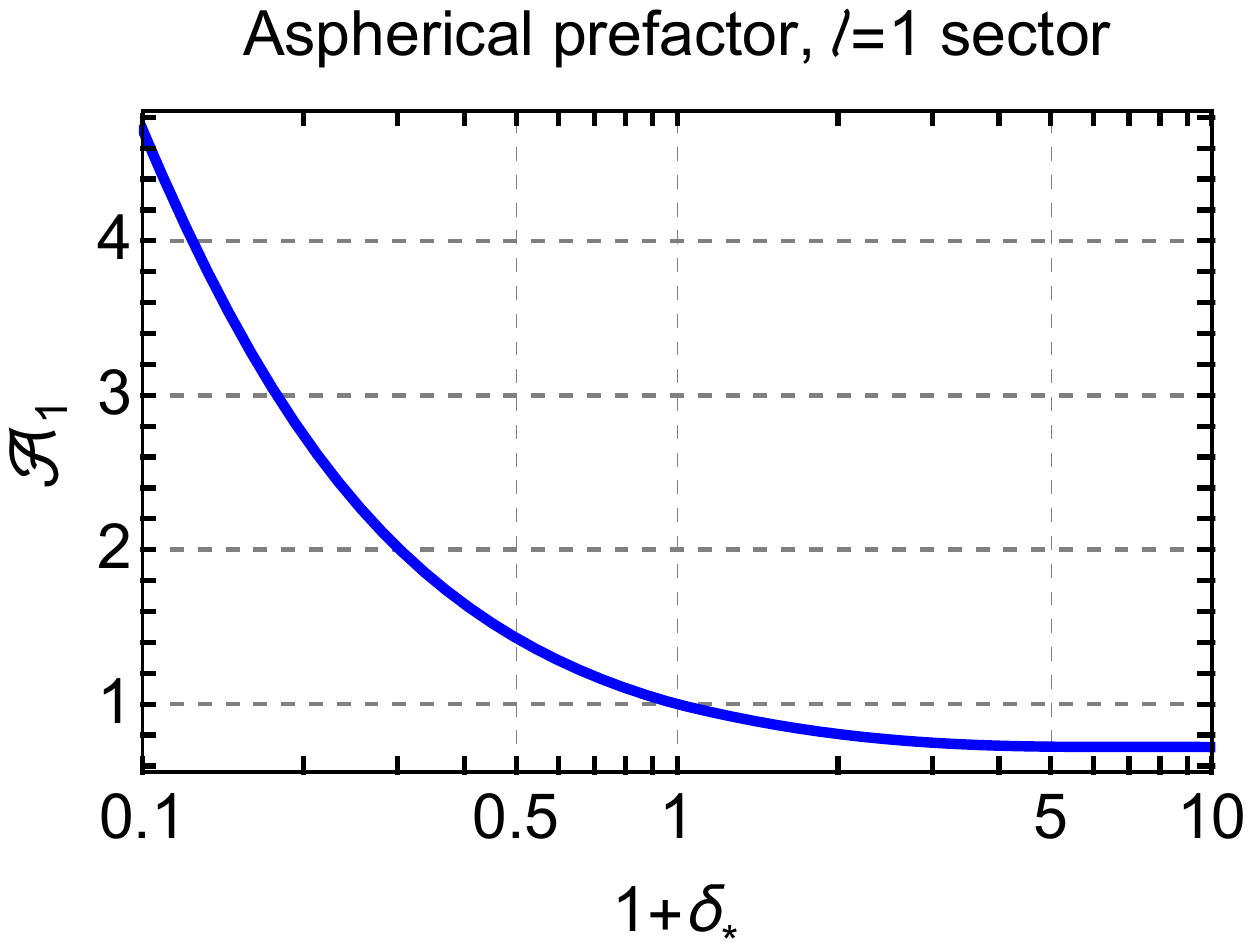}
\includegraphics[width=.50\textwidth]{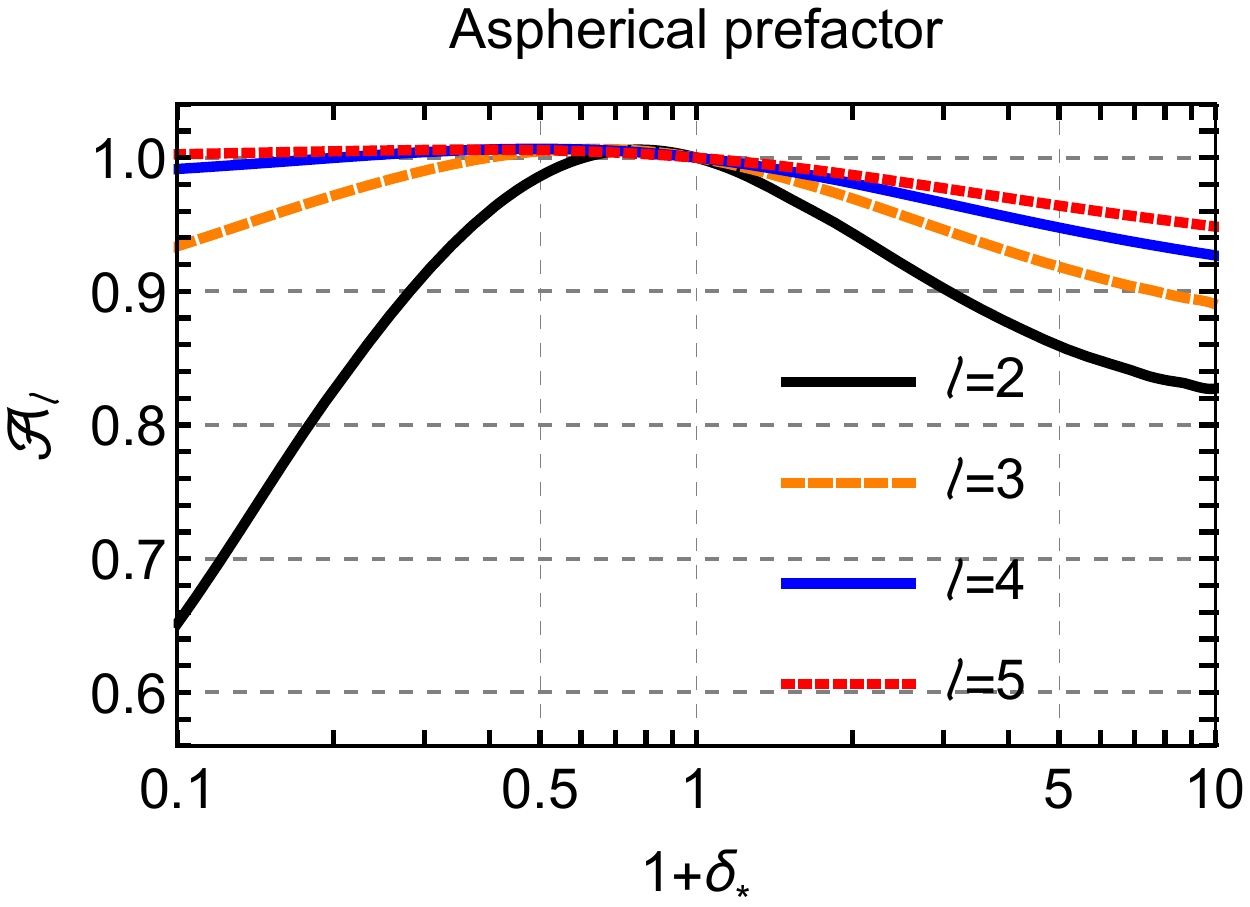}
\includegraphics[width=.50\textwidth]{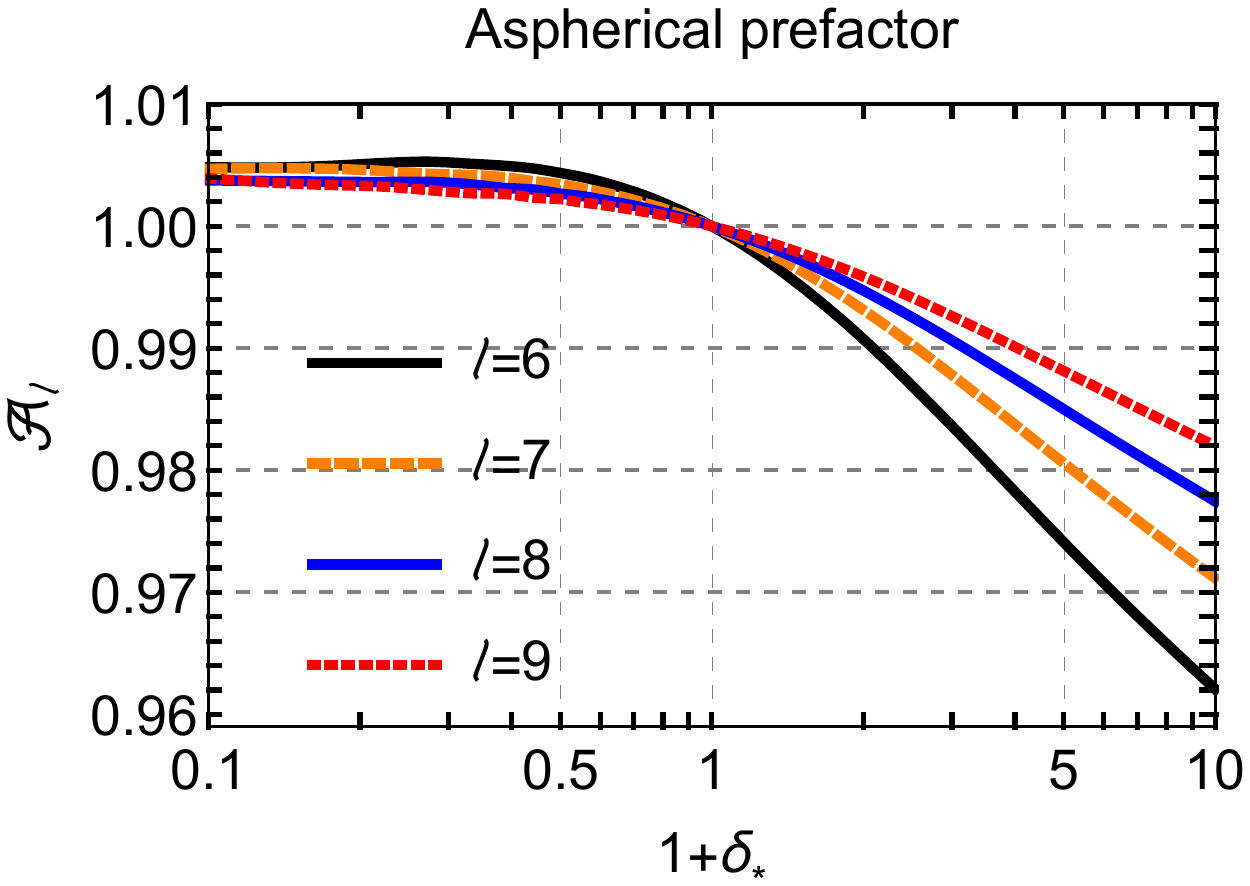}
\includegraphics[width=.49\textwidth]{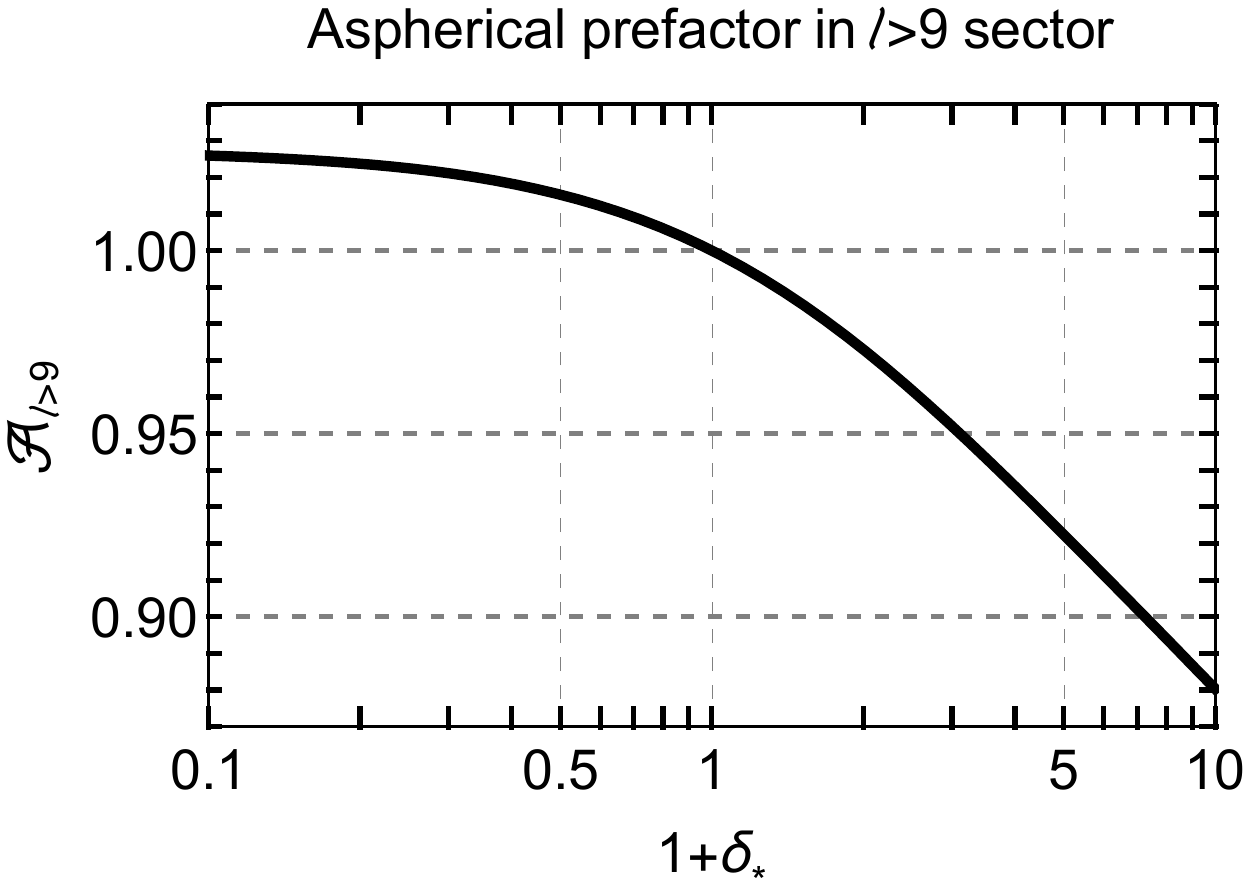}
\caption{
\label{fig:allmult} 
The prefactor of aspherical fluctuations in different orbital sectors.
Upper left panel: the dipole ($\ell=1$) sector.
Upper right panel: $\ell =2,3,4,5$ sectors.
Lower left panel:  $\ell =6,7,8,9$ sectors.
Lower right panel: the cumulative prefactor for orbital numbers $\ell >9$ computed in the WKB approximation. 
All results are shown for $r_*=10$ Mpc/$h$.
}
\end{center}
\end{figure}

Figure~\ref{fig:allmult} shows individual contributions of different
multipoles to the aspherical prefactor. We fix the cell size to
$r_*=10\,\text{Mpc}/h$; the results for $r_*=15\,\text{Mpc}/h$ are
similar. The most significant contribution comes from the dipole
sector and is shown in the upper left panel. We observe that it is a
decreasing convex function that changes by a
factor $\sim 0.2$ between $\delta_*=-0.9$ and $\delta_*=9$. At large
$\delta_*$ the curve flattens out. The contributions of the multipoles
with $2\leq \ell\leq 5$ and $6\leq \ell\leq 9$ are shown in the upper
right and lower left panels respectively. These curves are quite
different from the dipole: their deviation from unity in the explored
$\delta_*$-range is only $\sim 40\%$ for the quadrupole and even less
($\lesssim 10\%$) for the higher multipoles. The variation of ${\cal
  A}_\ell$ decreases with the multipole number. Note that in the case
of overdensities ($\delta_*>0$) all ${\cal A}_\ell$ are less than $1$ 
which is consistent with the expectation that any aspherical
fluctuation makes collapse less efficient. On the other hand, at
underdensities the partial contributions ${\cal A}_\ell$ can be both
larger or smaller than unity, depending on the value of $\ell$.

\begin{figure}[t]
\begin{center}
\includegraphics[width=.49\textwidth]{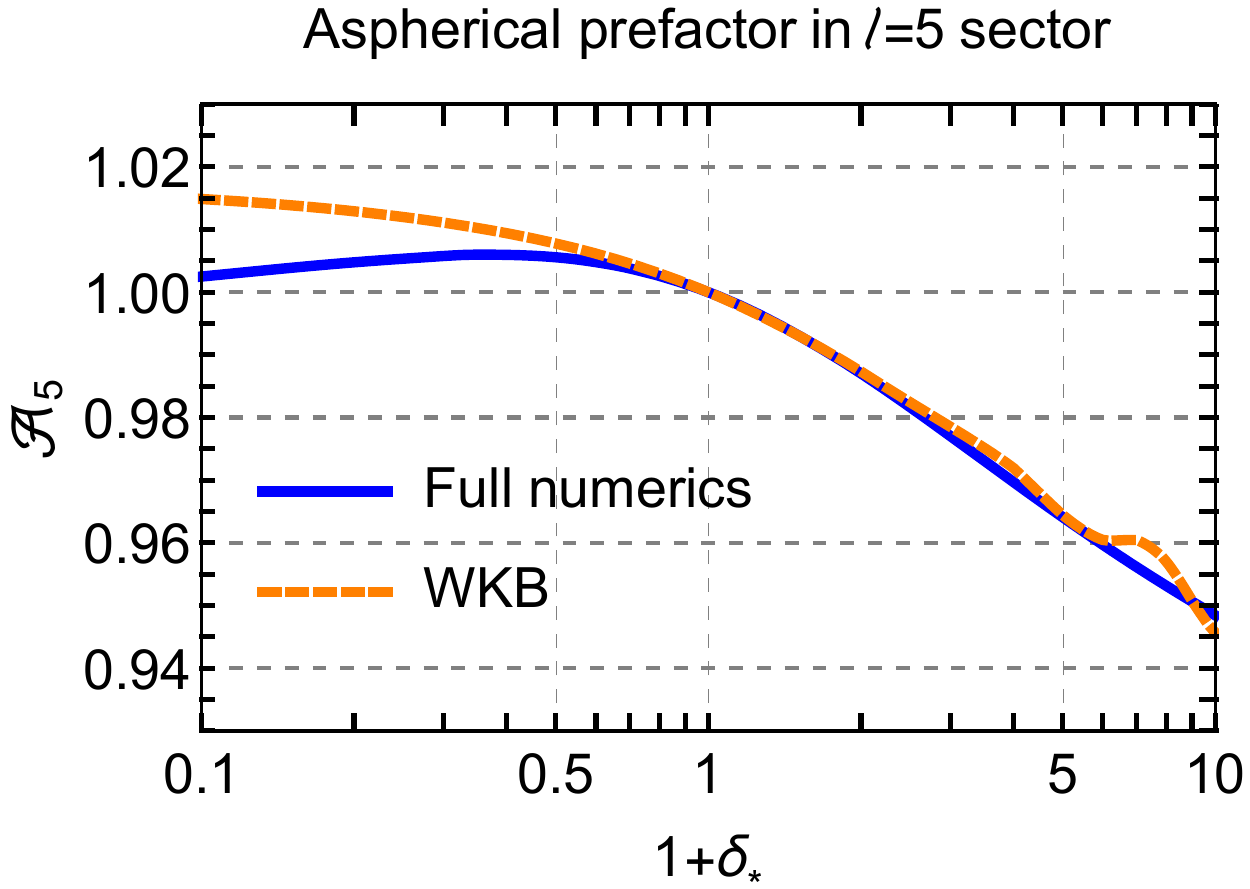}
\includegraphics[width=.49\textwidth]{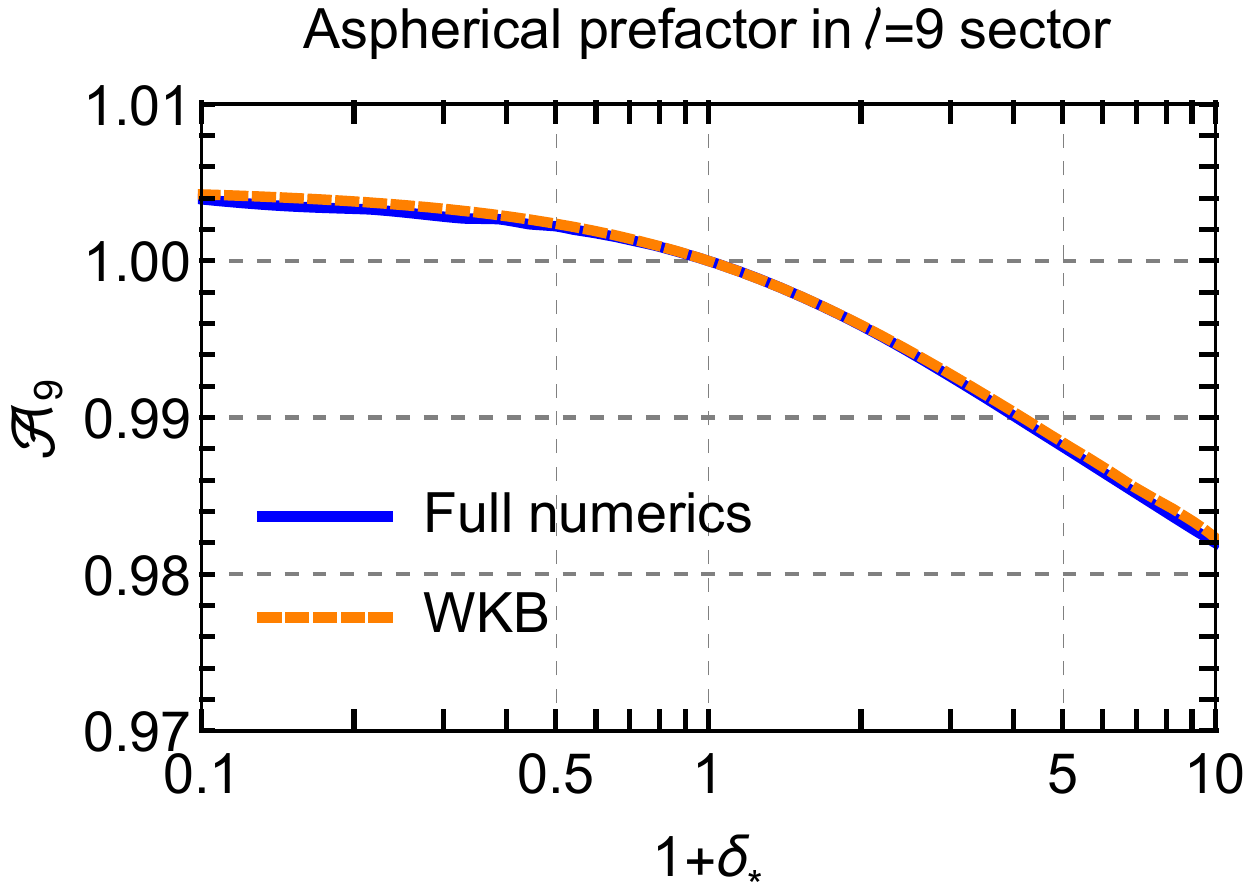}
\caption{\label{fig:wkbcomparison} 
Comparison between the WKB approximation and the full numerical calculation
for $\ell =5$ (left panel) and $\ell =9$ (right panel).
The results are shown for $r_*=10$\,Mpc/$h$.
}
\end{center}
\end{figure}

The aggregate contribution of all sectors with $\ell >9$ is shown in
the lower right panel of Fig.~\ref{fig:allmult}. It has been evaluated
using the WKB formula (\ref{Ahighl}). We test the validity of the WKB
approximation by comparing it to the results of the full numerical
routine in Fig.~\ref{fig:wkbcomparison}. The comparison is performed
for $\ell=5$ (left panel) and $\ell=9$ (right panel). For $\ell=5$
there is a significant difference between the full calculation and the
WKB approximation at strong underdensities. At overdensities the
WKB approximation exhibits spurious wiggles that can be traced back to
the baryon acoustic oscillations in the power spectrum\footnote{The
  WKB formula (\ref{Ahighl}) has an enhanced sensitivity to the shape
  of the power spectrum at $k\sim (\ell+1/2)/R_*$ due to the sharp
  increase of the function $q(\vk)$ in the vicinity of the point
  $\vk=1/R_*$. This unphysical sensitivity disappears for higher multipoles.}. 
However, already for $\ell=9$ the WKB approximation is in perfect
agreement with the full result.
We have checked that the relative error 
introduced in the aggregate contribution of $\ell >9$ 
by the use of the WKB approximation
does not exceed $10^{-3}$.
Given that this contribution itself is small compared to that of lower multipoles, the error in the whole prefactor is negligible.


\begin{figure}[t]
\begin{center}
\includegraphics[width=.49\textwidth]{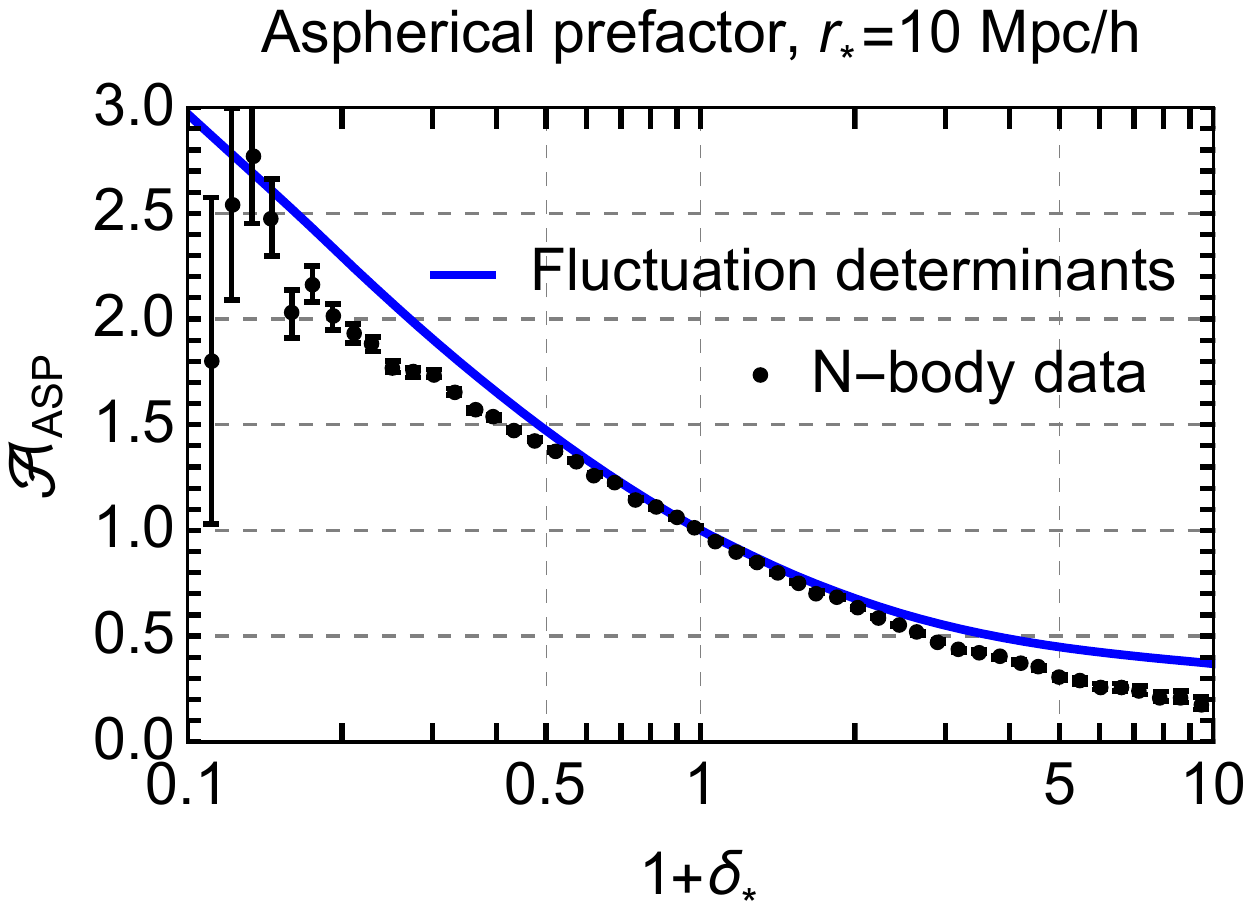}
\includegraphics[width=.49\textwidth]{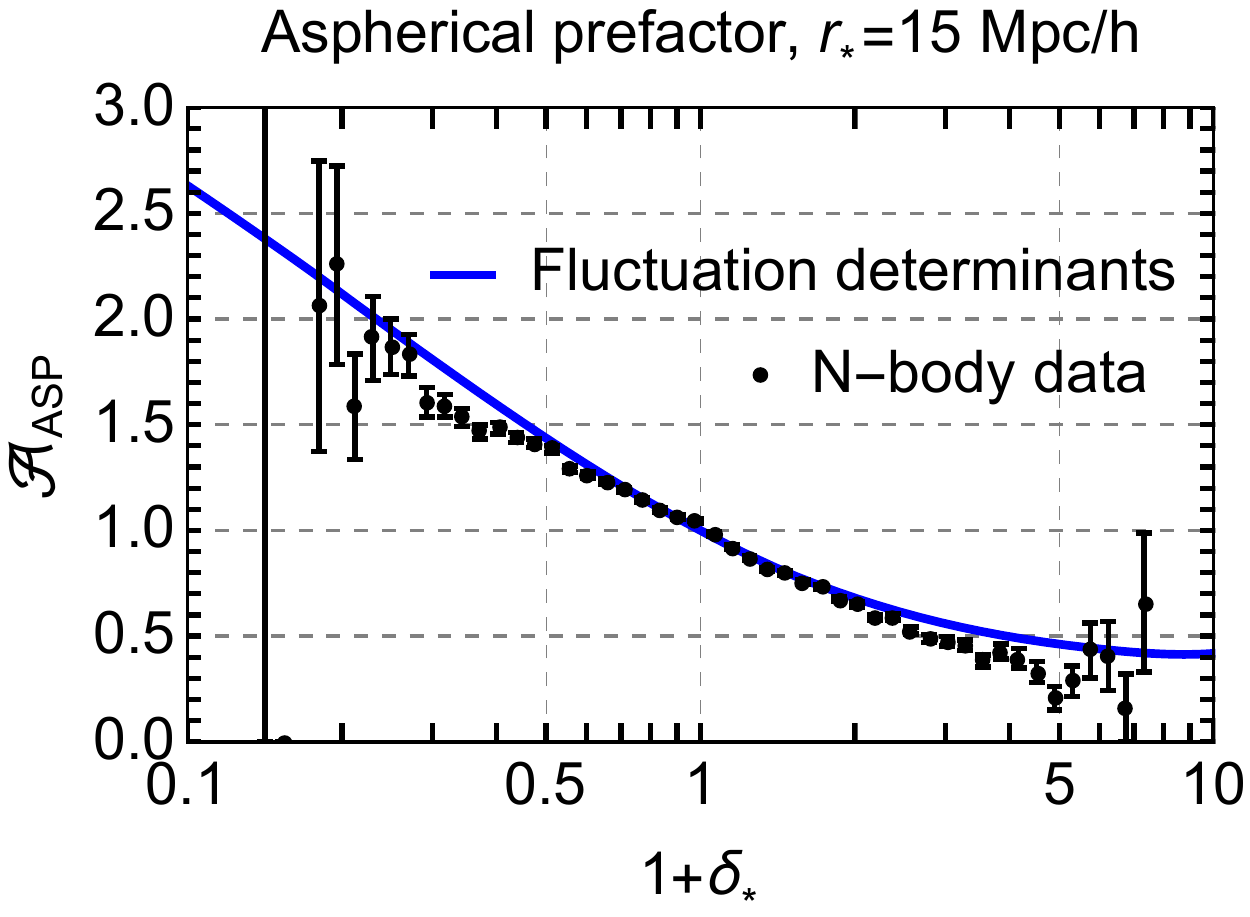}
\caption{\label{fig:final} 
The aspherical prefactor computed from fluctuation determinants
(solid blue curve)
against that extracted from the N-body simulations (black dots). The cell
radii are
$r_*=10\,\text{Mpc}/h$ (left panel)
and 
$r_*=15\,\text{Mpc}/h$ (right panel).
} 
\end{center}
\end{figure}

The total result for the aspherical prefactor 
obtained upon multiplying the contributions of all $\ell\geq 1$
is shown in Fig.~\ref{fig:final}, where it is compared with the
prefactor extracted from the N-body data (see
Sec.~\ref{sec:asppref}). One observes a good qualitative agreement
between the theoretical curve and the data. However, there is a clear
quantitative discrepancy which grows towards the edges of the
$\delta_*$-interval. The discrepancy is somewhat bigger for
$r_*=10\,\text{Mpc}/h$ than for $r_*=15\,\text{Mpc}/h$ and reaches
$30\%$ ($100\%$) for underdense (overdense) tail. We interpret this
discrepancy as the effect of short-scale physics that is not captured
by the perfect-fluid hydrodynamics. In the next subsection we show how
our results can be improved by renormalizing the 
contributions of short-scale modes.  

Let us make a comment. The fact that the fluctuation determinants found in
our calculation are always positive provides a consistency check of the
saddle-point approximation developed in Sec.~\ref{sec:spherical}. In
particular, it shows that there are no other saddle points of the path
integral (\ref{eq:pdf2}) that would branch off the spherical collapse
dynamics at any value of $\delta_*$ within the considered
range. Indeed, if it were the case the spectrum of fluctuations
around the spherical collapse at this value of $\delta_*$ would
contain a zero mode, and hence at least 
one of the determinants ${\cal D}_{\ell}$ 
would vanish, which
is not observed.

\subsection{Renormalization of short-scale contributions}
\label{sec:counterterm}

Up to this point we have worked within pressureless perfect fluid hydrodynamics,
which is known to break down at short scales. 
This introduces an error in our calculation that must be corrected. A
similar issue arises in the perturbative calculation of the density
correlation functions in the homogeneous background where a systematic
way to take into account the corrections due to UV modes is provided
by introduction of counterterms in the hydrodynamics equations. These
counterterms are constructed as a double expansion in the number of
spatial derivatives acting on the fields and in the powers of the
density contrast \cite{Abolhasani:2015mra}. We have encountered this
procedure in Sec.~\ref{sec:aspt} where we made contact between the
calculation of the prefactor at small density contrast and the
calculation of 1-loop corrections to the power spectrum. At that
level the sensitivity to the short-scale physics reduced to a single
counterterm $\gamma(z)$, see Eq.~(\ref{eq:dctr}). 

The situation is more complicated at large density contrasts
$\delta_*$ which we are interested in now. In this case, the
evaluation of the aspherical prefactor can be viewed as a 1-loop
calculation in the non-trivial background of the spherical collapse
solution. Then the counterterm is, in general, a functional of the
background, restricted by the symmetries of the problem, but otherwise
arbitrary. It is impossible to rigorously fix its form without going
beyond the EFT framework. In what follows we consider two schemes for
renormalization of the aspherical prefactor that are based on
reasonable physical assumptions. The difference between the two models
should be treated as an intrinsic theoretical uncertainty of our
current determination of ${\cal A}_{\rm ASP}$ due to the lack of
control over the UV physics.

We start by analyzing the UV sensitivity of the aspherical
prefactor. The contribution of modes with $k>k_{\rm UV}\gg 1/r_*$,
$\ell\gg 1$ is described by the WKB expression (\ref{Ahighl}). The sum
over $\ell$ in the exponent can be rewritten as an integral,  
\begin{align}
\int_{k_{\rm UV}}^\infty \frac{d k P(k)}{(2\pi)^3} \sum_{\ell}
\frac{1}{\ell+1/2}\;q\bigg(\frac{k}{\ell+1/2}\bigg)
&\simeq \int_{k_{\rm UV}}^\infty \frac{d k P(k)}{(2\pi)^3} 
\;\dashint 
\frac{d\ell}{\ell+1/2}\;q\bigg(\frac{k}{\ell+1/2}\bigg)\notag\\
&=\int_{k_{\rm UV}}^\infty \frac{d k P(k)}{(2\pi)^3} 
~\dashint 
\frac{d\vk}{\vk}\;q(\vk)\;.
\label{UVfact}
\end{align}
We observe that the integral over momenta and the background
dependence contained in the function $q$ factorize. In other words,
all high-$k$ modes contribute into ${\cal A}_{\rm ASP}$ in a universal
way. Of course, this is true only within the domain of validity of the
formula (\ref{Ahighl}) which neglects the interaction among the short
modes and the departures from the hydrodynamic description. Precisely
because of this inaccuracy, the integral over $k$ in (\ref{UVfact})
should be renormalized. 

\begin{figure}[t]
\begin{center}
\includegraphics[width=.49\textwidth]{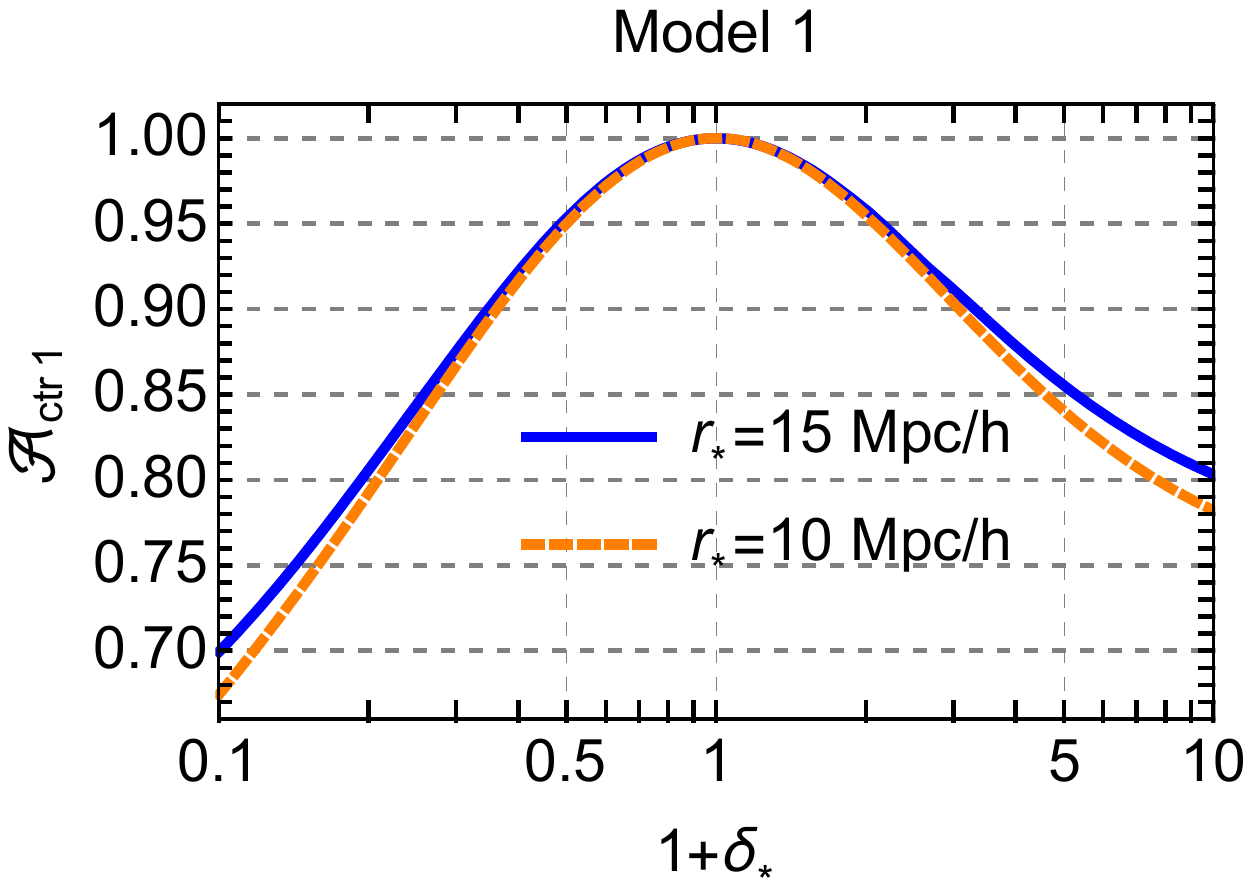}
\includegraphics[width=.49\textwidth]{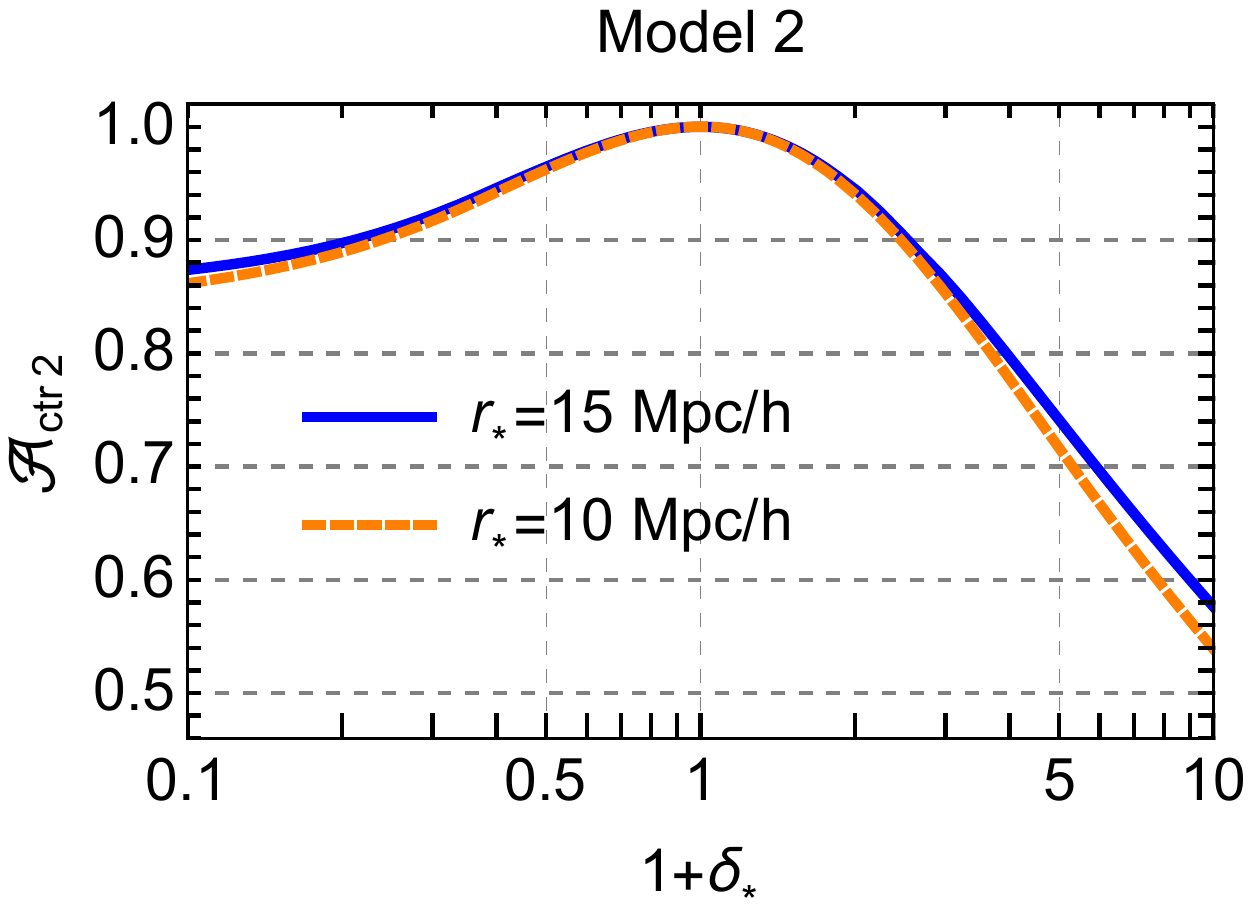}
\caption{\label{fig:ctr} 
The counterterm prefactor 
for model 1 (left panel) and model 2 (right panel)
evaluated for $\gamma_0=1.5$ (Mpc$/h$)$^2$,
$z=0$ and cell radii $r_*=10$ Mpc$/h$ and $r_*=15$ Mpc$/h$. 
} 
\end{center}
\end{figure}

The integral in (\ref{UVfact}) is
proportional to the high-$k$ contribution into the velocity dispersion 
\be
\label{sigmav}
\sigma_v^2\equiv \frac{1}{6\pi^2}\int dk\,P(k)\;.
\ee 
The same integral arises in the 1-loop
correction to the power spectrum (see (\ref{eq:1lpsUV})) where it 
is renormalized by the substitution 
\be 
\label{eftrepl}
\int \frac{dk P(k)}{(2\pi)^3} \mapsto 
\int \frac{dk P(k)}{(2\pi)^3}+\frac{315}{122\pi}\frac{\g(z)}{g^2(z)}\;.
\ee
We saw in Sec.~\ref{sec:EFT} that this substitution also works for the
aspherical prefactor at small $\delta_*$. Our first model for the
renormalization of ${\cal A}_{\rm ASP}$ is obtained by extending the
prescription (\ref{eftrepl}) to finite values of $\delta_*$. It
corresponds to an assumption that the main effect of renormalization
in all quantities is the replacement of the tree-level velocity
dispersion of high-$k$ modes with its renormalized value\footnote{This
assumption is supported by the observation \cite{Baldauf:2014qfa} that the N-body
data for bispectrum are well fitted by the EFT formula without any
additional counterterms beyond $\g$ (``0-parameter fit'' in
\cite{Baldauf:2014qfa}). Inclusion of further independent counterterms allowed by the
EFT framework does not significantly improve the quality of the fit.}.
For the redshift dependence of the counterterm 
we will use the scaling-universe
approximation, as we did in Sec.~\ref{sec:EFT}. In this way we arrive
at the following expression for the counterterm prefactor,
\be
\label{eq:Actr}
\mathcal{A}_{\text{ctr1}}=\exp\left(-\frac{315\g_0}{122\pi}
  \big(g(z)\big)^{-\frac{2(n+1)}{n+3}}\times
  2\hat{\lambda}\int \frac{d\vk}{\vk}q(\vk)\right) \,, 
\ee
where $\gamma_0$ is the 1-loop counterterm from the power spectrum and
$n$ is the slope of the power spectrum at the mildly non-linear scales. For
numerical estimates we will use $\g_0=1.5\,(\text{Mpc}/h)^2$,
$n=-1.5$. The final answer for the aspherical prefactor is given by
the product of (\ref{eq:Actr}) with the contribution obtained
from the fluctuation determinants and described in the previous
subsection. We will refer to the aspherical prefactor calculated using
the counterterm (\ref{eq:Actr}) as ``model 1''.

\begin{figure}[t]
\begin{center}
\includegraphics[width=.495\textwidth]{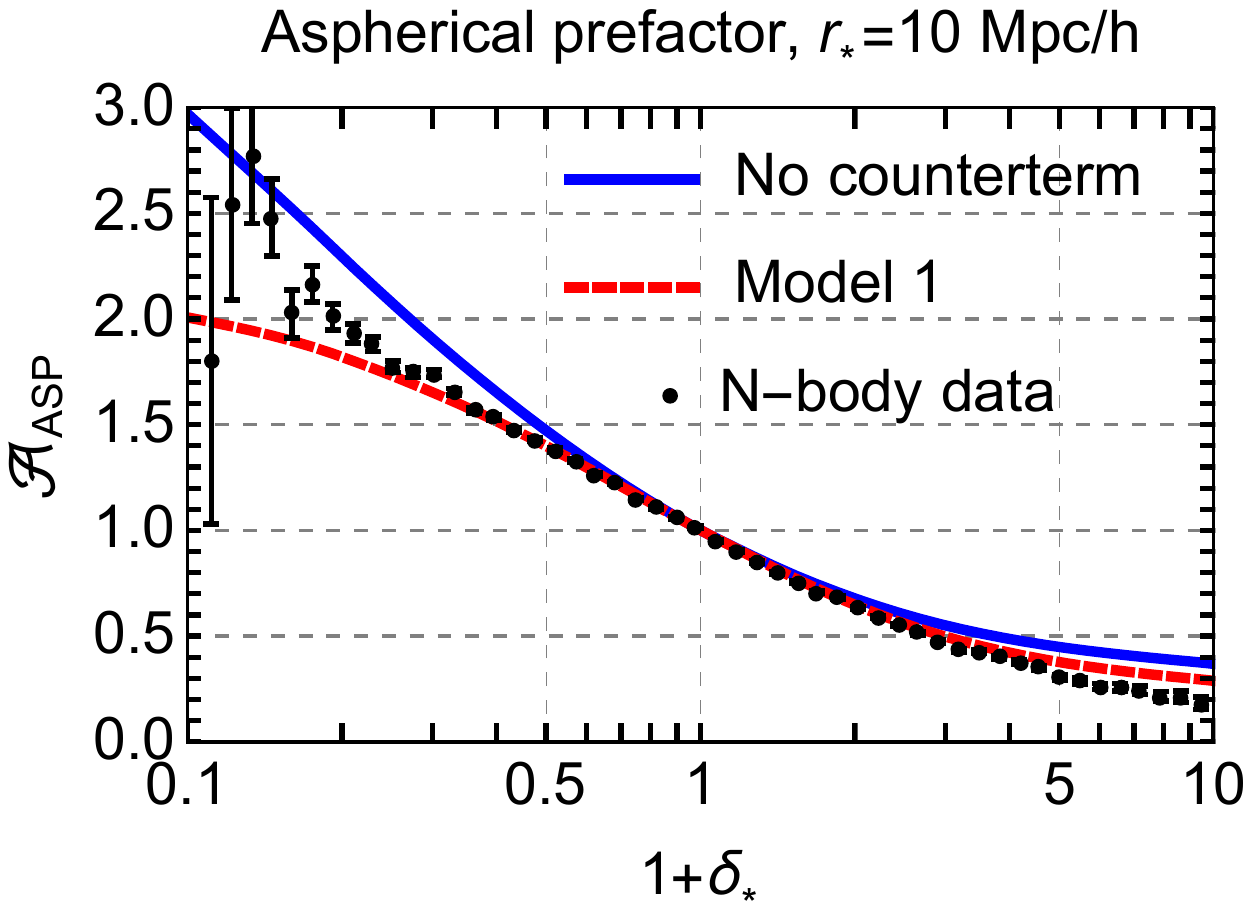}
\includegraphics[width=.49\textwidth]{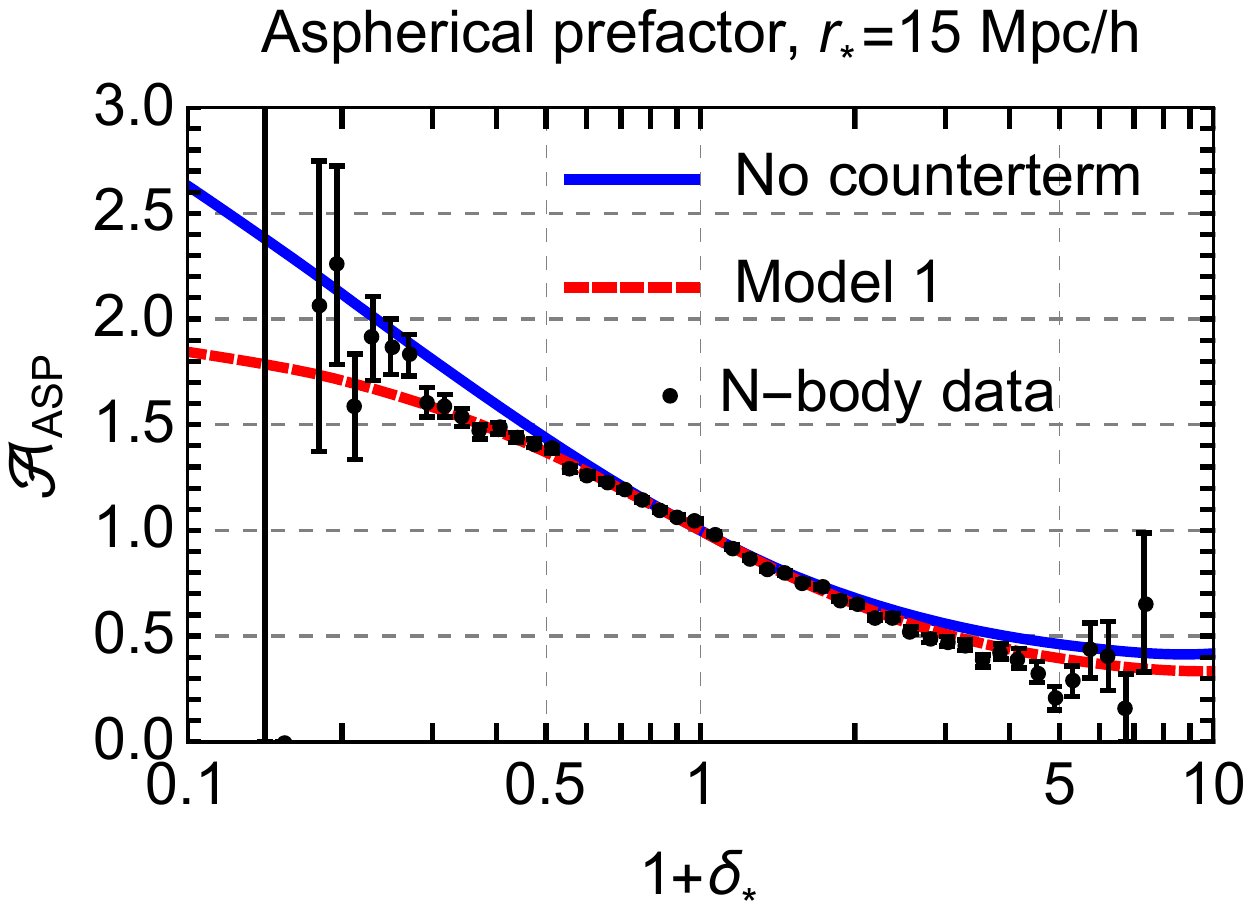}
\includegraphics[width=0.49\textwidth]{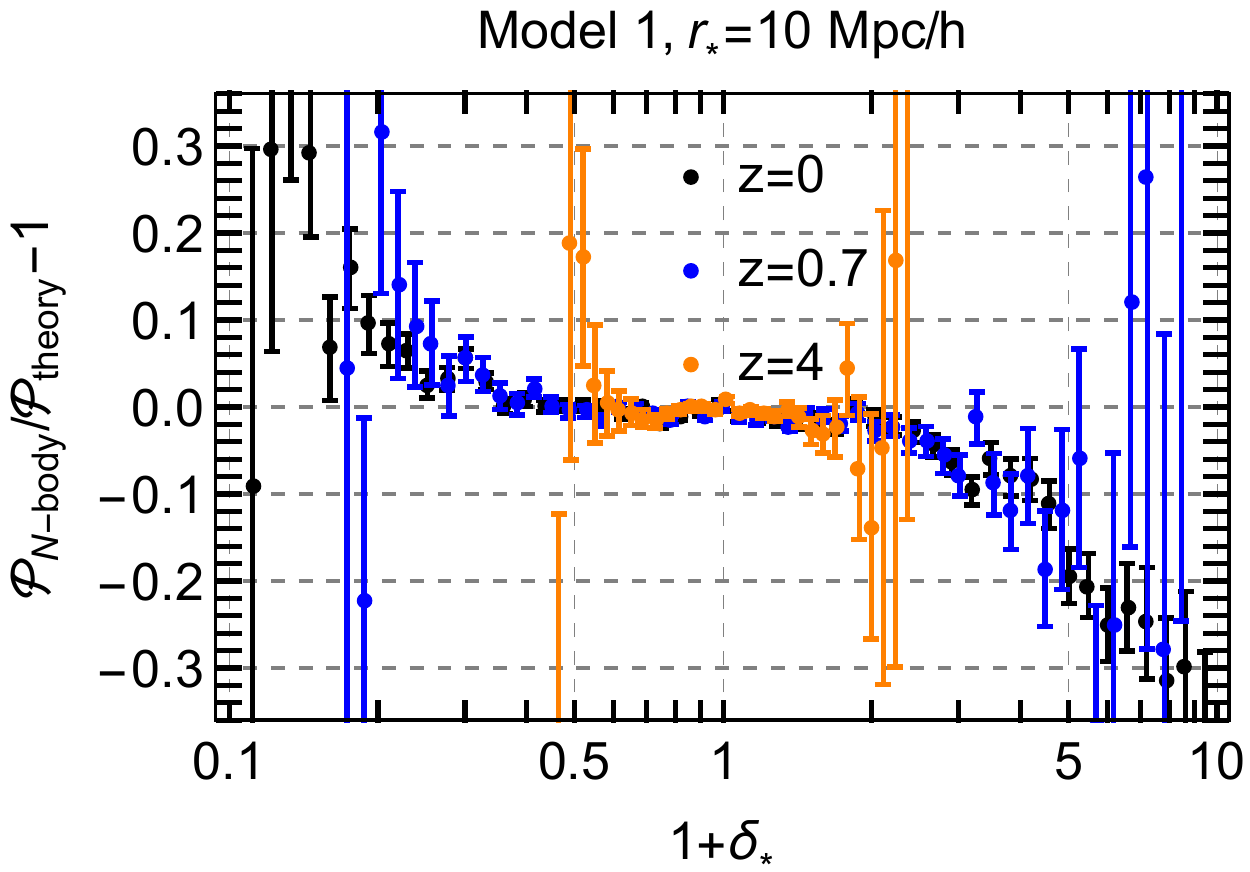}
\includegraphics[width=0.49\textwidth]{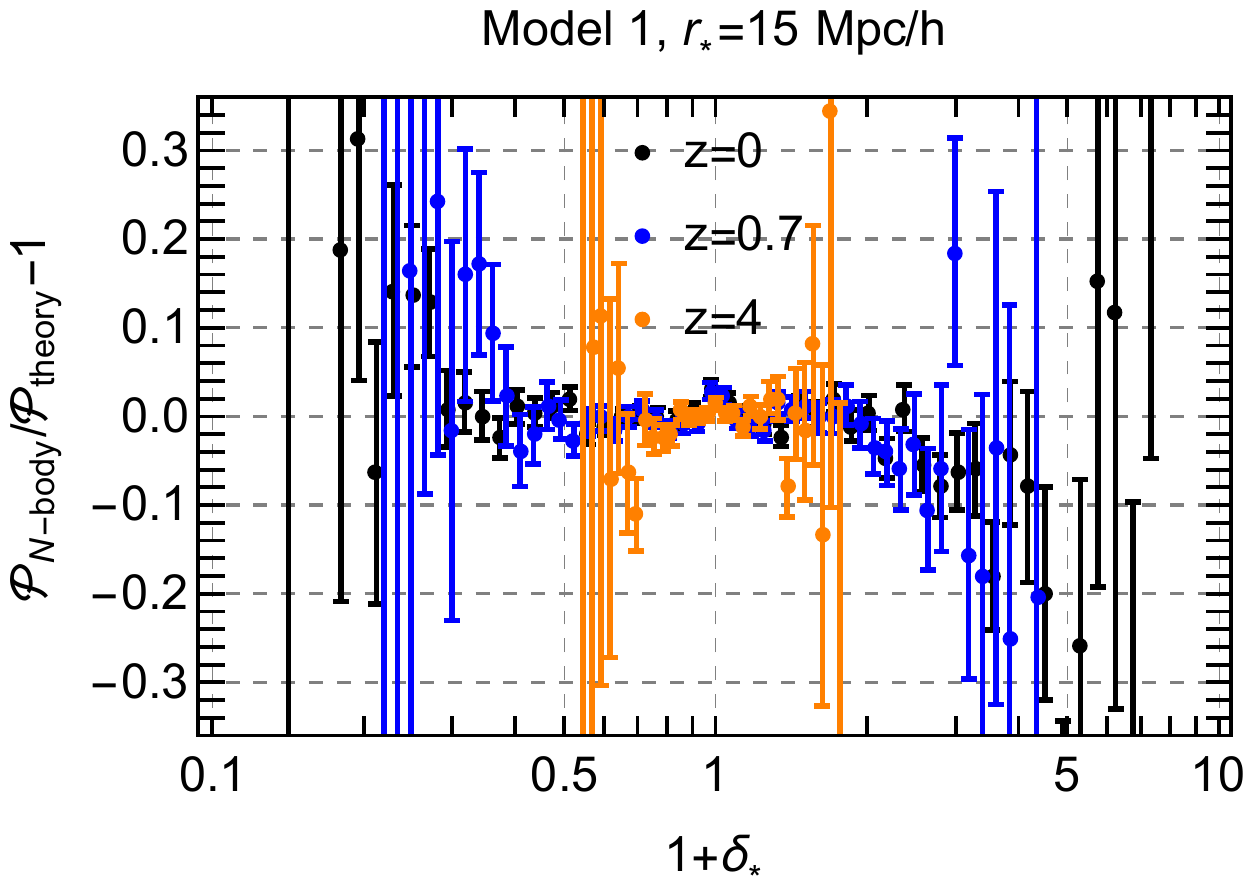}
\caption{\label{fig:ctrfinal} 
Upper panels: 
The aspherical prefactor in model 1 (dashed red curve) against the
N-body data (black dots) for cell radii $r_*=10$ Mpc$/h$ (left panel)
and $r_*=15$ Mpc$/h$ (right panel) at $z=0$. The aspherical prefactor
without the counterterm is reproduced for comparison (solid blue curve).
Lower panels: Residuals for the PDF extracted from the N-body data
compared to our theoretical prediction at several redshifts
for $r_*=10$ Mpc$/h$ (left panel) and $r_*=15$ Mpc$/h$ (right panel).
} 
\end{center}
\end{figure}

The counterterm prefactor (\ref{eq:Actr}) is plotted in the left panel
of Fig.~\ref{fig:ctr}. We see that it captures the 
main qualitative features: it has a zero derivative at the origin
where we expect the impact of shell-crossing to be negligible, and
suppresses the probability for big under- and overdensities. 
In the upper panels of Fig.~\ref{fig:ctrfinal} we plot 
the aspherical prefactor in model 1 
against the data. The aspherical prefactor without the counterterm is
also shown for comparison.
In the lower panels of Fig.~\ref{fig:ctrfinal} we show the residuals
between
the PDF measured from the N-body data and
our theoretical template for several values of redshift.
One observes a good agreement between the theory and the data. For
$r_*=10\,\text{Mpc}/h$ the residuals are at sub-percent level in the
range $-0.6<\delta_*<1$. They degrade to $10\%$ at
$-0.8<\delta_*<-0.6$ and $1<\delta_*<3$. Eventually they increase
to $\sim 30\%$ at the tails. Overall, the agreement is slightly better for
the underdensities than for the overdensities. Similar trends are
observed for $r_*=15\,\text{Mpc}/h$, though the precision of the
N-body data is too low to see them unambiguously. It is worth noting
that on general grounds one expects the effects of the UV physics to
be weaker for larger cells.

As clear from Fig.~\ref{fig:ctrfinal}, the model 1 systematically
underestimates the aspherical prefactor for underdensities and
overestimates for overdensities. This can be attributed to the
following deficiency. We have taken the counterterm $\g$ to be
independent of $\delta_*$. On the other hand, one expects the
overdense regions to be more non-linear than the underdense ones, so
that the effects of UV renormalization encapsulated by $\g$ should be
larger (smaller) at $\delta_*>0$ ($\delta_*<0$) than at
$\delta_*=0$. Comparing this with the formula (\ref{eq:Actr}) one sees
that qualitatively such a dependence would act in the right direction to
improve the agreement between the theory and the data.  

\begin{figure}[t]
\begin{center}
\includegraphics[width=.495\textwidth]{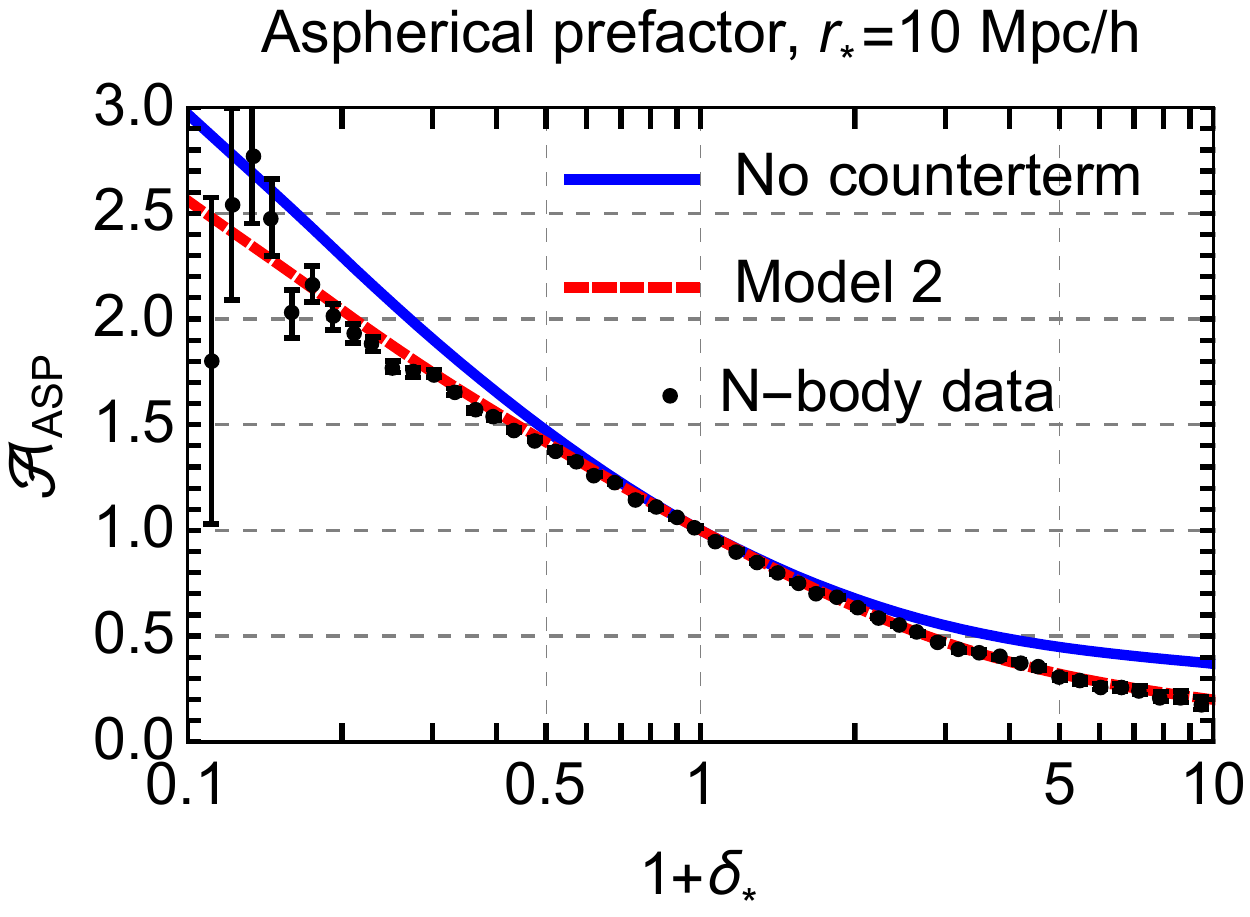}
\includegraphics[width=.49\textwidth]{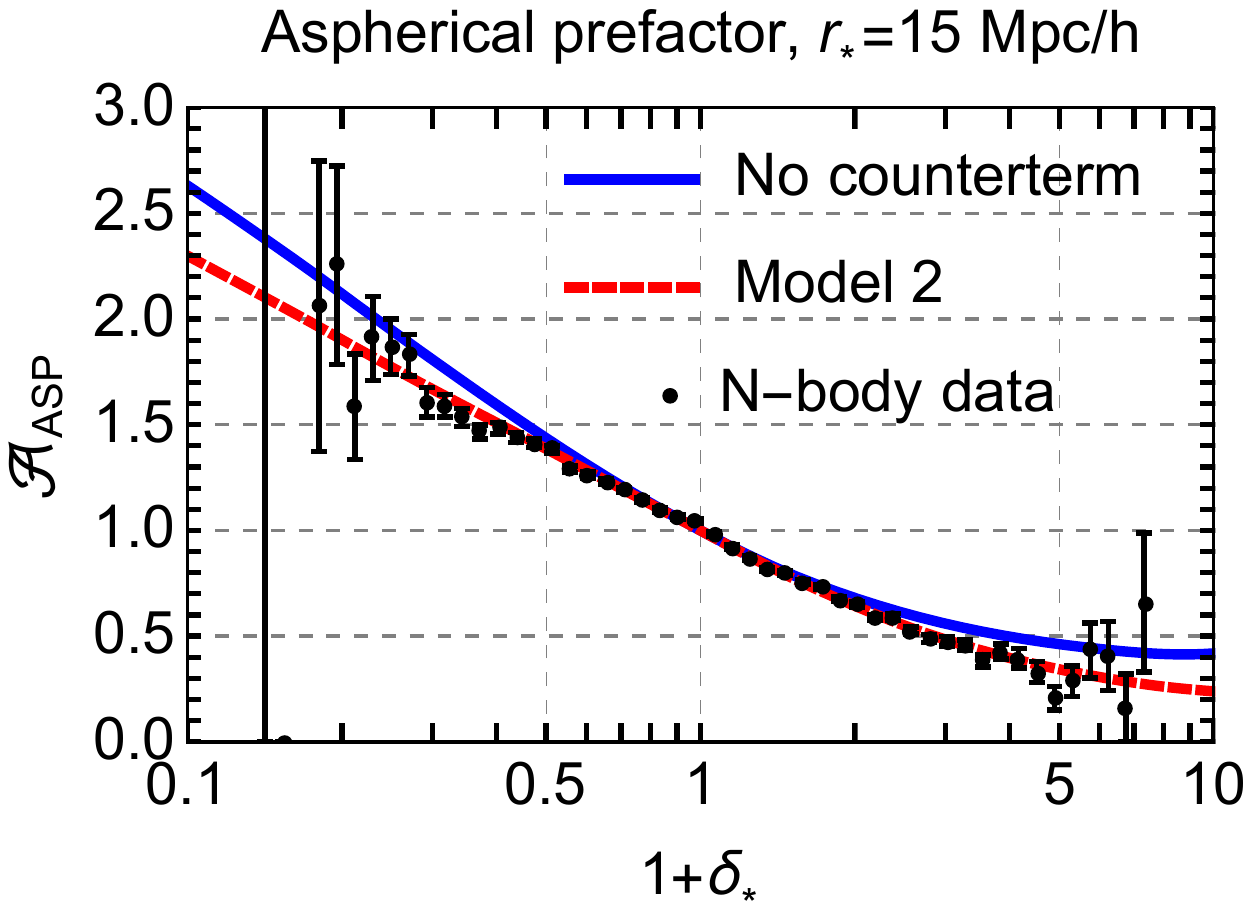}
\includegraphics[width=0.49\textwidth]{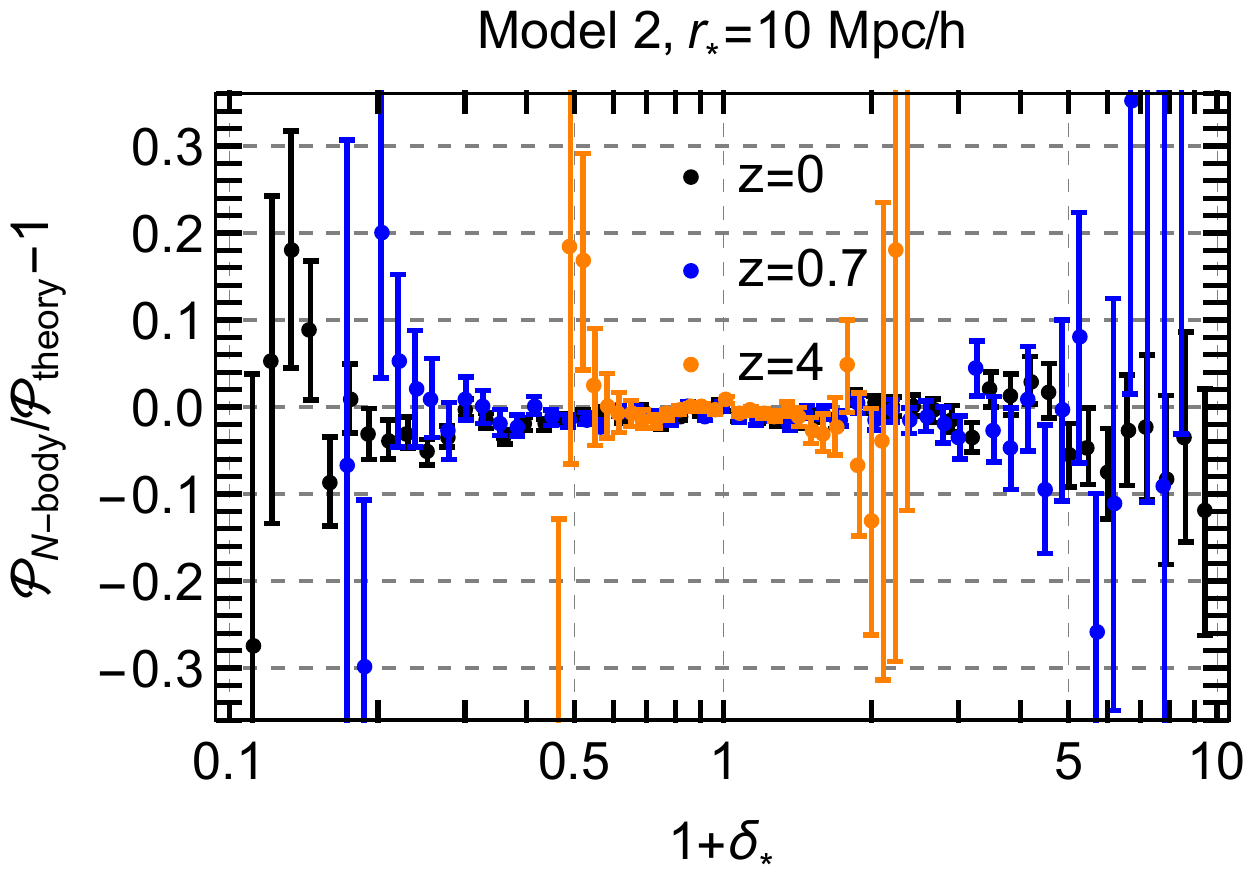}
\includegraphics[width=0.49\textwidth]{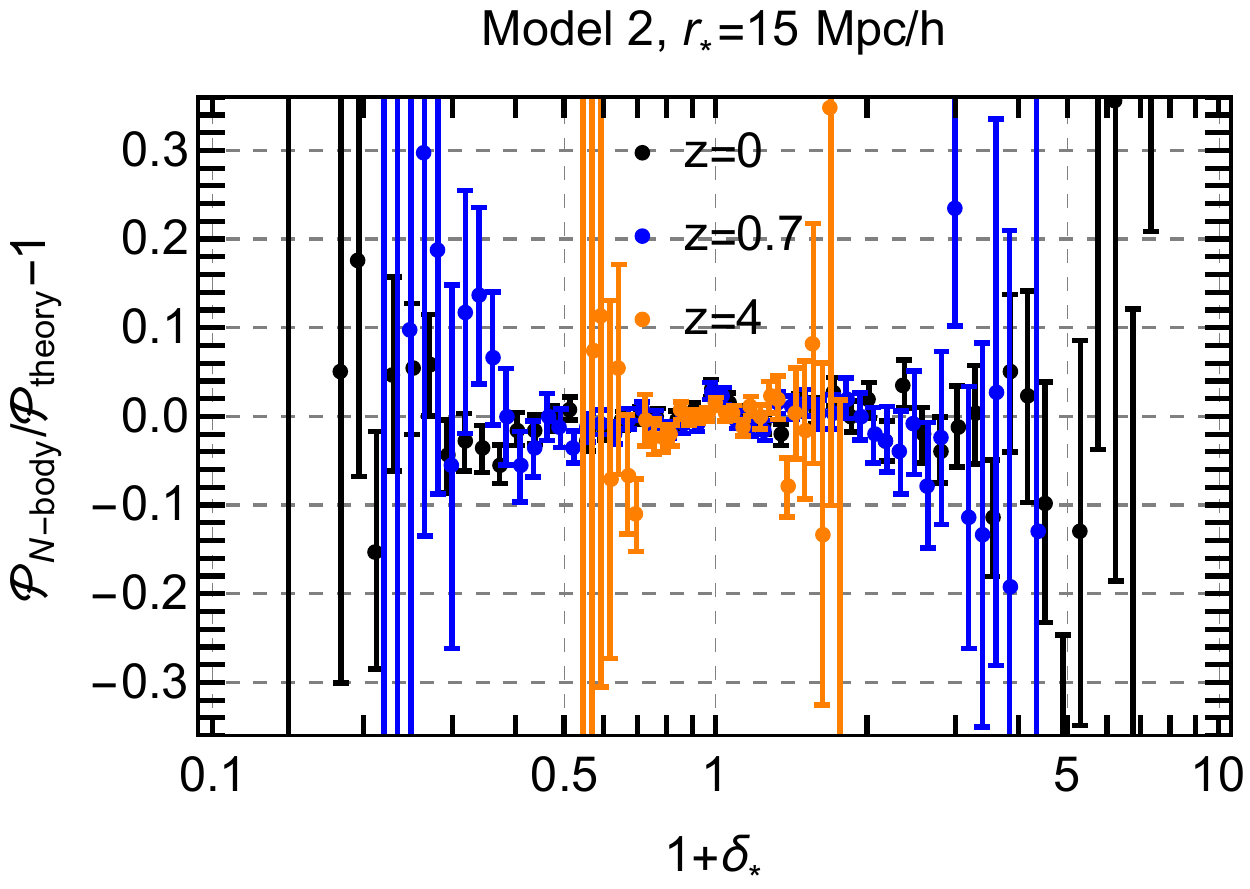}
\caption{\label{fig:ctrfinal2} 
Upper panels: 
The aspherical prefactor in model 2 (dashed red curve) against the
N-body data (black dots) for cell radii $r_*=10$ Mpc$/h$ (left panel)
and $r_*=15$ Mpc$/h$ (right panel) at $z=0$. The aspherical prefactor
without the counterterm is reproduced for comparison (solid blue curve).
Lower panels: Residuals for the PDF extracted from the N-body data
compared to our theoretical prediction at several redshifts
for $r_*=10$ Mpc$/h$ (left panel) and $r_*=15$ Mpc$/h$ (right panel).
} 
\end{center}
\end{figure}

To estimate a possible effect of the $\delta_*$-dependence of
$\gamma$, we use the following crude model. We approximate the spherical
collapse solution by top-hat density profile with the
final under-/over-density $\delta_*$. Treating such a profile as an
open/closed separate universe, we replace the counterterm and the
growth factor in
Eq.~(\ref{eftrepl}) by 
$\g(\delta_*,z)$ and $D(\delta_*,z)$ ---  the counterterm and the
growth factor in the separate universe. The latter is derived in
Appendix~\ref{app:Deff}. To estimate
$\g(\delta_*,z)$ we again use the power-law approximation for the
power spectrum and obtain 
$\g(\delta_*,z)\propto \big(D(\delta_*,z)\big)^{\frac{4}{n+3}}$. All
in all, this leads to the replacement of $g(z)$ in the counterterm
prefactor (\ref{eq:Actr}) by the density-dependent growth factor 
$D(\delta_*,z)$. Using the explicit expression for the latter, 
Eq.~(\ref{growthsepar}),
we obtain model 2 for the counterterm,
\be
\label{eq:Actr2}
\mathcal{A}_{\text{ctr2}}=\exp\left(-\frac{315\g_0}{122\pi}
  \Big(\frac{g(z)}{F'(\delta_*)(1+\delta_*)}\Big)^{-\frac{2(n+1)}{n+3}}\times
  2\hat{\lambda}\int \frac{d\vk}{\vk}q(\vk)\right) \,, 
\ee

This counterterm prefactor is shown in the right panel of 
Fig.~\ref{fig:ctr}. 
Compared to the model 1, it gives less suppression at underdensities
and stronger suppresses overdensities. 
In the upper panels of Fig.~\ref{fig:ctrfinal2} we compare 
the aspherical prefactor in model 2
with the N-body data and the prefactor without the counterterm. 
In the lower panels of Fig.~\ref{fig:ctrfinal2} we show the residuals
between the PDF measured from the N-body data and
our theoretical template for $z=0$, $0.7$, $4$. 
One observes an excellent agreement between the theory and
the data within the precision of the latter. This 
is striking given the crudeness of the model. 

We leave a detailed investigation 
of the counterterms in the spherical collapse background for future
and propose to treat  
the difference between the models 1 and 2 as a proxy for the
theoretical uncertainty. Notice that this uncertainty estimate is
internal to the theoretical approach and does not
require any comparison with N-body simulations.  
We also emphasize that none of the two counterterm models proposed in
this section introduces any additional fitting parameter, as the 
coefficient $\g_0$ entering in Eqs.~(\ref{eq:Actr}),
(\ref{eq:Actr2})
must be the same as the one measured from the dark matter power
spectrum.

\section{Summary and Discussion}
\label{sec:comp}

In this paper we computed the 1-point probability distribution function 
(PDF) of the cosmic matter density field in spherical cells. 
Our approach makes use of the path integral 
description of large-scale structure.
We identified the saddle point of the path integral that corresponds to 
the spherical collapse dynamics and yields the leading exponent of the PDF.
Then we computed the prefactor given by the determinant of the
quadratic fluctuations around the saddle-point solution. This
can be viewed as a 1-loop calculation in perturbation theory around
a fully non-linear background. 

We showed that the prefactor factorizes into the contributions of
fluctuations in different multipole sectors and evaluated the
monopole contribution exactly. Next we considered the contribution of
fluctuations with $\ell>0$ which we called `the aspherical
prefactor' ${\cal A}_{\rm ASP}$.
 We demonstrated that it is crucial for the consistency of
the PDF, in particular, for ensuring that the mean density contrast
evaluated using the PDF vanishes. Our final formula for the 1-point
PDF has the form,
\be
\label{Plast}
\P(\delta_*)={\cal A}_{\rm ASP}(\delta_*)
\frac{\hat C(\delta_*)}{\sqrt{2\pi g^2\sigma^2_{R_*}}}
\e^{-\frac{F^2(\delta_*)}{2g^2\sigma^2_{R_*}}} \,,
\ee
where $g(z)$ is the linear growth factor, $\sigma_{R_*}$ is the {\em
  linear} density variance at $z=0$ filtered at the Lagrangian radius
$R_*=r_*(1+\delta_*)^{1/3}$, $F(\delta_*)$ is the linear overdensity
corresponding to $\delta_*$ through the spherical collapse mapping,
and the function $\hat C(\delta_*)$ is defined by the formula
(\ref{eq:lambda}).    

We computed the aspherical prefactor
using several techniques. 
First, we treated the background perturbatively, 
which allowed us to capture the correct shape of the prefactor for
small averaged densities.  
Second, we
computed the partial contributions to the prefactor from sectors with  
high orbital numbers
treating the background non-perturbatively.
We showed that this limit allows one to use the WKB technique, which made 
possible a semi-analytic treatment of the problem. 
Finally, we developed a numerical procedure for a fully non-linear 
computation of the aspherical determinant on the grid. 
This procedure includes analytic factorization and cancellation of the
so-called `IR-divergences' --- spurious enhanced contributions that
appear in the dipole sector and are associated with large bulk flows.
We implemented this procedure in an open-source \texttt{Python} 
code \texttt{AsPy} available at the following link \cite{AsPy}.

\begin{figure}[t]
\begin{center}
\includegraphics[width=0.49\textwidth]{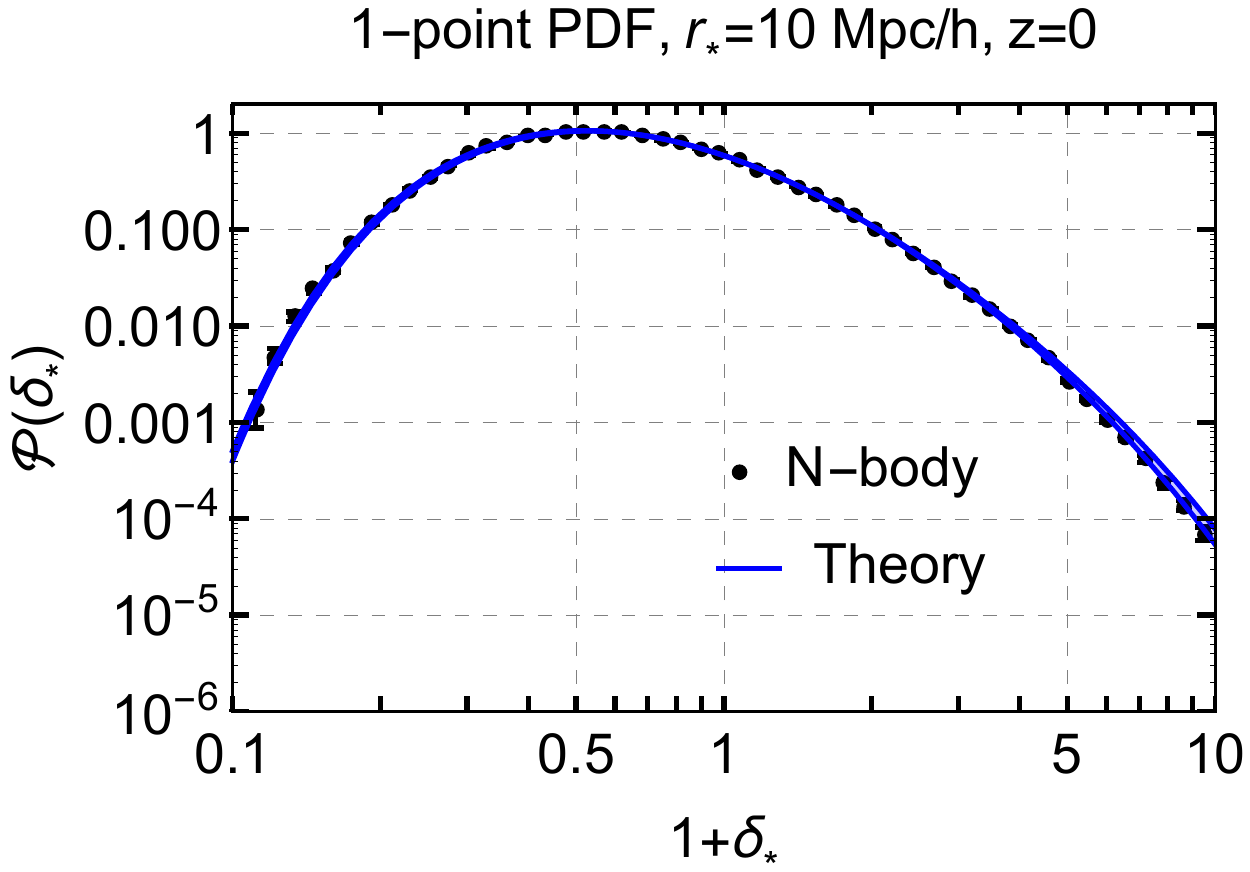}
\includegraphics[width=0.49\textwidth]{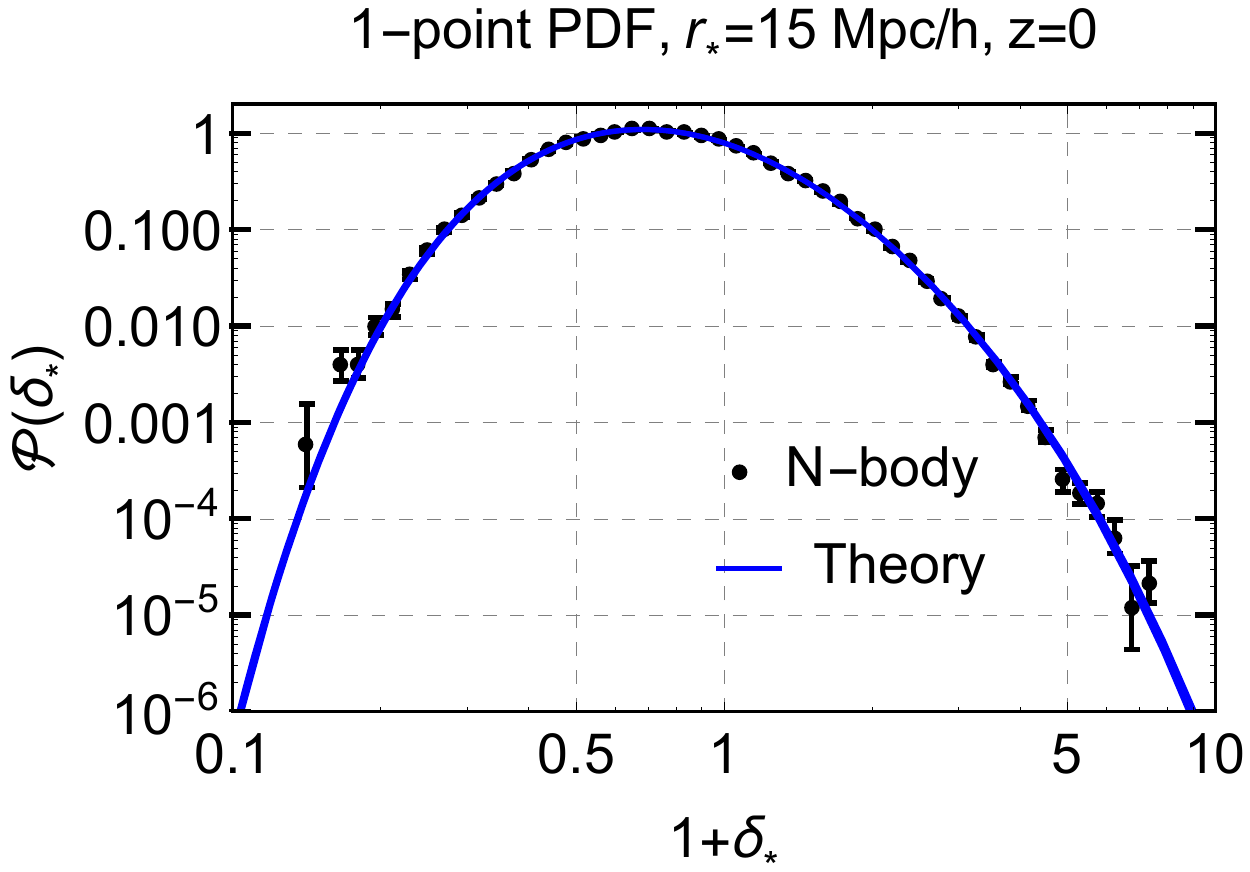}
\end{center}
\caption{\label{fig:finalPDF} 1-point probability distribution
  function computed in this work (blue band) against that extracted
  from our N-body data (black dots). The
  results are presented for redshift zero, $z=0$, and  
two cell radii, $r_*=10\,\text{Mpc}/h$
  (left panel) and $r_*=15\,\text{Mpc}/h$ (right panel). The width of
  the theoretical band is set by the uncertainty in modelling the
  short-distance physics; it exceeds the line width only at the tails
  of the distribution. 
}
\end{figure}

We compared the results of our computation to the N-body data.
Despite a qualitative agreement, we observed a sizable quantitative
discrepancy, which we attributed to the failure of the pressureless fluid
approximation at short scales.
We proposed two models
for renormalization of the short-scale contributions 
in the spirit of the EFT of
LSS.
The two models agree at the percent level at moderate density
contrasts and deviate by at most $30\%$ at the tails of the
distribution. We have suggested to use the difference between the two
models as an estimate of the theoretical uncertainty of our approach
stemming from the lack of control over the short-distance physics. 

The resulting theoretical PDFs for cells with radii
$r_*=10\,\text{Mpc}/h$ and
$15\,\text{Mpc}/h$ at $z=0$ are shown against N-body data in
Fig.~\ref{fig:finalPDF}. The lines corresponding to the two
counterterm models are almost indistinguishable. We see that the
theory and the data are in excellent agreement. The theoretical
uncertainty is smaller for the larger radius, which is consistent with
the expectation that the UV effects should be suppressed at large
distances. 

The 1-point PDF has a very distinctive sensitivity to the dynamics and
initial statistics  
of the matter density perturbations. 
One observes from Eq.~(\ref{Plast}) that 
the dependence of the PDF on the filtered 
linear density variance $g\sigma_{R_*}$
factorizes.
By 
varying $\delta_*$ one effectively changes the filtering
radius, and thus probes the variance of the density field at different
scales.  
Going to the underdense tail allows one to test the linear power
spectrum at very small Lagrangian  
radii, i.e. at the scales which are beyond the regime of validity
of the standard cosmological perturbation theory. 

We have shown that the prefactor in (\ref{Plast}) has only a weak
dependence on cosmology. Thus, any variation of the cosmological
parameters or extension of $\L$CDM is expected to affect the PDF
primarily through the leading exponent. Even a small change in the
growth factor or the linear variance can have a strong effect on the
PDF. On the other hand, the sensitivity to the non-linear dynamics 
at leading order 
is encoded in the spherical collapse mapping 
$F(\delta_*)$. It will be interesting to understand to which extent
this property of the 1-point PDF can be used to constrain non-standard
dark matter scenarios or modifications of gravity.  

The expansion parameter in our approach is the linear density variance
smoothed at the scale of the window function. 
Thus, corrections to our
result for the aspherical prefactor are expected to scale as
$(g\s_{r_*})^2$, 
c.f.
Eq.~\eqref{eq:scaling}.
On the other hand, our comparison with the
N-body data has not revealed any presence of such corrections for  
$(g\s_{r_*})^2$ as large as $\sim 0.5$ (for $z=0$, $r_*=10$Mpc$/h$,
see Table~\ref{tab:sigmas}). This indicates that the coefficient in
front of the correction is suppressed. Nevertheless, as one decreases
the cell radius,
the corrections will grow and eventually the `semiclassical'
approximation is expected to break down. Another limitation of our
method in its present form is its reliance on the existence of an
analytic spherical collapse saddle-point solution. As discussed in
Sec.~\ref{sec:SP}, this assumption is actually violated for large
overdensities $\delta_*\gtrsim 7$ where the saddle-point solution
exhibits shell crossing. Remarkably, the N-body data still obey the
`semiclassical' scaling up to the maximal value $\delta_*=9$ that we
were able to explore. We have interpreted it as a consequence of the
slow signal propagation in dark matter which implies that the
information about the shell crossing in the inner part of the density
profile does not have enough time to reach the boundary of the
cell. However, an extension to yet higher overdensities will likely
require a modification of the semiclassical method to properly take
the shell crossing into account, cf.~\cite{Pajer:2017ulp}. 
It would be highly instructive to map the domain of validity of the
`semiclassical' formula (\ref{Plast}) in the space of cell radii $r_*$
and densities $\delta_*$ using high-precision counts-in-cells
statistics obtained from state of the art cosmological simulations. 

Let us briefly comment on the relation between the PDF derived in this
work and the log-normal distribution that has been widely used in the
literature to model the counts-in-cells statistics. As discussed in
Appendix~\ref{app:ln}, the success of the log-normal model does not
appear to have any physical meaning, but is a consequence of an
accidental conspiracy between the spherical collapse dynamics and the
shape of the power spectrum in our universe, that makes the
combination $F(\delta_*)/\sigma_{R_*}$ entering in the exponent of
(\ref{Plast}) look similar to $\ln(1+\delta_*)/\sigma_{\rm ln}$, where
$\sigma_{\rm ln}$ is the log-density variance. 
A change of the slope
of the power spectrum would destroy this conspiracy. Even for the
standard $\L$CDM 
the approximation of $F/\sigma_{R_*}$ by the logarithm does not work
for large under- and over-densities. Moreover, the log-normal model does not
incorporate the correct prefactor. As a result, it 
significantly deviates from the N-body data in the tails of the
distribution (cf. Refs.~\cite{Klypin:2017jjg,Uhlemann:2016liz}).

Before concluding, we summarize several key features of our approach:
\begin{enumerate}
\item It clearly separates the leading exponent from the
  prefactor. This allows us to keep  
the saddle-point expansion under control and disentangle the
cosmology-dependent effects  
from those of non-linear clustering.

\item We use the exact $\L$CDM mapping for spherical collapse. 
This is crucial for the accuracy of our calculation, as
the PDF is exponentially sensitive to the mapping.

\item It explicitly takes into account aspherical fluctuations
along with the contributions beyond the single-stream
pressureless perfect fluid approximation.

\item It is based on the first principles and does not introduce
  any fitting parameters.   

\item It provides an intrinsic estimate of the theoretical uncertainty
  that does not require an input from the N-body data. 

\item It increases the range of agreement 
between the analytic theory and N-body simulations 
compared to previous approaches.

\end{enumerate}

In this paper we have studied the simplest case of non-perturbative
cosmological statistics: 1-point PDF 
of dark matter in real space for Gaussian adiabatic initial
conditions. 
Applications to realistic observations, such as galaxy surveys,
Lyman-$\a$ forest or
21~cm intensity mapping will require extension of the method to the biased
tracers in redshift space. Another line of research is the statistics
of the 2-dimensional 
projected density field and weak lensing convergence. 
Last but not
least, a generalization to the 2-point PDF will be 
very interesting as a way to
probe the primordial non-Gaussianity.
We believe that our study paves the way for a systematic investigation
these non-perturbative statistics as
potentially powerful cosmological probes. 

\paragraph*{Acknowledgments}

We are indebted to Cora Uhlemann for
stimulating discussions and comments on the draft. 
We thank 
D.~Blas, 
S.~Dubovsky,
S.~Foreman,
M.~Garny,
J.~Garcia-Bellido,
M.~Mirbabayi,
L.~Hui,
F.~Popov,
F.~Schmidt,  M.~Schmittfull, 
M.~Simonovi\'c,
A.~Shkerin,
P.~Valageas,
Z.~Vlah
and M.~Zaldarriaga
for fruitful conversations and ecouraging interest. 
The numerical part of this work was done on the Helios cluster 
at the Institute for Advanced Study.
The work of M.I. and S.S. is supported by the Swiss National Science
Foundation.  
A.K. is supported by the William D. Loughlin Membership Fund.
M.I. also acknowledges a partial support by the RFBR grant No. 17-02-01008.
S.S. is supported by the RFBR grant No. 17-02-00651.

\appendix

\section{Conventions}
\label{sec:notations}

In this Appendix we summarize our main notations and conventions. 
The Fourier transform is defined as, 
\be
\delta(\x)=\int_\k \delta(\k)
\e^{i\k\cdot\x}\,, 
\ee
where the integration measure in momentum space is
\be
\label{measurek}
\int_{\bf k} = \int \frac{d^3k}{(2\pi)^3}\;.
\ee
We also use the shorthand notation for the radial integral in momentum
space, 
\be
\label{kintradial}
\int [dk]=\int_0^\infty \frac{k^2 dk}{(2\pi)^3}\;,
\ee 
and its generalization to several wavenumbers,
\be
\label{kintradialmany}
\int [dk]^n=\int\prod_{i=1}^n \frac{k^2_i dk_i}{(2\pi)^3}\;,
\ee 
The power spectrum is defined as,
\be
\langle \delta(\k)\delta(\k')\rangle = (2\pi)^3\delta_{\rm D}^{(3)}(\k+\k')P(k)\,,
\ee
where $\delta_{\rm D}^{(d)}(\k)$ is the Dirac delta-function in a
$d$-dimensional space.

We use the following definition for the spherical harmonics:
\bseq
\label{sphharm}
\begin{align}
\label{sphharm0}
&Y_0(\theta,\phi)=1\;,\\
&Y_{\ell m}(\theta,\phi)=\frac{(-1)^{\ell+m}}{2^\ell\ell !}
\bigg[\frac{2\ell+1}{4\pi}\frac{(\ell-|m|)!}{(\ell+|m|)!}\bigg]^{1/2}
\e^{im\phi}
(\sin\theta)^{|m|}\bigg(\frac{d}{d\cos\theta}\bigg)^{\ell+|m|}(\sin\theta)^{2\ell},
\notag\\
&\qquad\qquad\qquad\ell>0\;,~~~~-\ell<m<\ell\;.
\label{sphharmlm}
\end{align}
\eseq
They obey the relations, 
\be
\label{Yprop}
\Delta_{\Omega}Y_{\ell m}=-\ell (\ell+1)Y_{\ell m}~,~~~~ 
Y_{\ell m}(-{\bf n})=(-1)^\ell Y_{\ell m}({\bf n})~,~~~~
Y_{\ell m}^*({\bf n})=Y_{\ell,-m}({\bf n})\;,
\ee
where $\Delta_{\Omega}$ is the Laplacian on a unit 2-dimensional
sphere.
All harmonics are orthogonal and normalized to 1 when integrated over
a 2d sphere, except the monopole that has the norm $4\pi$,
\be
\label{harmnorm}
\int d\Omega\, Y_{\ell
  m}\,Y^*_{\ell'm'}=(4\pi)^{\delta_{0\ell}}\delta_{\ell
  \ell'}\delta_{mm'}\,, 
\ee
where $\delta_{ij}$ is the Kronecker delta symbol. Note that our
definition (\ref{sphharmlm}) differs by a factor
$(-1)^{\frac{m-|m|}{2}}$ from the standard conventions \cite{DLMF}.

We expand the fields over spherical harmonics in position and Fourier
space as,
\bseq
\begin{align} 
&\delta(\x)=\delta_0(r)+\sum_{\ell >0}\sum_{m=-\ell}^\ell \delta_{\ell
  m}(r)\, Y_{\ell m}(\x/r)\;,\\
&\delta(\k)=\delta_0(k)+\sum_{\ell >0}\sum_{m=-\ell}^\ell (-i)^\ell \,
\delta_{\ell m}(k)\, Y_{\ell m}(\k/k)\;.
\label{eq:Ylk}
\end{align}
\eseq
Due to the relations (\ref{Yprop}) we have,
\be
\label{deltalmprop}
\big(\delta_{\ell m}(r)\big)^*=\big(\delta_{\ell,-m}(r)\big)~,~~~~~~
\big(\delta_{\ell m}(k)\big)^*=\big(\delta_{\ell,-m}(k)\big)\;.
\ee
The coefficient functions in the above expansions are related by,
\be
\label{eq:dlfourier}
 \delta_{\ell m}(r)=4\pi \int [dk] \; 
j_\ell (kr)\,\delta_{\ell m}(k)\,,
\ee
where $j_\ell(x)$ is the spherical Bessel function of order $\ell$. It
is related to the  
Bessel function of the first kind via
\be 
\label{jJ}
j_\ell(x)=\sqrt{\frac{\pi}{2x}}J_{\ell+1/2}(x)\,.
\ee
The first few functions are,
\be
\label{j012}
j_0(x)=\frac{\sin x}{x}~,~~~~
j_1(x)=\frac{\sin x}{x^2}-\frac{\cos x}{x}~,~~~~
j_2(x)=\bigg(-\frac{1}{x}+\frac{3}{x^3}\bigg)\sin x-\frac{3}{x^2}\cos x\;.
\ee
Spherical Bessel functions $j_\ell(kr)$ with different arguments $k$
form an orthogonal basis on the half-line with the normalization 
\be
\label{Besselnorm}
\int_0^{\infty} dr \,r^2j_\ell(k'r) j_\ell(kr)
=\frac{\pi}{2k^2}\delta_{\rm D}^{(1)}(k-k')\,.
\ee
They are eigenmodes of the radial part of the Laplace operator,
\be
\label{Besseleq}
\d_r^2 j_\ell(kr)+\frac{2}{r}\d_rj_\ell(kr)-\frac{\ell(\ell+1)}{r^2}
 j_\ell(kr)=-k^2 j_\ell(kr)\;.
\ee

\section{Description of N-body data}
\label{app:Nbody}

We use the data on the counts-in-cells 1-point PDF extracted from
the N-body simulations run with the \texttt{FastPM} code 
\cite{Feng:2016yqz}.
To generate these data we simulated
$300$ boxes with $L_{\text{box}}=256$ Mpc/$h$ on a side, totaling the
volume of $5~(\text{Gpc}/h)^3$. 
The number of dark matter particles per box is
$1024^3$, corresponding to 
mass resolution $1.1\cdot 10^9~ M_\odot/h$. 
The adopted force resolution
is half the mean of the dark matter particle separation $\sim 125$ kpc$/h$.
The boxes are initialized at $z=99$ with second-order Lagrangian
perturbation theory (2LPT), and then evolved to $z=0$ using the
\texttt{FastPM} integration scheme and 40 linearly spaced in scale
factor time steps. 
We assumed a flat $\L$CDM with $\Omega_m = 0.26$, $\Omega_b=0.044$, $h = 0.72$,
$n_s=0.96$, Gaussian initial conditions, $\sigma_8=0.794$.
The input linear power spectrum
was generated with the
Boltzmann code \texttt{CLASS} \cite{Blas:2011rf}.

For each box we saved snapshots taken at $z=0,~0.7,~4$ and extracted
$12^3$ non-overlapping $r_*=10$ Mpc$/h$ (518400 in total) and $8^3$
$r_*=15$ Mpc$/h$ spheres (153600 in total) centered on a regular grid
with $2r_*$ spacing. 
The data shown in this paper were binned in $\sim 50$ 
logarithmically-spaced intervals spanning the range 
$1+\delta_*=[0.1,10]$.
When comparing the data against theory, we also integrate 
the theoretical predictions within corresponding bin intervals.
The data errors shown correspond to the Poissonian standard deviation.

The counterterm $\g_0$ is measured from the non-linear 
dark matter power spectrum of the Horizon Run 2 simulation (HR2)
\cite{Kim:2011ab}, whose cosmology is identical to ours. 
The HR2 simulation box ($L_{\text{box}}=7.2$ Gpc/$h$) is significantly
larger than the one of our simulations, and therefore
 allows for a very precise measurements of $\g_0$, 
which is important for the accuracy of our theoretical prediction.
We also estimated the non-linear dark matter power spectrum of our
simulations using  
the \texttt{nbodykit} toolkit \cite{Hand:2017pqn}, and found that it
is consistent with HR2 within 
statistical errors within interesting range of $k$.

The simulations used in this paper compromise on accuracy 
in order to produce large statistics given limited computational
resources available to us.  
Thus, they are much less accurate than 
the state-of-the-art simulations such as
reported in Ref.~\cite{Klypin:2017jjg}. 
In particular, our experiments show that numerical effects such as force resolution, super-sample variance and particle counts are not negligible. They have a noticeable effect on the tails of the counts-in-cells probability distribution. 
Our experiments suggest that the systematic numerical errors 
are comparable to the statistical ones
for the most over(under)-dense bins, whose comparison
with theory should be taken with a grain of salt.
We plan to use more accurate N-body simulations
for an exhaustive precision comparison in the future.

\section{Dynamics of spherical collapse}
\label{sec:SC}

\subsection{Spherical collapse in Einstein--de Sitter universe}
\label{sec:back}

Consider a spherically symmetric density perturbation in a spatially flat
universe filled with non-relativistic matter. For concreteness, we
focus on the case of an overdensity. We study the motion of a
spherical shell of matter enclosing the total mass $M$. Before the
onset of shell-crossing the mass within the shell is conserved. Due to 
Newton's theorem (or Birkhoff's theorem in general relativity)
the mechanical energy of the shell is conserved, so we write,
\be 
\label{energy}
\frac{1}{2}\left(\frac{dy}{d\tau}\right)^2-\frac{GM}{y}={\cal E}\,,
\ee
where $y$ is the physical radius of the shell and $\tau$ is the
physical time. The conserved energy ${\cal E}$ is negative for the
case of an overdensity. It is straightforward to obtain the solution
to (\ref{energy}) in a parametric form, 
\bseq
\label{eq:ytau}
\begin{align}
\label{eq:y}
&y=-\frac{GM}{2{\cal E}}(1-\cos\theta)\;,\\
\label{eq:tau}
&\tau=\frac{GM}{(-2{\cal E})^{3/2}}(\theta-\sin\theta)\;.
\end{align}
\eseq
Next, we switch from the variables $y$, $\tau$ to the comoving
radius of the shell $r=y/a$ and the scale factor $a$. We use,
\begin{align}
\label{eq:a}
&a=\bigg(\frac{8\pi
  G}{3}\rho_ia_i^3\bigg)^{1/3}\bigg(\frac{3}{2}\tau\bigg)^{2/3}\;,\\
\label{eq:M}
&M=\frac{4\pi}{3}\rho_i a_i^3 R^3\;,
\end{align}
where $\rho_i$, $a_i$ and $R$ are the matter density, the scale factor
and the comoving radius of the shell at some early time when the
universe was almost homogeneous. Note that $R$ has a finite limit at
$a_i\to 0$ which coincides with the Lagrangian radius of the
shell. Substitution of (\ref{eq:a}), (\ref{eq:M}) into Eqs.~(\ref{eq:ytau}),
gives,
\bseq
\label{sphercol}
\begin{align}
\label{sphercol1}
&r=R \bigg(\frac{2}{9}\bigg)^{1/3}\frac{1-\cos\theta}{(\theta-\sin\theta)^{2/3}}\;,\\
\label{sphercol2}
&a=\bigg(\frac{9}{2}\bigg)^{1/3}\frac{4\pi G}{3(-2{\cal
    E})}\rho_ia_i^3R^2
(\theta -\sin\theta)^{2/3}\;,
\end{align} 
\eseq 

We now recall the definition of the spherically averaged density
contrast (\ref{deltabar}). Expressing it through the enclosed mass $M$, the
radius of the shell and the mean density of the universe $\rho_{\rm
  univ}$ we obtain,
\be
\label{masscons}
1+\bar\delta(r)=\frac{3M}{4\pi y^3\rho_{\rm
    univ}}=\bigg(\frac{R}{r}\bigg)^{3}\;, 
\ee
where in the second equality we used that $\rho_{\rm
  univ}=\rho_ia_i^3/a^3$. This gives the relation (\ref{eq:LagR})
between the Lagrangian and
Eulerian radii of the shell. Besides, we have from (\ref{sphercol1}),  
\be
\label{fdef}
 \bar\delta={\cal F}(\t)~,~~~~\text{where}~~
{\cal F}(\t)\equiv \frac{9}{2}\frac{(\t-\sin\t)^2}{(1-\cos\t)^3}-1\,.
\ee
It remains to relate the constant ${\cal E}$ to the
initial overdensity. To this end, we consider Eqs.~(\ref{sphercol2}),
(\ref{fdef}) at the initial time. The parameter 
$\t$ is initially small, so we can
expand, 
\be
a_i=\frac{\t_i^2}{2}\frac{4\pi G}{3(-2{\cal E})}\rho_i a_i^3 R^2~,~~~~~~~
\bar\delta_i(R)=\frac{3}{20}\theta_i^2\;,
\ee 
which gives
\be
\label{Edeltai}
{\cal E}=-\frac{5}{3}\,\frac{\bar\delta_i(R)}{a_i}\,\frac{4\pi
  G}{3}\rho_ia_i^3R^2\;. 
\ee
Substituting ${\cal E}$ back into (\ref{sphercol2}) and introducing
the rescaled linear density contrast $\bar\delta_L(R)\equiv
a\bar\delta_i(R)/a_i$ we arrive at
\be
\label{gdef}
\bar\delta_L(R)={\cal G}(\t)\equiv
\frac{3}{20}[6(\theta-\sin\theta)]^{2/3}\,.
\ee
Equations (\ref{fdef}), (\ref{gdef}) together provide a mapping
between the linear and non-linear averaged density contrasts at a
given moment of time 
expressed parametrically through the so-called development angle $\t$.
The functions $f$ and $F$ used in the main text (see
Eq.~(\ref{eq:scmap})) are the superpositions
\be
f={\cal F}\circ{\cal G}^{-1}~,~~~~~
F={\cal G}\circ{\cal F}^{-1}\;.
\ee

We now derive several useful expressions for the fields characterizing
the spherical collapse that are required for the calculation of
linear fluctuations around the spherical collapse saddle point in
Sec.~\ref{sec:mainEq}. It is convenient to choose the logarithm of the
growth factor as a new time variable,
\be
\eta=\ln a\;.
\ee
The key object is the linear density profile which we rescale to zero
redshift. This will be denoted by $\delta_{L|0}(R)$. All other
quantities are sourced by it and should be understood as functions of
$\eta$ and $R$. We first rewrite (\ref{gdef})
\bseq
\be
\label{sphercol2a}
\theta-\sin\theta=\frac{\e^{3\eta/2}}{6}\bigg(\frac{20}{3}\,
\bar\delta_{L|0}(R)\bigg)^{3/2}\;,
\ee
which implicitly defines the function $\t(\eta,R)$. Then we obtain the
relations, 
\begin{align}
&\frac{\d\theta}{\d R}=\frac{3(\theta-\sin\theta)}{2(1-\cos\theta)}\cdot
\frac{\bar\delta'_{L|0}(R)}{\bar\delta_{L|0}(R)}~,~~~~~~~~~~~
\frac{\d\theta}{\d\eta}=
\frac{3(\theta-\sin\theta)}{2(1-\cos\theta)}~,\\ 
&\frac{\d r}{\d R}=
\bigg(\frac{2}{9}\bigg)^{1/3}\frac{1-\cos\theta}{(\theta-\sin\theta)^{2/3}}
\bigg[1+R\frac{\bar\delta_{L|0}'}{\bar\delta_{L|0}}
\bigg(\frac{3(\theta-\sin\theta)\sin\theta}{2(1-\cos\theta)^2}-1\bigg)
\bigg]\;,
\end{align}
which yield the overdensity field,
\be
\label{deltahat}
\delta=\bigg[\frac{r^2}{R^2}\frac{\d r}{\d R}\bigg]^{-1}-1
=\frac{9(\theta-\sin\theta)^2}{2(1-\cos\theta)^3}
\bigg[1+R\frac{\bar\delta_{L|0}'}{\bar\delta_{L|0}}
\bigg(\frac{3(\theta-\sin\theta)\sin\theta}{2(1-\cos\theta)^2}-1\bigg)\bigg]^{-1}-1\,.
\ee
We also need the velocity potential $\Psi$ defined as
\[
\d_r\Psi=-u_r/{\cal H}\;,
\]
where $u_r$ is the radial velocity of collapsing matter,
$u_r=\frac{\d r}{\d t}$, and ${\cal H}=\frac{1}{a}\frac{d a}{d
  t}$. Here $t$ is the conformal time. We obtain,
\be
\label{Psihat}
\d_r\Psi=-\frac{\d r}{\d \eta}=
-R\bigg(\frac{2}{9}\bigg)^{1/3}
\frac{1-\cos\theta}{(\theta-\sin\theta)^{2/3}}
\bigg[\frac{3(\theta-\sin\theta)\sin\theta}{2(1-\cos\theta)^2}-1\bigg]\;.
\ee
Finally, the rescaled velocity divergence is
\be
\label{Thetahat}
\Theta\equiv -\frac{\d_i u_i}{\cal H}
=\bigg(\frac{\d r}{\d R}\bigg)^{-1}\frac{\d}{\d R}\d_r\Psi
+\frac{2}{r}\d_r\Psi\;.
\ee
\eseq

In the case of an underdensity, 
the spherically symmetric dynamics is similar with only minor
modifications. Without repeating the analysis, we summarize the
relevant expressions, 
\bseq
\begin{gather}
{\cal G}(\t)= \frac{3}{20}[6(\sh\t-\t)]^{2/3}\,,
\qquad{\cal F}(\t)= \frac{9(\sh\theta-\theta)^2}{2(\ch\theta-1)^3}-1\,,\\
\sh\theta-\theta=\frac{\e^{3\eta/2}}{6}
\bigg(-\frac{20}{3}\bar\delta_{L|0}(R)\bigg)^{3/2}\;,\\
\frac{\d\theta}{\d R}=\frac{3(\sh\theta-\theta)}{2(\ch\theta-1)}
\frac{\bar\delta_{L|0}'}{\bar\delta_{L|0}}\;,
\qquad\frac{\d\theta}{\d \eta}=\frac{3(\sh\theta-\theta)}{2(\ch\theta-1)}\;,
\\
r=R\bigg(\frac{2}{9}\bigg)^{1/3}
\frac{\ch\theta-1}{(\sh\theta-\theta)^{2/3}}\;,\\ 
\frac{\d r}{\d R}=\bigg(\frac{2}{9}\bigg)^{1/3}
\frac{\ch\theta-1}{(\sh\theta-\theta)^{2/3}}
\bigg[1+R\frac{\bar\delta_{L|0}'}{\bar\delta_{L|0}}\bigg(
\frac{3(\sh\theta-\theta)\sh\theta}{2(\ch\theta-1)^2}-1\bigg)\bigg]\;,\\
\delta=\frac{9(\sh\theta-\theta)^2}{2(\ch\theta-1)^3}
\bigg[1+R\frac{\bar\delta_{L|0}'}{\bar\delta_{L|0}}\bigg(
\frac{3(\sh\theta-\theta)\sh\theta}{2(\ch\theta-1)^2}-1\bigg)\bigg]^{-1}-1\;,\\
\d_r\hat\Psi=-R\bigg(\frac{2}{9}\bigg)^{1/3}
\frac{\ch\theta-1}{(\sh\theta-\theta)^{2/3}}
\bigg[\frac{3(\sh\theta-\theta)\sh\theta}{2(\ch\theta-1)^2}-1\bigg]\;.
\end{gather}
\eseq

\subsection{Spherical collapse in $\L$CDM}

Here we discuss how the previous results are modified in $\L$CDM. In
the presence of a cosmological constant $\L$ the equation (\ref{energy})
for the trajectory of a spherical shell 
is replaced by \cite{Peebles:1984ge,Lahav:1991wc}, 
\be
\label{energyLambda}
\frac{1}{2}\left(\frac{dy}{d\tau}\right)^2-\frac{GM}{y}
-\frac{\L y^2}{6}={\cal E}\,.
\ee
Unlike Eq.~(\ref{energy}), 
this cannot be solved analytically, so one has to resort to
numerical integration. It is convenient to use the scale
factor\footnote{We choose the scale factor to be normalized to 1 at
  the present epoch.} $a$ as
the time variable and switch from $y$ to the variable
\be
\label{zeta}
\zeta\equiv R/r\;,
\ee
where $r=y/a$ and $R=\lim_{a\to 0}r$. One uses the Hubble equation,
\be
\label{eq:Hubble}
\frac{1}{a^2}\bigg(\frac{da}{d\tau}\bigg)^2=
\frac{8\pi G}{3}\rho_{\rm univ}+\frac{\L}{3}\;,
\ee
and the relations $\rho_{\rm univ}=\rho_0/a^3$, $\L=8\pi
G\rho_0\Omega_\L/\Omega_m$, with $\rho_0$ the present-day average {\em
  matter} density and $\Omega_\L=1-\Omega_m$. Then
Eq.~(\ref{energyLambda}) takes the following form,
\be
\label{eqzeta1}
\bigg(1-\frac{d\ln \zeta}{d\ln a}\bigg)^2=
\bigg(1+\frac{\Omega_\L}{\Omega_m}a^3\bigg)^{-1}
\bigg(\zeta^3+\frac{\Omega_\L}{\Omega_m}a^3+\frac{3{\cal E}}{4\pi
  G\rho_0 R^2}\,a\zeta^2\bigg)\;.
\ee
To fix the value of the energy ${\cal E}$, we observe that
Eq.~(\ref{masscons}) still applies in $\L$CDM, so we have,
\be
\label{bardeltazeta}
1+\bar\delta(r)=\zeta^3\;,
\ee
which at early times gives
$\zeta=1+\big(a\bar\delta_i(R)\big)/(3a_i)$. Substituting this into
(\ref{eqzeta1}) and matching terms linear in $a$ at $a\to 0$, we
recover the same expression for ${\cal E}$, as in the EdS case,
\be
\label{ELambda}
\frac{3{\cal E}}{4\pi
  G\rho_0R^2}=-\frac{5}{3}\frac{\bar\delta_i(R)}{a_i}\;. 
\ee
The next step is to express the initial overdensity in terms of the
linear density contrast 
$\bar\delta_L(R)$ at the redshift $z$, at which we want to establish the
spherical collapse mapping. To this end we write,
\be
\label{deltaLLambda}
\frac{\bar\delta_i(R)}{a_i}=\frac{g(z)}{g(z_i)}\bar\delta_i(R)\,
\frac{g(z_i)}{a_i}\,\frac{1}{g(z)}
=\frac{g_\L}{g(z)}\bar\delta_L(R)\;.
\ee
In the last equality
we have used that at early times the growth factor is
proportional to $a$,
\be
\label{gearly}
g(z_i)=g_\L\cdot a_i\,
\ee
where $g_\L$ is a constant\footnote{Recall that we normalize $g(z)$ to
be 1 at $z=0$, which leads to a constant offset between $g$ and $a$ in
the matter-dominated era. For our reference cosmology
$g_\L=1.328$.}. Collecting the relations (\ref{ELambda}),
(\ref{deltaLLambda}) and inserting them into Eq.~(\ref{eqzeta1}) we
cast the latter in the form,
\be
\label{eqzeta2}
\bigg(1-\frac{d\ln \zeta}{d\ln a}\bigg)^2=
\bigg(1+\frac{\Omega_\L}{\Omega_m}a^3\bigg)^{-1}
\bigg(\zeta^3
-\frac{5g_\L}{3g(z)}\bar\delta_L(R)\,a\zeta^2
+\frac{\Omega_\L}{\Omega_m}a^3\bigg)\;.
\ee
With this in hand, the algorithm to construct the spherical collapse
mapping goes as follows:
\begin{itemize} 
\item[({\sf i})] Fix a value $\bar\delta_L(R)$ of the spherically
  averaged
linear overdensity at redshift $z$;
\item[({\sf ii})] Solve Eq.~(\ref{eqzeta2}) from $a=0$ to
  $a=(1+z)^{-1}$ with the initial condition\\ $\zeta\big|_{a=0}=1$;
\item[({\sf iii})] Compute $f$ as 
$f\big(\bar\delta_L(R);z\big)=\zeta^3\big((1+z)^{-1}\big)-1$.
\end{itemize}
The function $F$ is then found as the inverse of $f$.

As discussed in Sec.~\ref{sec:SP}, the functions $f$ and $F$ computed
in this way are very weakly depending on the redshift and, somewhat
surprisingly, coincide with the corresponding functions in EdS
cosmology at the level of a few per mil in the relevant range of
density contrasts.

Before concluding this section, we note that the formulas derived
above can be used to obtain a first order differential
equation for the growth factor $g$ as a function of $a$ in the $\L$CDM
universe. To this end, we assume that the overdensity is small at all
times, so that we can linearize Eq.~(\ref{bardeltazeta}),
\[
\zeta=1+\frac{g(a)}{3g(a_i)}\bar\delta_i(R)\;.
\] 
Substituting this into (\ref{eqzeta2}) and also linearizing it in
$\bar\delta_i(R)$ we arrive at,
\be
\label{gLambda}
\frac{dg}{da}=\bigg(1+\frac{\Omega_\L}{\Omega_m}a^3\bigg)^{-1}
\bigg(-\frac{3g(a)}{2a}+\frac{5g_\L}{2}\bigg)\;,
\ee 
which is to be integrated with the boundary condition $g\big|_{a=1}=1$.

\subsection{Monopole response matrix}
\label{sec:spher}

In this section we derive analytic expressions for the monopole
response matrix $Q_0(k_1,k_2)$ introduced in Eq.~(\ref{perturbQl}) and
the monopole fluctuation determinant (\ref{Dl0}). These results are used
in Sec.~\ref{sec:aspt} for the perturbative calculation of the
aspherical prefactor and for validating our numerical code (see
Appendix~\ref{sec:aspy}). The starting point of the derivation is the
relation provided by the spherical collapse mapping,
\be
\label{FdeltaW}
F(\bar\delta_W)=\bar\delta_L\big(r_*(1+\bar\delta_W)^{1/3}\big)\;.
\ee
We consider a monopole fluctuation on top of the saddle-point
configuration (\ref{eq:deltahatlin}), (\ref{eq:deltahatlinprofile}),
so we write
\be
\delta_L(R)=\hat\delta_L(R)+\delta_{L,0}^{(1)}(R)~,~~~~~~~
\bar\delta_W=\delta_*+\bar\delta_W^{(1)}+\bar\delta_W^{(2)}\;,
\ee
where the terms $\bar\delta_W^{(1)}$ and $\bar\delta_W^{(2)}$ are
linear and quadratic in $\delta_{L,0}^{(1)}$
respectively. Substituting these expressions into (\ref{FdeltaW}),
Taylor expanding the two sides and grouping the terms of linear and
quadratic order, we obtain two equations,
\bseq
\begin{align}
\label{linresponse}
&F'(\delta_*)\,\bar\delta_W^{(1)}=\frac{R_*\bar{\hat\delta}_{L}'(R_*)}{3(1+\delta_*)}
\,\bar\delta_W^{(1)}+\bar\delta_{L,0}^{(1)}(R_*)\;,\\
&F'(\delta_*)\,\bar\delta_W^{(2)}
+\frac{F''(\delta_*)}{2}\big(\bar\delta_W^{(1)}\big)^2
=\frac{R_*\bar{\hat\delta}_{L}'(R_*)}{3(1+\delta_*)}
\,\bar\delta_W^{(2)}
+\frac{R_*^2}{18(1+\delta_*)^2}\Big(\bar{\hat\delta}_L''(R_*)
-2\bar{\hat\delta}_L'(R_*)\Big)\big(\bar\delta_W^{(1)}\big)^2\notag\\
&\qquad\qquad\qquad\qquad\qquad\qquad~~
+\frac{R_*}{2(1+\delta_*)}\,\bar\delta_W^{(1)}\,
\big(\bar\delta_{L,0}^{(1)}\big)'(R_*)\;,
\label{quadresponse}
\end{align} 
\eseq
where $R_*$ is defined in (\ref{Rstar}). Next we use the expressions
\bseq
\begin{align}
\label{dhatprime}
&\bar{\hat\delta}'_L(R_*)=-\frac{3F(\delta_*)}{R_*}
\bigg(1-\frac{\xi_{R_*}}{\sigma^2_{R_*}}\bigg)\;,\\
\label{dhatdprime}
&\bar{\hat\delta}''_L(R_*)=\frac{12F(\delta_*)}{R_*^2}
\bigg(1-\frac{\xi_{R_*}}{\sigma^2_{R_*}}\bigg)
-F(\delta_*)\frac{\Sigma^2_{R_*}}{\sigma^2_{R_*}}\;,
\end{align}
\eseq
where $\sigma_{R_*}^2$, $\xi_{R_*}$ are defined in Sec.~\ref{sec:SP}
and 
\be
\label{defSSigma}
\Sigma^2_{R_*}=4\pi\int [dk]\,k^2 |W_{\rm th}(kR_*)|^2 P(k)\,.
\ee
Substituting (\ref{dhatprime}) into (\ref{linresponse}) we get,
\be
\label{deltaW1}
\bar\delta_W^{(1)}=\frac{\bar\delta_{L,0}^{(1)}(R_*)}{\hat C(\delta_*)}\;,
\ee
where $\hat C(\delta_*)$ is introduced in Eq.~(\ref{eq:lambda}). We
note in passing that this relation implies an expression
for the linear monopole response kernel $S(k)$ (see
Eq.~(\ref{perturbQl})),
\[
S(k)=\frac{W_{\rm th}(kR_*)}{\hat C(\delta_*)}
\]
From (\ref{quadresponse}) we further obtain,
\be
\label{deltaW2}
\bar\delta_W^{(2)}=-\frac{\hat E(\delta_*)}{\hat C^3(\delta_*)}
\Big(\bar\delta_{L,0}^{(1)}(R_*)\Big)^2
+\frac{1}{(1+\delta_*)\hat C^2(\delta_*)}
\,\delta_{L,0}^{(1)}(R_*)\,\bar\delta_{L,0}^{(1)}(R_*)\;,
\ee
where
\be
\label{Edef}
\hat E(\delta_*)=\frac{F''(\delta_*)}{2}
+\frac{F'(\delta_*)}{1+\delta_*}+\frac{F(\delta_*)}{(1+\delta_*)^2}
\frac{R_*^2\Sigma^2_{R_*}}{18\sigma^2_{R_*}}\,,
\ee
and we have used the identity,
\[
\big(\bar\delta_{L,0}^{(1)}\big)'(R_*)=\frac{3}{R_*}
\Big(\bar\delta_{L,0}^{(1)}(R_*)-\delta_{L,0}^{(1)}(R_*)\Big)\;.
\]
Finally, switching from position to momentum space,
\[
\bar\delta_{L,0}^{(1)}(R_*)=4\pi\!\!\int [dk]\,W_{\rm
  th}(kR_*)\,\delta_{L,0}^{(1)}(k)~,
~~~~~~~\delta_{L,0}^{(1)}(R_*)=4\pi\!\!\int [dk]\,
\frac{\sin(kR_*)}{kR_*}\,\delta_{L,0}^{(1)}(k)\;,
\]
and comparing (\ref{deltaW2}) to Eq.~(\ref{perturbQl}) we arrive at
the following expression for the monopole response matrix,
\be
\label{Q0}
\begin{split}
Q_{0}(k_1,k_2)=&-\frac{4\pi \hat E}{\hat C^3}W_{\rm th}(k_1 R_*)W_{\rm
th}(k_2 R_*)\\
&+\frac{2\pi}{(1+\delta_*)\hat C^2}
\left[W_{\rm th}(k_1\hat R_*)\frac{\sin(k_2R_*)}{k_2 R_*}+
\frac{\sin(k_1 R_*)}{k_1 R_*} W_{\rm th}(k_2 R_*)\right] \,,
\end{split}
\ee

To evaluate the monopole fluctuation determinant ${\cal D}_0$
defined in (\ref{Dl0}), we observe that the matrix $\mathbb{1}+2\hat\l
\sqrt{P}Q_0\sqrt{P}$ can be written as
\[
\mathbb{1}(k_1,k_2)+a(k_1)b(k_2)+b(k_1)a(k_2)\;
\] 
with
\begin{align}
&a(k)=2\hat\l\sqrt{P(k)}\,W_{\rm th}(kR_*)\;,\notag\\
&b(k)=\bigg[-\frac{2\pi\hat E}{\hat C^3}W_{\rm th}(kR_*)
+\frac{2\pi}{(1+\delta_*)\hat C^2}\frac{\sin kR_*}{kR_*}\bigg]\sqrt{P(k)}\;.\notag
\end{align}
The general formula for the determinant of a matrix of this form is
derived in Appendix~\ref{app:determinant}. Applying it to the case at
hand and using the expression (\ref{eq:lambda}) for $\hat\l$ 
gives,
\be
\label{eq:detl0}
\begin{split}
 \D_{0}=1+\frac{2F}{\hat C^2}\left[\hat E-\frac{\hat C}{(1+\delta_*)}
\frac{\xi_{R_*}}{\sigma^2_{R_*}}\right]+
\frac{F^2}{(1+\delta_*)^2\hat C^2}
\left[\left(\frac{\xi_{R_*}}{\sigma^2_{R_*}}\right)^2
-\frac{\sigma^2_{1\,R_*}}{\sigma^2_{R_*}}\right]\,,
\end{split} 
\ee
where we have defined
\be
\sigma_{1\,R_*}^2=
4\pi\int[dk]
\left(\frac{\sin(k R_*)}{k R_*}\right)^2P(k)\,.
\ee

\begin{figure}[t]
\begin{center}
\includegraphics[width=.6\textwidth]{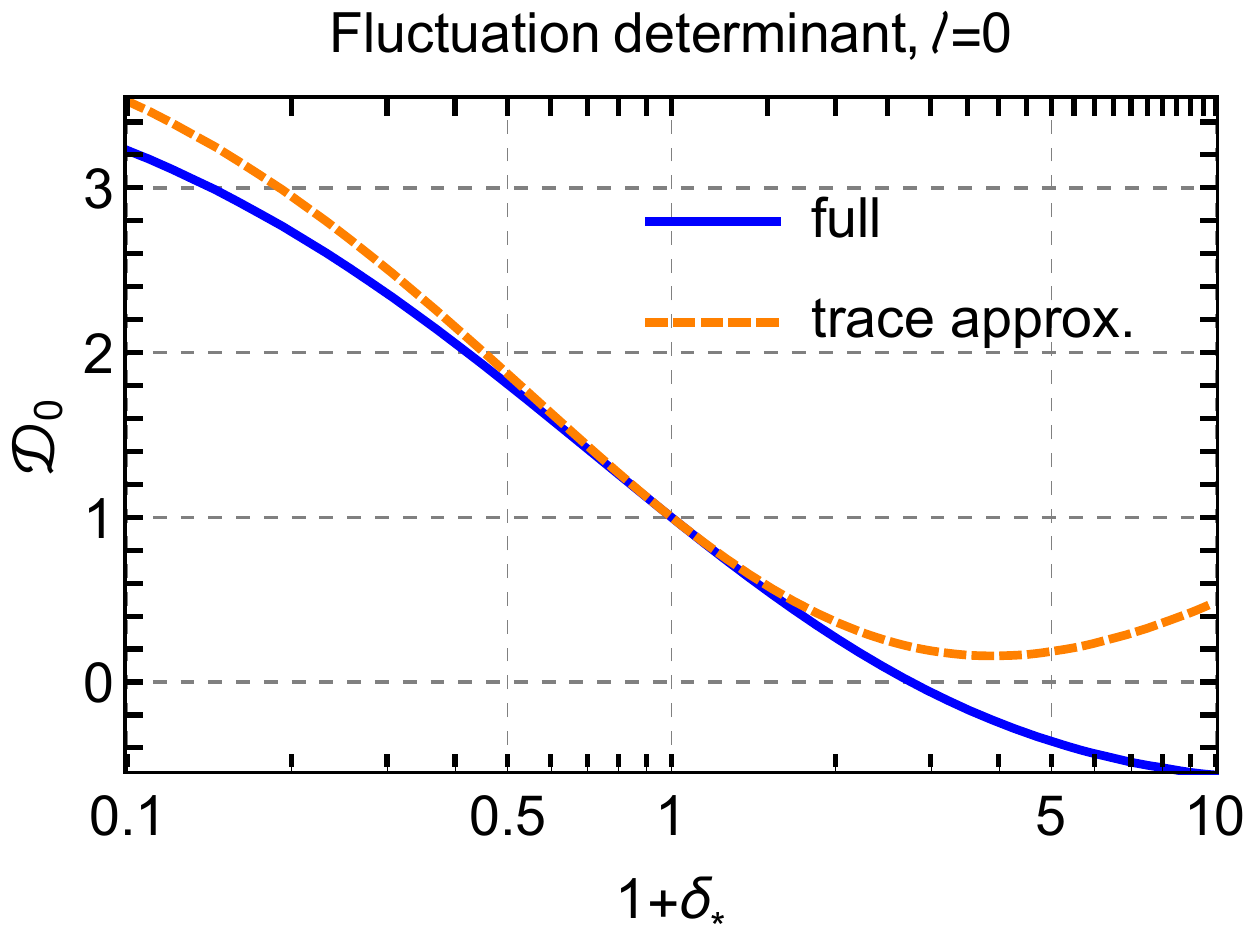}
\caption{\label{fig:detl0} 
Monopole fluctuation determinant ${\cal D}_0$ (blue, solid) and it
trace approximation (orange, dashed). Note that the determinant
crosses zero at $\delta_*\approx 1.75$. 
}
\end{center}
\end{figure}

It is instructive to compare the full result (\ref{eq:detl0}) to a
trace approximation which treats the matrix $2\hat\l
\sqrt{P}Q_0\sqrt{P}$ as small,
\eqref{eq:detl0}, 
\be
\label{eq:tracel0}
\begin{split}
\D_{0} &= \exp\left\{\text{Tr}\ln(\mathbb{1}+2\hat\l \sqrt{P} Q_{0}
\sqrt{P})\right\}\\
&
\approx  1+2\hat\l\text{Tr}(\sqrt{P}Q_{0}\sqrt{P})
= 1+\frac{2F}{\hat C^2}\left[\hat E-\frac{\hat C}{1+\delta_*}
\frac{\xi_{R_*}}{\sigma^2_{R_*}}\right]\,.
\end{split}
\ee
We see that it reproduces the first two terms in (\ref{eq:detl0}), but
misses the third one. 
In Fig.~\ref{fig:detl0} we display the trace approximation versus the
full result (\ref{eq:detl0}) for our reference cosmology. We observe
that, though the trace approximation is, strictly speaking, applicable
only for $\delta_*\ll 1$, it works quite well in the range
$\delta_*\in [-0.9,1]$.
Still, at larger overdensities it deviates significantly from the true
result. 

\subsection{Growth factor in a spherically-symmetric separate universe}
\label{app:Deff}

To estimate the dependence of the UV
counterterm on the density in Sec.~\ref{sec:counterterm}, 
we need the linear growth factor for
perturbations in the background of a spherical top-hat
overdensity\footnote{We are talking about overdensity for concreteness. For
  an underdensity the reasoning is exactly the same.}. Due to the
Birkhoff theorem, such an overdensity can be treated as a separate
closed universe. Then the linear growth factor does not depend on the
wavenumber of the mode and can be derived by considering spherically
symmetric top-hat perturbations. 

Consider a spherically symmetric lump of matter with a top-hat profile,
whose final density contrast w.r.t. the unperturbed cosmology is equal
to $\delta$.  
Let us additionally
perturb this lump by a linear fluctuation $\delta_L^{(1)}$. 
According to the spherical collapse mapping,
this fluctuation produces the following 
perturbation of the non-linear density $\delta^{(1)}$,
\be
\delta^{(1)}=\frac{\delta_L^{(1)}}{F'(\delta_*)}\;.
\ee 
The density contrast in the separate universe should be normalized to
the background density of that universe,  
\be
\delta_{\rm su}^{(1)}=\frac{\delta^{(1)}}{1+\delta_*}\;.
\ee
This gives for the growth factor in the separate universe:
\be
\label{growthsepar}
D(\delta_*,z)=\frac{g(z)}{F'(\delta_*)(1+\delta_*)}\;.
\ee 
Note that the dependence of the growth factor on redshift and density
factorize. 

\section{Determinant of a matrix made of  two
  vectors}
\label{app:determinant}
In this Appendix we derive the following formula:
\be
\label{detvec}
\det(\delta_{ij}+a_i b_j+b_i a_j)=1+2(a\cdot b)+(a\cdot b)^2-a^2\, b^2\;,
\ee
where dot stands for the scalar product, $a\cdot b=\sum_i a_ib_i$,
etc.
We start with the trace representation of the determinant,
\be
\label{detlog}
\begin{split}
\det(\delta_{ij}+a_i b_j+b_i a_j)&=
\exp\big[\Tr\ln(\delta_{ij}+a_i b_j+b_i a_j)\big]\\
&=\exp\bigg[\sum_{n=1}^\infty
\frac{(-1)^{n-1}}{n}
\Tr\Big((a_i b_j+b_i a_j)^n\Big)\bigg].
\end{split}
\ee
Next, we write an Ansatz,
\be
\label{matrn}
(a_i b_j+b_i a_j)^n=U_n a_ia_j+V_n b_ib_j+W_n(a_ib_j+b_ia_j)\;,
\ee
where the coefficients obey the recursion relations,
\bseq
\label{recrel}
\begin{align}
&U_{n+1}=U_n(a\cdot b)+W_nb^2\;,\\
&V_{n+1}=V_n(a\cdot b)+W_na^2\;,\\
&W_{n+1}=U_na^2+W_n(a\cdot b)=V_nb^2+W_n(a\cdot b)\;.
\end{align}
\eseq
The last equality implies,
\be
\label{UnVn}
U_n=C_n a^2~,~~~~V_n=C_n b^2
\ee
and the system (\ref{recrel}) simplifies,
\bseq
\label{recrel2}
\begin{align}
&C_{n+1}=C_n (a\cdot b)+W_n\;,\\
&W_{n+1}=C_n a^2b^2+W_n(a\cdot b)\;.
\end{align}
\eseq
These are solved by the Ansatz,
\be
\begin{pmatrix} 
C_n\\
W_n
\end{pmatrix}
=
\begin{pmatrix} 
C_0\\
W_0
\end{pmatrix}
\; \a^n\;.
\ee
Substituting this into (\ref{recrel2}) one obtains two lineraly
independent solutions; the general solution is their sum,
\be
\begin{pmatrix} 
C_n\\
W_n
\end{pmatrix}
=C_0^+\begin{pmatrix} 
1\\
\sqrt{a^2b^2}
\end{pmatrix}\a_+^{n-1}
+
C_0^-\begin{pmatrix} 
1\\
-\sqrt{a^2b^2}
\end{pmatrix}\a_-^{n-1}\;,
\ee
where 
\be
\label{alphapm}
\a_{\pm}=(a\cdot b)\pm\sqrt{a^2b^2}\;.
\ee
Imposing the initial conditions $C_1=0$, $W_1=1$ fixes
\be
C_n=\frac{1}{2\sqrt{a^2b^2}}(\a_+^{n-1}-\a_-^{n-1})\;,~~~~
W_n=\frac{1}{2}(\a_+^{n-1}+\a_-^{n-1})\;.
\ee
Substituting these expressions into (\ref{matrn}) and taking the trace we find,
\be
\label{Trn}
\Tr\Big((a_i b_j+b_i a_j)^n\Big)=\a_+^n+\a_-^n\;.
\ee
Finally, inserting it into (\ref{detlog}) we get,
\be
\begin{split}
\exp\bigg[\sum_n\frac{(-1)^{n-1}}{n}(\a_+^n+\a_-^n)\bigg]
&=\exp\big[\ln(1+\a_+)+\ln(1+\a_-)\big]\\
&=\exp\big[\ln\big((1+(a\cdot b))^2-a^2b^2\big)\big],
\end{split}
\ee
which leads to (\ref{detvec}).

\section{Perturbation equations in $\L$CDM}
\label{sec:LCDM}

In the real universe the linear growth factor deviates quite significantly from the scale factor, see
the right panel of Fig.~\ref{fig:sigmasq_g}.
Thus, it is desirable to compute the prefactor for the exact cosmological model. 
In this section we present the generalization of Eqs.~(\ref{EPppp}), 
(\ref{mu2r2})
to the case of the $\L$CDM universe.

\begin{figure}[t]
\begin{center}
\includegraphics[width=.49\textwidth]{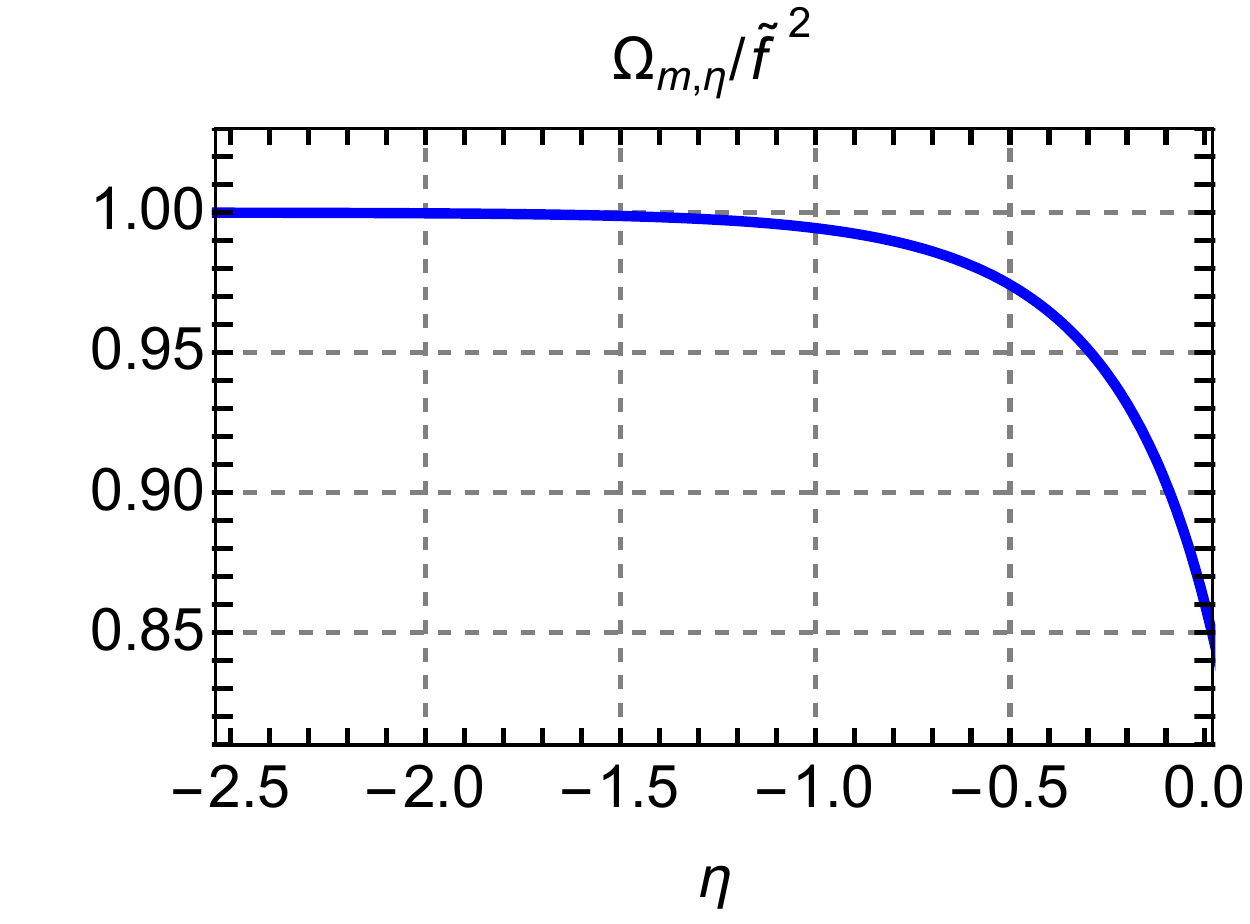}
\caption{\label{fig:growth} 
The ratio $\O_{m,\eta}/\tilde f^2$ as a function of $\eta=\ln g$.
}
\end{center}
\end{figure}

The departures from the EdS approximation are parametrized 
by the logarithmic growth factor\footnote{We use the notation $\tilde
  f$ for the logarithmic growth factor to avoid confusion with the
  function $f$ appearing in the spherical collapse mapping (\ref{eq:scmap}).}
\be
\tilde f(\eta)\equiv \frac{d \ln g}{d \ln a}\,,
\ee
where $g(\eta)$ is the linear growth factor in the $\L$CDM cosmology,
and $\eta$ now is defined as 
\be
\eta \equiv \ln g\,. 
\ee
Starting from the fluid 
equations for the $\L$CDM universe and proceeding as in
Sec.~\ref{sec:5.1}  
one obtains the set of equations for linearized fluctuations
with angular number $\ell$ on the spherical collapse solution:
\bseq
\label{EPpppL}
\begin{align}
\label{EPpppL1}
&\dot{\delta}_\ell-\Theta_\ell-\d_r\hat\Psi\,\d_r\delta_\ell
-\hat\Theta\,\delta_\ell-\d_r\hat\delta\,\d_r\Psi_\ell
-\hat\delta\,\Theta_\ell=0\;,\\
\label{EPpppL2}
&\dot{\Theta}_\ell+\left(\frac{3\O_{m,\eta}}{2\tilde f^2}-1\right)\Theta_\ell
-\frac{3\O_{m,\eta}}{2\tilde f^2}\delta_\ell
-\d_r\hat\Psi\,\d_r\Theta_\ell
-\d_r\hat\Theta\,\d_r\Psi_\ell
-2\d_r^2\hat\Psi\,\Theta_\ell\notag\\
&\qquad\qquad\qquad+2\bigg(\d_r^2\hat\Psi-\frac{\d_r\hat\Psi}{r}\bigg)
\,\bigg(\frac{2}{r}\d_r\Psi_\ell-\frac{\ell(\ell+1)}{r^2}\Psi_\ell\bigg)=0\;,\\
\label{EPpppL3}
&\d_r^2\Psi_\ell+\frac{2}{r}\d_r\Psi_\ell-\frac{\ell(\ell+1)}{r^2}\Psi_\ell
=\Theta_\ell\;, 
\end{align}
\eseq
where 
\be
\O_{m,\eta}\equiv\frac{\O_m}{\O_m+\O_\L a^3(\eta)}
\ee
is the time-dependent matter density fraction and the relations
between $\Psi$, $\T$ and the fluid velocity is now modified,
\be
\d_i\Psi=-\frac{u_i}{\tilde f\HH}\;,\qquad 
\T=-\frac{\d_i u_i}{\tilde f\HH}\;.
\ee
Equations (\ref{EPpppL}) differ from the EdS case only in the second
and third terms in (\ref{EPpppL2}). 
The departures are captured by the fraction $\O_{m,\eta}/\tilde f^2$, 
which is displayed in Fig.~\ref{fig:growth}. One observes that this
fraction is quite close to 1 (its value in the EdS universe) until
a very recent epoch.

Proceeding along the lines of Sec.~\ref{2ndorder}
it is straightforward to derive the equations that replace (\ref{mu2r2})
in the $\L$CDM case,
\bseq
\label{mustarandr2eqsL}
\begin{align}
&\dot\mu^{(2)}+\dot r_\eta^{(2)}
\hat r_\eta^2\big(1+\hat\delta(\hat r_\eta)\big)
+r_\eta^{(2)} \frac{d}{d\eta}\Big(
\hat r_\eta^2\big(1+\hat\delta(\hat r_\eta)\big)\Big)
=\hat r_\eta^2 \Upsilon_\delta(\hat r_\eta)\;,\\
&\ddot r^{(2)}_\eta+\left(\frac{3\O_{m,\eta}}{2\tilde f^2}-1\right)\dot r^{(2)}_\eta
+\frac{\O_{m,\eta}}{\tilde f^2}
\bigg(1+\frac{3}{2}\hat\delta(\hat r_\eta)
-\frac{\hat R_*^3}{\hat r_\eta^3}\bigg) r^{(2)}_\eta
+\frac{3\O_{m,\eta}}{2\tilde f^2\hat r_\eta^2}\mu^{(2)}
=-\Upsilon_\Theta(\hat r_\eta)\;.
\end{align}
\eseq
Again, the only difference from the EdS equations is the  
presence of factors $\O_{m,\eta}/\tilde f^2$.

We have computed  the aspherical prefactor 
for the exact reference cosmological model using Eqs.~(\ref{EPpppL}),
(\ref{mustarandr2eqsL}) and found that its deviation from the EdS
approximation remains at sub-per cent level.

\section{Regularization of the WKB integral}
\label{app:WKB}

\subsection{Boundary term in the WKB integral}
\label{app:BT}
Here we derive Eq.~(\ref{qint}). The linear relation
(\ref{linsource}) implies that it is sufficient to prove the corresponding
formulae for the $\vk$-integrals of the sources 
$\Upsilon(\hat r_\eta)$. To be concrete, let us focus on the source
$\Upsilon_\delta$, the reasoning for $\Upsilon_\Theta$ is the same. 
We start with the asymptotic expression
for the Bessel function in the vicinity of
the turning point, Eq.~(10.19.8) from \cite{DLMF},
\be
J_{\nu}(\nu+a\nu^{1/3})=\frac{2^{1/3}}{\nu^{1/3}}{\rm Ai}(-2^{1/3}a)+O(1/\nu)\;,
\ee
where ${\rm Ai(z)}$ is the Airy function. Comparing it with
(\ref{linit}) and recalling the definition of spherical Bessel
functions (\ref{jJ}) we find the initial conditions for the perturbations at
$\eta\to -\infty$ in the vicinity of the point $\vk r=1$,
\bseq
\label{initAi}
\begin{align}
\label{initAi1}
&\delta_\ell=\Theta_\ell=\frac{\e^\eta}{(\ell+1/2)^{5/6}}\cdot(2\pi)^{3/2}2^{1/3}
{\rm Ai}\big[-2^{1/3}(\ell+1/2)^{2/3}(\vk r-1)\big]\;,\\
\label{initAi2}
&\Psi_l=-\frac{\e^\eta}{(\ell+1/2)^{17/6}}\cdot\frac{(2\pi)^{3/2}2^{1/3}}{\vk^2}
{\rm Ai}\big[-2^{1/3}(\ell+1/2)^{2/3}(\vk r-1)\big]\;.
\end{align}
\eseq
From these expressions we infer the scaling of the fields and their
derivatives at all moments of time,
\be
\label{derPsil}
\delta_\ell,\Theta_\ell=O(\ell^{-5/6})~,~~~~\Psi_\ell=O(\ell^{-17/6})~,~~~~
\d_r\Psi_\ell=O(\ell^{-13/6})~,~~~~\d^2_r\Psi_\ell=O(\ell^{-3/2})\;.
\ee
This implies that the terms with derivatives in the Poisson equation
(\ref{EPppp3}) are subdominant 
compared to the last term on the l.h.s., so that the
relation between $\Psi_\ell$ and $\Theta_\ell$ becomes very simple,
\be
\label{PsiThetAi}
\Psi_\ell=-\frac{r^2}{(\ell+1/2)^2}\Theta_\ell\;.
\ee
Inspecting the magnitude of various terms in Eqs.~(\ref{EPppp1}),
(\ref{EPppp2}) we find that they also simplify, 
\bseq
\label{Aidyn}
\begin{align}
\label{Aidyn1}
&\frac{d\delta_\ell}{d\eta}\bigg|_{\rm flow}-\hat\Theta\delta_\ell-
(1+\hat\delta)\Theta_\ell=0\;,\\
\label{Aidyn2}
&\frac{d\Theta_\ell}{d\eta}\bigg|_{\rm flow}-\frac{3}{2}\delta_\ell
+\bigg(\frac{1}{2}-2\frac{\d_r\hat\Psi}{r}\bigg)\Theta_\ell=0\;.
\end{align}
\eseq
Similarly to Eqs.~(\ref{WKBNLO}) they have the ultralocal form (no
$r$-derivatives of the perturbations), so that the fields evolve
independently along different flow lines. 
Besides, in the small vicinity of the `turning flow line' 
$R= 1/\vk$ the background functions can be considered as
$r$-independent\footnote{Of course, they still have a non-trivial time
  dependence that must be taken into account.}. Thus, in the comoving
frame the
time-evolution of the perturbations factorizes from their spatial
dependence, and we obtain,
\bseq
\label{evolAi}
\begin{align}
\label{evolAi1}
&\delta_\ell=\frac{\alpha(\eta)}{(\ell+1/2)^{5/6}}\cdot(2\pi)^{3/2}2^{1/3}
{\rm Ai}\big[-2^{1/3}(\ell+1/2)^{2/3}(\vk R-1)\big];,\\
\label{evolAi2}
&\Theta_\ell=\frac{\beta(\eta)}{(\ell+1/2)^{5/6}}\cdot(2\pi)^{3/2}2^{1/3}
{\rm Ai}\big[-2^{1/3}(\ell+1/2)^{2/3}(\vk R-1)\big];,\\
\label{evolAi3}
&\Psi_\ell=\frac{\gamma(\eta)}{(\ell+1/2)^{17/6}}\cdot\frac{(2\pi)^{3/2}2^{1/3}}{\vk^2}
{\rm Ai}\big[-2^{1/3}(\ell+1/2)^{2/3}(\vk R-1)\big]\;,
\end{align}
\eseq
where $\alpha,\beta,\gamma$ are some functions whose precise form 
is not important to us\footnote{It follows from (\ref{PsiThetAi}) that
  $\gamma(\eta)=-\vk^2 \beta(\eta) r^2(\eta,R=1/\vk)$.}.

Let us evaluate the following integral:
\be
\label{Upsint}
\begin{split}
&\int_{\frac{1-\epsilon}{R_*}}^{\frac{1+\epsilon}{R_*}}
d\vk\,\varphi(\vk)\Upsilon_\delta(\hat r_\eta,\eta; \vk)=
-\frac{1}{4\pi}\frac{\alpha(\eta)\gamma(\eta)}{(\ell+1/2)^3}\cdot
2(2\pi)^3\frac{\d R}{\d r}\bigg|_{\hat r_\eta}\\
&\times\int_{\frac{1-\epsilon}{R_*}}^{\frac{1+\epsilon}{R_*}}
d\vk\frac{\varphi(\vk)}{\vk}
{\rm Ai}\big[-2^{1/3}(\ell+1/2)^{2/3}(\vk R_*-1)\big]
{\rm Ai}'\big[-2^{1/3}(\ell+1/2)^{2/3}(\vk R_*-1)\big]\\
&=\frac{1}{8\pi}\frac{\alpha(\eta)\gamma(\eta)}{(\ell+1/2)^{11/3}}\cdot
2^{2/3}(2\pi)^3\frac{\d R}{\d r}\bigg|_{\hat r_\eta}
\varphi(1/R_*) \Big[{\rm
  Ai}\big(-2^{1/3}(\ell+1/2)^{2/3}\epsilon\big)\Big]^{2}\\
&=\frac{1}{8\pi}\frac{\d R}{\d r}\bigg|_{\hat r_\eta}
\frac{\varphi(1/R_*)}{R_*^2}
\;\delta_\ell\,\Psi_\ell\Big|_{\hat r_\eta,\,\vk=(1+\epsilon)/R_*}\;,
\end{split}
\ee 
where we have substituted Eq.~(\ref{Upsd}) and 
in passing to the third line used that the Airy function is
exponentially suppressed at positive values of its argument. At 
$\ell^{-2/3}\ll \epsilon \ll 1$ we can use the WKB form
(\ref{WKBform}) for $\delta_\ell$ and $\Psi_\ell$. Substituting it into
the last expression in (\ref{Upsint}) and averaging away the oscillating
pieces\footnote{These pieces, if kept, would cancel the integral over 
  $\vk>(1+\epsilon)/R_*$ of the  
  oscillating part of $\Upsilon_\delta$, which we neglect anyway.} we obtain,
\be
\label{Upsint1}
\int_{(1-\epsilon)/R_*}^{(1+\epsilon)/R_{*}}
d\vk\,\varphi(\vk)\Upsilon_\delta(\hat r_\eta, \eta; \vk)=
\frac{1}{4\pi k^2}
\frac{\varphi(1/R_{*})}{R_{*}^2}
\frac{\d R}{\d r}\bigg|_{\hat r_\eta}
\;\tilde\delta_{\ell 1}\tilde\Psi_{\ell 1}
\Big|_{\hat r_\eta,\,\vk=(1+\epsilon)/R_*}
\;.
\ee
Using the initial conditions 
(\ref{dl1init}), (\ref{dl2init}) and the ultralocality of the
evolution in the vicinity of the turning point it is straightforward
to show that
\bseq
\begin{align}
&\tilde\delta_{\ell 1}\tilde\Psi_{\ell 1}\Big|_{\hat r_\eta,\;
  \vk=(1+\epsilon)/R_*}=-\frac{4 R_*}{\sqrt{\epsilon}}
\lim_{\epsilon'\to
  0}(\epsilon')^{3/2}\tilde\delta_{\ell 1}\frac{\d\tilde\Psi_{\ell
    1}}{\d R}
\Big|_{\hat r_\eta,\;\vk=(1+\epsilon')/R_*}\;,\\
&(\tilde\delta_{\ell 1}\tilde\Psi_{\ell 2}-\tilde\delta_{\ell
  2}\tilde\Psi_{\ell 1})\Big|_{\hat r_\eta,\;\vk=(1+\epsilon)/R_*}
=O(\epsilon^{-1})\;.
\end{align}
\eseq
The latter combination appears in $\Upsilon_\delta$ multiplied by
$\frac{\d S_\ell}{\d R}$, see Eq.~(\ref{UpsdWKB}). Taking into account
the expression
\be
\frac{\d S_\ell}{\d R}=\frac{\sqrt{(\vk R)^2-1}}{\vk R}=O(\sqrt{\epsilon})\;,
\ee
one concludes that the corresponding term does not contribute into 
$\lim_{\epsilon\to 0}\epsilon^{3/2}\Upsilon_\delta$.
Then Eq.~(\ref{Upsint1}) can be cast into the form,
\be
\label{Upsint2}
\int_{(1-\epsilon)/R_*}^{(1+\epsilon)/R_*}
d\vk\,\varphi(\vk)\Upsilon_\delta(\hat r_\eta, \eta; \vk)=
-\frac{2\varphi(1/R_*)}{R_*\sqrt{\epsilon}}
\lim_{\epsilon'\to 0}(\epsilon')^{3/2}
\Upsilon_\delta\big(\hat r_\eta, \eta;\vk=(1+\epsilon')/R_*\big)\;.
\ee
This proves the expression of the type (\ref{qint}) for the integrals
involving $\Upsilon_\delta$. The argument for the integrals involving
$\Upsilon_\Theta$ is completely analogous. This completes the
derivation of Eq.~(\ref{qint}).

\subsection{Evaluation of the $\vk$-integral}
\label{app:integral}
Numerical evaluation of the WKB integral (\ref{qint}) presents a
non-trivial challenge. Indeed, in the limit $\epsilon\to 0$ 
the expression on the r.h.s. contains a
difference between two large numbers that must be evaluated with very
high accuracy, which may be impractical. 
On the
other hand, at finite values of $\epsilon$ the error is estimated
as $O(\sqrt{\epsilon})$. Thus, to reach an acceptable level of
accuracy of, say, $1\%$ one would have to go down to $\epsilon\sim
10^{-4}$. To improve the convergence of the numerical procedure,
we derive here an expression that explicitly takes into account the
$O(\sqrt{\epsilon})$ corrections.

Let us introduce dimensionless variables\footnote{There should be no
  confusion with different uses of the notations $x$, $y$ in other
  sections, as the alternative usage does not appear in this appendix.},
\be
\label{xy}
x=\vk R_*-1~,~~~~~ y=R/R_*-1\;.
\ee
At small $x$ we have
\be
\label{qfsmallx}
q(x)=q_{3\over 2}\, x^{-3/2}+q_{1\over 2}\, x^{-1/2} +O(\sqrt{x})~,~~~~~
\varphi(x)=\varphi_0+\varphi_1\,x+O(x^2)\;,
\ee
And the integral takes the form,
\be
\label{dashint}
\dashint d\vk\,q(\vk)\varphi(\vk)=\frac{1}{R_*}\bigg[
-\frac{2q_{3\over 2}\varphi_0}{\sqrt{\epsilon}}+
2(q_{3\over 2}\varphi_1+q_{1\over 2}\varphi_0)\sqrt{\epsilon}+
\int_\epsilon^\infty dx\, q(x)\varphi(x)+O(\epsilon^{3/2})\bigg]\;.
\ee
We expand the sources $\Upsilon_i$, $i=\delta,\T$ in a similar way,
\be
\label{Upssmallx}
\Upsilon_i(\hat r_\eta)
=\frac{1}{k^2(\ell+1/2)^2}\big(\Upsilon_{i,{3\over 2}}\,x^{-3/2} 
+\Upsilon_{i,{1\over 2}}\,x^{-1/2}+O(\sqrt{x})\big)\;.
\ee
Substituting this into (\ref{linsource}) and comparing with (\ref{Qq})
we find,
\bseq
\label{qUps*}
\begin{align}
\label{qUps3/2}
&q_{3\over 2}=\int_{-\infty}^0d\eta\;
\big(K_\delta(\eta)\Upsilon_{\delta,{3\over2}}(\eta)
+K_\Theta(\eta)\Upsilon_{\T,{3\over2}}(\eta)
\big)\;,\\
\label{qUps1/2}
&q_{1\over 2}=\int_{-\infty}^0d\eta\;
\big(K_\delta(\eta)\Upsilon_{\delta,{1\over2}}(\eta)
+K_\Theta(\eta)\Upsilon_{\T,{1\over2}}(\eta)
\big)\;.
\end{align}
\eseq
Here the kernels $K_i$ as well as the sources 
$\Upsilon_{i,{3\over 2}}$, 
$\Upsilon_{i,{1\over 2}}$
are regular functions of time (and are independent of $x$), so that 
$q_{3\over 2}$, $q_{1\over 2}$ can be obtained by a straightforward
numerical integration of eqs.~(\ref{mu2r2}) with the
corresponding sources.  
Our task is to derive expressions for $\Upsilon_{i,{3\over 2}}$, 
$\Upsilon_{i,{1\over 2}}$.

For $y\ll x\ll 1$ we write,
\bseq
\label{dTl1}
\begin{align}
\tilde\delta_{\ell 1}(\eta,y;x)=\big(\a_0(\eta)+\a_1(\eta)x+\a_2(\eta)y
+O(x^2,xy)\big)\,\tilde\delta_{\ell 1}^{\rm sing}(y;x)\;,\\
\tilde\T_{\ell 1}(\eta,y;x)=\big(\b_0(\eta)+\b_1(\eta)x+\b_2(\eta)y
+O(x^2,xy)\big)\,\tilde\delta_{\ell 1}^{\rm sing}(y;x)\;.
\end{align}
\eseq
where
\be
\label{deltal10}
\begin{split}
\tilde\delta_{\ell 1}^{\rm sing}(y;x)&\equiv\frac{2\pi}{(\ell+1/2)}
\cdot\frac{1}{\sqrt{\vk R}\,[(\vk R)^2-1]^{1/4}}\\
&=\frac{2\pi}{(\ell+1/2)}\cdot\frac{1}{2^{1/4}x^{1/4}}
\bigg[1-\frac{5}{8}x-y\bigg(\frac{1}{4x}+\frac{23}{32}\bigg)\bigg]\;.
\end{split}
\ee
In the last expression we expanded to the first subleading order in
$x$ and kept only up to linear order in
$y$. This is sufficient for our purposes as 
we will only need the values of the fields and their 
first derivatives at $y=0$. The functions $\a_0(\eta)$ etc. can be
found by integrating Eqs.~(\ref{WKBNLO}) in the vicinity of the point
$x=y=0$ with smooth $x$- and $y$-independent initial conditions
$\tilde \delta_{\ell 1},\tilde\T_{\ell 1}=\e^{\eta}$ at $\eta\to-\infty$.
From (\ref{dTl1}) one 
reads off the values of the fields at $y=0$,
\bseq
\label{dTl1y0}
\begin{align}
&\tilde\delta_{\ell 1}(\eta,0;x)=\frac{2\pi}{(\ell+1/2)}\cdot\frac{1}{2^{1/4}x^{1/4}}
\bigg[\a_0+x\,\bigg(-\frac{5}{8}\a_0+\a_1\bigg)\bigg]\;,\\
&\tilde\T_{\ell 1}(\eta,0;x)=\frac{2\pi}{(\ell+1/2)}\cdot\frac{1}{2^{1/4}x^{1/4}}
\bigg[\b_0+x\,\bigg(-\frac{5}{8}\b_0+\b_1\bigg)\bigg]\;.
\end{align}
\eseq
Next, we have,
\be
\label{S1S2}
\frac{\d S_\ell}{\d r}\Big|_{y=0}=\frac{\d R}{\d r}\Big|_{\hat
  r_\eta}\cdot\sqrt{2x} +O(x^{3/2})\;,~~~~
\frac{\d^2 S_\ell}{\d r^2}\Big|_{y=0}=\frac{1}{R_{*}}\cdot
\bigg(\frac{\d R}{\d r}\bigg)^2\Big|_{\hat
  r_\eta}\cdot\frac{1}{\sqrt{2x}} +O(x^{1/2})\;.
\ee
Substituting this into (\ref{PsiLO}) we obtain,
\bseq
\label{Psil1*}
\begin{align}
\label{Psil11}
\tilde \Psi_{\ell 1}\Big|_{y=0}=&-\frac{2\pi \hat r^2_\eta}{(\ell+1/2)
  R_*^2}
\cdot\frac{1}{2^{1/4}x^{1/4}}\bigg[\b_0+x\,
\bigg(\frac{11\b_0}{8}-2\b_0\bigg(\frac{\d\ln r_{in}}{\d\ln
  r}\bigg)^2
+\b_1\bigg)\bigg]\;,\\
\tilde\Psi_{\ell 1}'\Big|_{y=0}=&-\frac{2\pi \hat r^2_\eta}{(\ell+1/2)
  R_*^3}\frac{\d R}{\d r}\cdot\frac{1}{2^{1/4}x^{5/4}}
\bigg[-\frac{\b_0}{4}\notag\\
&\qquad+x\,\bigg(-\frac{39}{32}\b_0
-\frac{3\b_0}{2}\bigg(\frac{\d\ln R}{\d\ln r}\bigg)^2
+2\b_0\bigg(\frac{\d\ln R}{\d\ln r}\bigg)^{-1}
-\frac{\b_1}{4}+\b_2\bigg)\bigg]\;.
\label{Psil12}
\end{align}
\eseq

We now show that the second-order WKB perturbations do not contribute
at the order we are interested in. First, we demonstrate that the r.h.s. of
Eqs.~(\ref{WKBNNLO}) evaluated at $y=0$ is of order
$O(x^{1/4})$. Indeed, using (\ref{S1S2}), (\ref{Psil11}) we find for
Eq.~(\ref{WKBNNLO1}), 
\be
S_\ell'\Psi_{\ell 1}\Big|_{y=0}\sim \sqrt{x}\cdot x^{-1/4} \sim x^{1/4}\;.
\ee
On the r.h.s. of (\ref{WKBNNLO2}) the only term that can potentially
be of order $O(x^{-3/4})$ has the form,
\be
-\frac{4S_\ell'\tilde\Psi_{\ell 1}'+2S_l''\tilde\Psi_{\ell 1}}{1+(\vk rS_\ell')^2}
\bigg(\d^2_r\hat\Psi-\frac{\d\hat\Psi}{r}\bigg)\bigg|_{y=0}\;.
\ee
However, from (\ref{S1S2}), (\ref{Psil1*}) we find that the divergent
terms cancel out, so that
\be
\label{Psicancel}
(2S_\ell'\tilde\Psi_{\ell 1}'+S_l''\tilde\Psi_{\ell 1})\Big|_{y=0}=O(x^{1/4})\;.
\ee
Thus, working only up to order $O(x^{-3/4})$ we can neglect the r.h.s. in
Eqs.~(\ref{WKBNNLO}) and write,
\bseq
\label{dTl2}
\begin{align}
&\tilde\delta_{\ell 2}(\eta,0;x)=\big(\a_0(\eta)+\a_1(\eta)x\big)
\,\tilde\delta_{\ell 2}^{\rm sing}(0;x)\;,\\
&\tilde\T_{\ell 2}(\eta,0;x)=\big(\b_0(\eta)+\b_1(\eta)x\big)
\,\tilde\delta_{\ell 2}^{\rm sing}(0;x)\;,
\end{align}
\eseq
where
\be
\begin{split}
\tilde\delta_{\ell 2}^{\rm sing}(0;x)\equiv&\frac{\pi}{4(\ell+1/2)}
\cdot\sqrt{\frac{\vk}{R_*}}\,
\bigg(\frac{5}{3[(\vk R_*)^2-1]^{7/4}}+
\frac{1}{[(\vk R_*)^2-1]^{3/4}}\bigg)\\
=&\frac{\pi}{4(\ell+1/2)R_*}\cdot\frac{1}{2^{7/4}x^{7/4}}
\bigg[\frac{5}{3}+\frac{11}{8}x+O(x^2)\bigg]\;.
\end{split}
\ee
Next, from (\ref{PsiNLO}) and using (\ref{Psicancel}) we get,
\be
\tilde \Psi_{\ell 2}\Big|_{y=0}=-\frac{\tilde\T_{\ell 2}}{(S_\ell')^2+(\vk
  r)^{-2}}\Big|_{y=0}
+O(x^{1/4})\;.
\ee
Then the term in the source
$\Upsilon_\delta$ containing second-order WKB perturbations is (see
Eq.~(\ref{UpsdWKB})), 
\be
S_\ell'(\tilde\delta_{\ell 1}\tilde\Psi_{\ell 2}-\tilde\delta_{\ell 2}\tilde\Psi_{\ell 1})\Big|_{y=0}
=S_\ell'\;\frac{-\tilde\delta_{\ell 1}\tilde\T_{\ell 2}
+\tilde\delta_{\ell 2}\tilde\T_{\ell 1}}{(S_\ell')^2+(\vk
r)^{-2}}\bigg|_{y=0}+O(\sqrt x)=O(\sqrt x)\;,
\ee
where in the last equality we have used the expressions (\ref{dTl1}),
(\ref{dTl2}). Similarly, one shows that the second-order WKB
contribution in $\Upsilon_\T$ is also of order $O(\sqrt{x})$. Given
that we are keeping only terms up to order $O(x^{-1/2})$, we conclude
that the second-order WKB contributions can be omitted altogether.

It remains to substitute the expressions (\ref{dTl1y0}),
(\ref{Psil1*}) into (\ref{UpsWKB}). A straightforward calculation
yields,
\bseq
\begin{align}
\Upsilon_{\delta,{3\over 2}}=&\frac{\pi \hat r^2_\eta}{2\sqrt{2} R_*^3}
\frac{\d R}{\d r}\bigg|_{y=0}\a_0(\eta)\b_0(\eta)\;,\\
\Upsilon_{\T,{3\over 2}}=&-\frac{\pi \hat r^2_\eta}{2\sqrt{2} R_*^3}
\frac{\d R}{\d r}\bigg|_{y=0}\b_0^2(\eta)\;,\\
\Upsilon_{\delta,{1\over 2}}=&\frac{\sqrt{2}\pi \hat r^2_\eta}{R_*^3}
\frac{\d R}{\d r}\bigg|_{y=0}\bigg[\a_0\b_0
\bigg(\frac{17}{16}+\frac{3}{2}\bigg(\frac{\d\ln R}{\d\ln
  r}\bigg)^2
-2\bigg(\frac{\d\ln R}{\d\ln
  r}\bigg)^{-1}\bigg)\notag\\
&\qquad\qquad\qquad\qquad
+\frac{1}{4}(\a_0\b_1+\a_1\b_0)-\a_0\b_2\bigg]\;,\\
\Upsilon_{\T,{1\over 2}}=&\frac{\sqrt{2}\pi \hat r^2_\eta}{R_*^3}
\frac{\d R}{\d r}\bigg|_{y=0}\bigg[\b_0^2
\bigg(-\frac{17}{16}-\frac{1}{2}\bigg(\frac{\d\ln R}{\d\ln
  r}\bigg)^2
+\bigg(\frac{\d\ln R}{\d\ln
  r}\bigg)^{-1}\bigg)
-\frac{\b_0\b_1}{2}+\b_0\b_2\bigg]\;.
\end{align}
\eseq
These are the final expressions for the sources to be used in
Eqs.~(\ref{qUps*}).

\section{Numerical procedure}
\label{sec:aspy}

In this Appendix we discuss the details of our numerical method, 
which is implemented in the open-source 
code \texttt{AsPy} \cite{AsPy} written in \texttt{Python} using
\texttt{scipy} and  
\texttt{numpy} libraries. 

We first cast the partial differential equations~\eqref{EPppp} in the
form suitable for numerical solution using
finite differences.  
In this section we will omit the subscript $\ell$ denoting linear
aspherical perturbations. 
We switch to the Lagrangian coordinate $R$ comoving with the
background flow, which  
allows us to absorb the shift terms into the time derivative,
\be
\frac{\d }{\d \eta}-\d_r\hat{\Psi}\frac{\d}{\d r}\equiv  \frac{d }{d
  \eta}\bigg|_{\text{flow}}\,. 
\ee 
Equations (\ref{EPppp1}), (\ref{EPppp2}) take the form,
\bseq
\label{eqsFD}
\begin{align}
& \frac{d\delta}{d\eta} =A_1(\eta,R)  \delta 
+A_2(\eta,R)\T+A_3(\eta,R)\d_{R}\Psi\,,\\
& \frac{d\T}{d\eta}=\frac{3}{2}\delta +A_4(\eta,R)  \T 
+A_5(\eta,R)\d_{R}\Psi+
\ell(\ell+1)A_6(\eta,R)\Psi \,,
\end{align}
\eseq
where we defined the following background functions:
\begin{subequations}
\label{eq:backgroundFD}
\begin{align}
&A_1=\hat{\T}\,,&& A_4 = -\frac{1}{2} + 2\left(\hat{\T}-\frac{2}{r}\d_r\hat{\Psi}\right)\,, \\
&A_2 = 1+\hat{\delta}\,,
&& A_5 =\frac{1}{\frac{\d r}{\d R}}\left(\frac{1}{\frac{\d r}{\d R}}\d_{R}\hat{\T} - \frac{4}{r}\left(\hat{\T}-3\frac{\d_r\hat{\Psi}}{r}\right)\right) \,, \\
&A_3 =\frac{1}{\left(\frac{\d r}{\d R}\right)^{2}}\d_{R}\hat{\delta}\,,
&&
A_6 = 2\left(\hat{\T}-3\frac{\d_r\hat{\Psi}}{r}\right)\frac{1}{r^2}\,.
\end{align}
\end{subequations}
The initial conditions for the density and velocity fields 
are given by Eqs.~\eqref{linit1}.
Note that Eqs.~(\ref{eqsFD}) do not contain spatial derivatives of
$\delta$ or $\T$, so we do not need to impose any 
boundary conditions on them.

The Euler and continuity equations are supplemented by the Poisson equation
\be 
\label{EqF4}
 \left(\d^2_r+\frac{2\d_r}{r}-\frac{\ell(\ell+1)}{r^2}\right)\Psi(\eta,R)
=\T(\eta,R)\,. 
\ee
The boundary conditions for the velocity potential
are given by,
\be
\begin{split}
&\Psi(\eta,R)\propto r^\ell (R)\,,\quad \text{at} \quad R\to 0\\
& \Psi(\eta,R_{max})=\e^{(\eta-\eta_{min})} \Psi(\eta_{min},R_{max})\,.
\end{split}
\ee
The boundary condition at the origin is dictated by the structure of
the Poisson  
equation~\eqref{EPppp3}.
The second condition comes from the assumption that at spatial
infinity the velocity potential follows the linear evolution, which is
justified since 
the background profile 
falls off quickly outside the window
function. 

We work on an equally-spaced rectangular lattice with $N_R\times N_\eta$
nodes and physical size
$[R_{min},R_{max}]\times[\eta_{min},0]$. 
We implement an implicit second-order Runge-Kutta scheme (RK2) for 
the Euler and continuity equations.
For the Poisson equation we use an implicit second-order finite
difference scheme. 

We use the discrete version of the fluctuation operator obtained by 
rewriting the integrals in the exponent of \eqref{Al1} in a discrete form 
and taking the corresponding Gaussian integral.
This yields,
\be
\label{discr}
\mathcal{O}_\ell=\delta_{ij} + 2\hat \lambda  \frac{\Delta k}{(2\pi)^3} k_ik_j Q_{\ell}(k_i,k_j)\sqrt{P(k_i)P(k_j)} \,,
\ee
with $i,j=0,\ldots, N$; $\Delta k = (k_N-k_0)/N$. 
One can check that this definition 
gives the correct continuous limit for the 
trace\footnote{We do not assign the weight $1/2$ to the boundary
  values, but choose IR and UV cutoffs to make 
sure that the results are independent of them.}:
\be
\label{tracemeas}
 \sum_{i=0}^{N}\frac{\Delta k k^2_i}{(2\pi)^3}\,Q_\ell(k_i,k_i)P(k_i)
\xrightarrow{N\to \infty} \int_0^\infty [dk]\, Q_\ell(k,k)P(k)=\Tr\, Q_\ell P\;. 
\ee

We implement the algorithm for computing the aspherical determinant
from Sec.~\ref{sec:algorithm}.
At the first step the code computes the background functions \eqref{eq:backgroundFD}
required for solving the fluid equations on the grid. 
To this end we make a sample of $\sim 20$
values of $\delta_*$ in the range $[-0.9,9]$ and use the spherical
collapse linear profile \eqref{eq:deltahatlinprofile} 
to compute the non-linear background configuration 
defined by the equations from Sec.~\ref{sec:back}.

At the second step we sample the momentum space and compute the evolution
of linear fluctuations given by the finite difference approximation to the equations
\eqref{eqsFD}, (\ref{EqF4})
with appropriate initial
and boundary conditions
for each momentum $k_i$ from the sample.
We use the sample of $N=200$ wavenumbers which we found sufficient for
our purposes.

We found that the following grid parameters lead to a good convergence
for most of the multipoles
in the $\delta_*$-range of interest: 
\be
\begin{split}
& R_{min}=10^{-2}\,\text{Mpc}/h\,,\quad  
R_{max}=10\cdot R_*\,,\quad N_R=1000\,,\\
&\eta_{min}=-7\,,\quad N_\eta=500\,.
\end{split}
\ee
For the dipole we increased the spatial extent of the grid to
$R_{max}=15\cdot R_*$, $N_R=1500$.
We have run several tests and found that increasing
the grid resolution further or moving the box boundaries 
can only change the final results at the $0.1\%$ level. 

At the third step we use the linear mode functions computed in step 2
to construct the sources $\Upsilon_{\Theta,\delta}$ for all different
pairs of momenta $(k_i,k_j)$, $i,j=0,\dots,N$.
For the dipole sector, we also construct the sources $B_{\Theta,\delta}$

At the fourth step we solve the finite difference equations for the
time evolution of  
$\mu^{(2)}$ and $r^{(2)}_\eta$ obtained by discretizing Eqs.~(\ref{mu2r2}). 
This yields the matrix $Q_{\ell}(k_i,k_j)$, which is used to obtain the
desired determinant. 
In the dipole sector we separately compute $\breve Q_{\ell}(k_i,k_j)$, 
$B(k_i)$, and $A$
using (\ref{mu2r2}) with the corresponding sources
\eqref{Upscoeff}. 
The final determinant is obtained upon multiplying the IR-sensitive determinant 
$\D_{IR}$
and the IR-safe determinant $\det(\mathbb{1}+2\hat\l 
\sqrt{P}\breve Q_{\ell}\sqrt{P})$.
As a cross-check, we computed $A$ both by solving Eqs.~(\ref{mu2r2})
and from the relation \eqref{detcancel}, which yielded results that
agree at the per mil level. 

\begin{figure}[h!]
\begin{center}
\includegraphics[width=.6\textwidth]{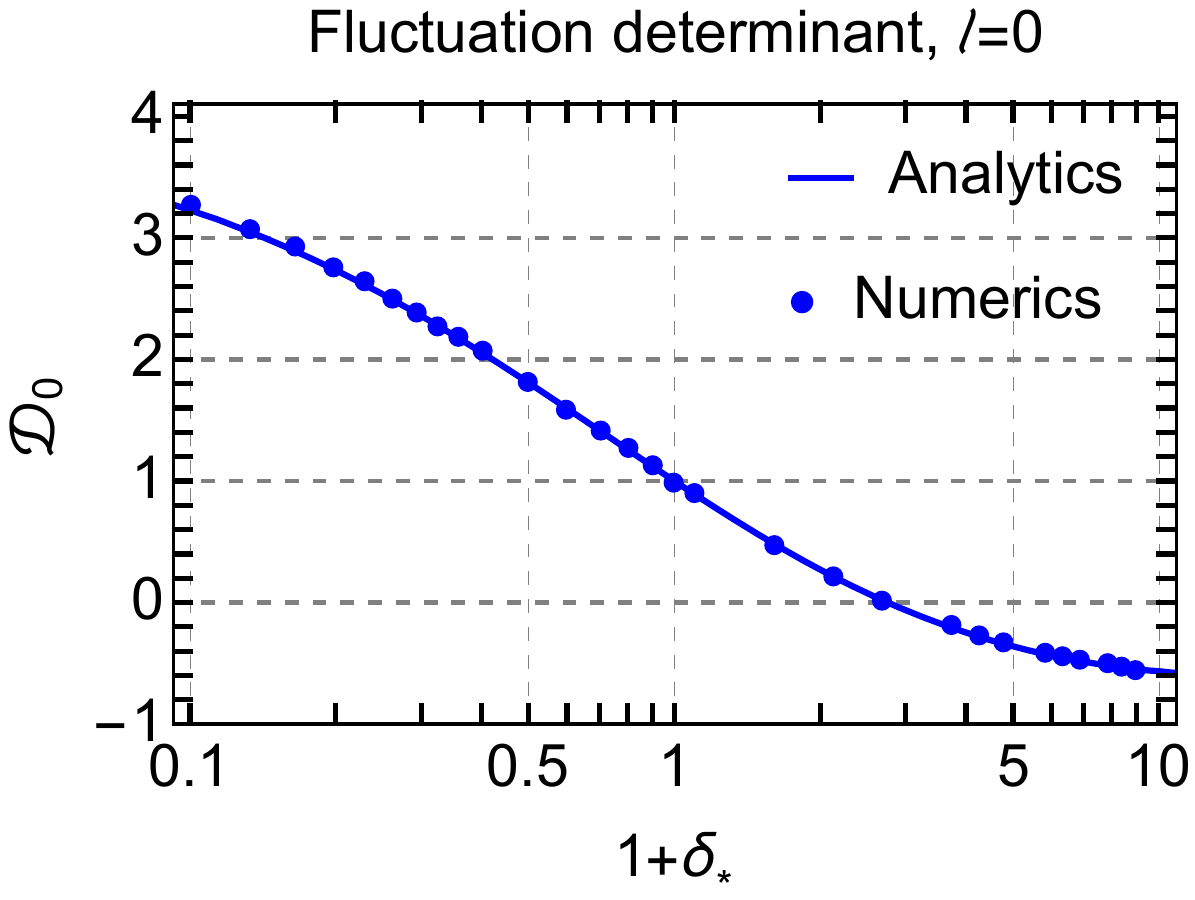}
\caption{\label{fig:l0} 
The fluctuation determinant in the monopole sector: 
the result of our numerical procedure (dots) vs.
Eq.~\eqref{eq:detl0} (line). 
}
\end{center}
\end{figure}

We have validated our code 
by computing the determinant of the monopole fluctuations and
comparing it to the analytic expression \eqref{eq:detl0}. The 
results of this test are displayed in Fig.~\ref{fig:l0}. Numerical
procedure agrees with the analytic formula at per mil
level. 

\section{A comment on log-normal model}
\label{app:ln}

\begin{figure}[t]
\begin{center}
\includegraphics[width=.46\textwidth]{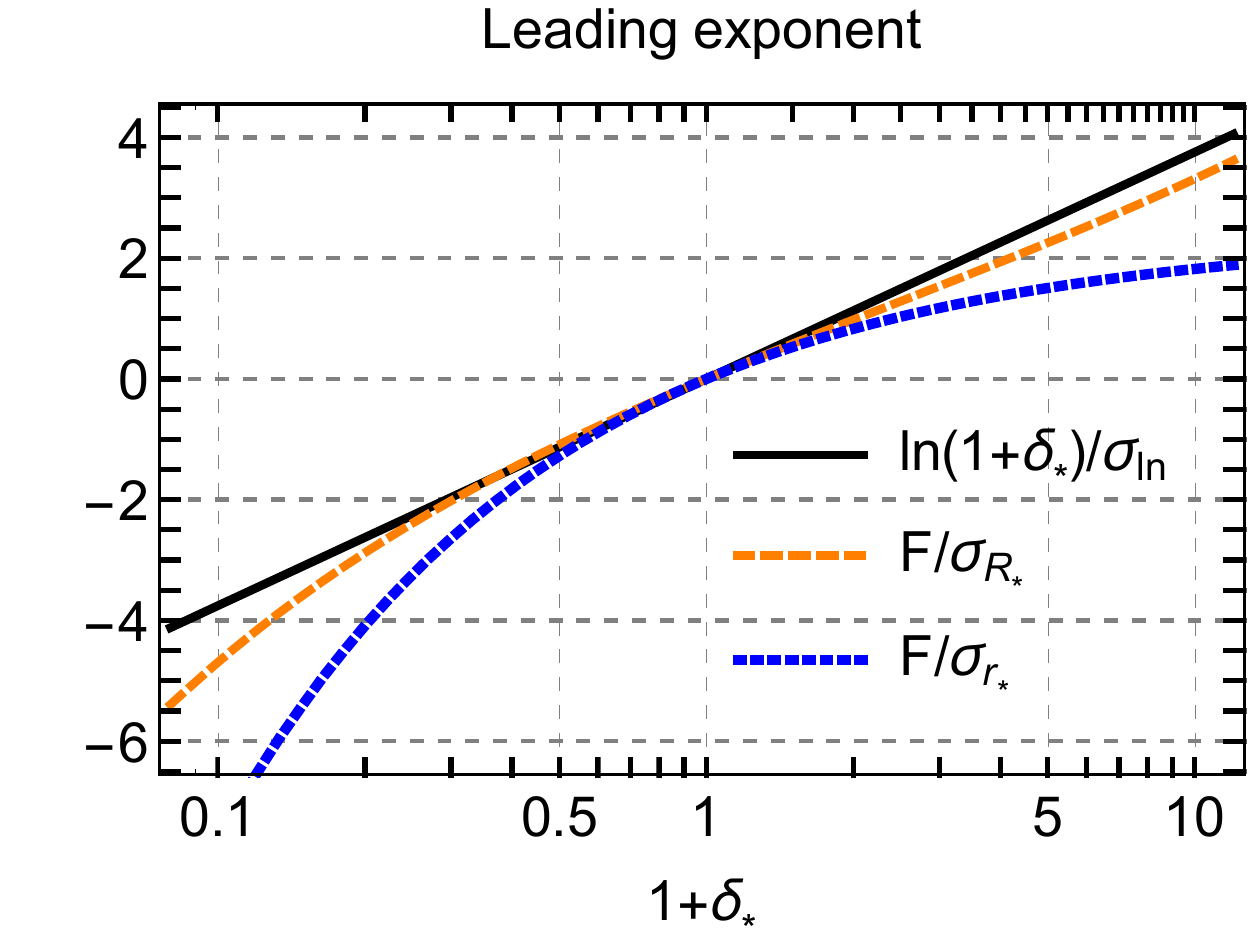}
\hspace{0.5cm}
\includegraphics[width=.49\textwidth]{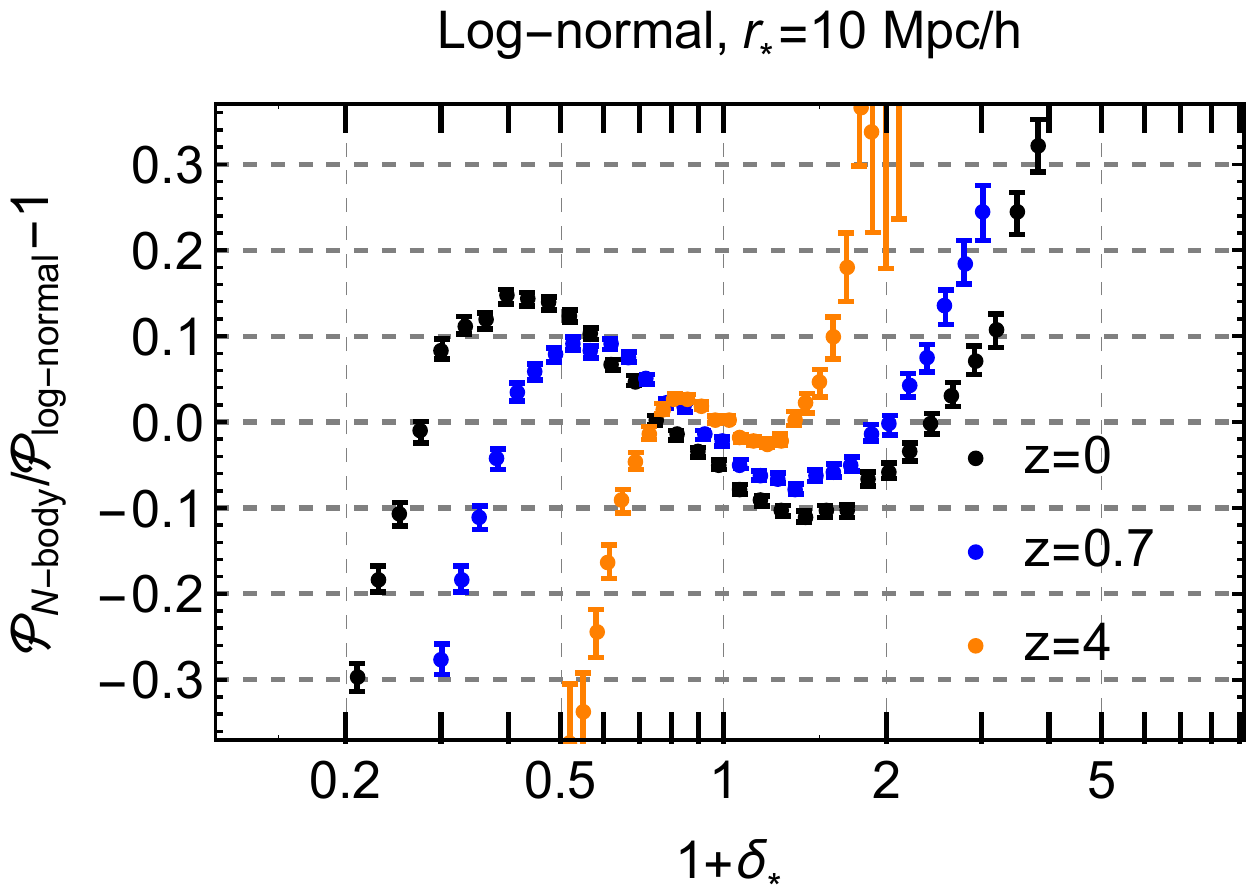}
\caption{\label{fig:lognorm} 
Left panel: 
The functions defining the leading exponential behavior of the
log-normal PDF \eqref{lognorm} and
 our theoretical PDF \eqref{leadingexp}. The log-variance
$\sigma_{\rm ln}$ is found from the fit to N-body data. Right panel:
Residuals of the N-body data with respect to the best fit log-normal
models at different redshifts. $\sigma_{\rm ln}$ is refitted for each
$z$ independently. The cell radius is $r_*=10\,\text{Mpc}/h$. 
} 
\end{center}
\end{figure}

It has long been known that the observed counts-in-cells distribution
can be well approximated by log-normal 
\cite{Coles:1991if,Kofman:1993mx,Kayo:2001gu,Wild:2004me},
\be
\label{lognorm}
\P_{\text{log-normal}}(\delta_*)=\frac{1}{\sqrt{2\pi \sigma^2_{\rm
      ln}}(1+\delta_*)} \exp\left\{
-\frac{\big(\ln(1+\delta_*)+\sigma^2_{\rm ln}/2\big)^2}{2\sigma^2_{\rm
  ln}}
\right\}\,,
\ee
where $\sigma_{\rm ln}^2=\langle[\ln(1+\delta_*)]^2\rangle$ 
is the log-density variance, to be fitted from the data.
The mean of the distribution (\ref{lognorm}) is adjusted to ensure
$\langle\delta_*\rangle=0$. The success of this model is partially due
to the fact that the spherical collapse mapping $F(\delta_*)$ is close
to $\ln(1+\delta_*)$ for moderate density contrasts, see the left
panel of 
Fig.~\ref{fig:lognorm}.
The difference grows for bigger $|\delta_*|$, but, curiously enough,
gets largely compensated by the scale dependence of
$\sigma_{R_*}$. This compensation is a mere coincidence due to the
shape of the power spectrum at mildly non-linear scales
\cite{Bernardeau:1994zd,Bernardeau:1994aq}. Indeed,
consider, for example, a universe with a power-law power spectrum
$P(k)\propto k^n$. In such a universe the variance scales as 
$\sigma_{R_*}^2\propto (1+\delta_*)^{-1-n/3}$, which clearly depends
on the slope $n$. On the other hand, spherical collapse mapping is
determined exclusively by dynamics and is insensitive to the
statistics of the initial conditions. One concludes that changing the
slope of the power spectrum would destroy the consipracy and the
log-normal model would fail.

Although the log-normal PDF gives a good leading order approximation,
it does not incorporate the correct prefactor. As a consequence, it
is unable to describe the data with the accuracy better than $\sim 10\%$
even at moderate densities and quickly deviates from the data in the
tails \cite{Uhlemann:2016liz,Klypin:2017jjg}. 
This is illustrated in the right panel
of Fig.~\ref{fig:lognorm} where we show the residuals of the N-body
data with respect to the best fit log-normal model.

The agreement between the log-normal model and the data can be improved  by artificially allowing the minimal value of $\delta_{*}$ to be different from $-1$. 
This was observed e.g. for the case of the projected density and convergence fields in Refs.~\cite{Hilbert:2011xq,Friedrich:2017qfe}.
However, allowing $\delta_{*,min}$ 
to be different from $-1$ appears hard to justify on the physical grounds.


\end{document}